\newcounter{toce}
\catcode`\@=11
\newif\if@fewtab\@fewtabtrue
{\count255=\time\divide\count255 by 60 \xdef\hourmin{\number\count255}
\multiply\count255 by-60\advance\count255 by\time
\xdef\hourmin{\hourmin:\ifnum\count255<10 0\fi\the\count255}}
\def\ps@draft{\let\@mkboth\@gobbletwo \def\@oddhead{}
    \def\@oddfoot{\hbox to 7 cm{\tiny \versionno
       \hfil}\hskip -7cm\hfil\rm\thepage \hfil {\tiny\draftdate}}
    \def\@evenhead{}\let\@evenfoot\@oddfoot}
\def\draftcite#1{\ifnum\draftcontrol=1#1\else{}\fi}
\newcommand\A[2]   {A^{#1#2}_{}}
\def\abar          {\mbox{$\bar\alpha$}}
\newcommand\ad[1]  {{\rm ad}_{#1}^{}}
\def\aff           {affine Lie algebra}
\def\ai            {\mbox{$\alpha^{(i)}$}}
\def\aI            {{\alpha^{(i)}}}
\def\aibar         {\mbox{$\bar\alpha^{(i)}$}}
\def\aiv           {\mbox{$\alpha^{(i)^\Vee}$}}
\def\Aiv           {\mbox{$a_i^\Vee$}}
\def\aim           {\mbox{$\alpa^i_I$}}
\def\aj            {\mbox{$\alpha^{(j)}$}}

\def\ajv           {\mbox{$\alpha^{(j)^\Vee}$}}
\newcommand\al[1]  {\mbox{$\alpha^{(#1)}$}}
\newcommand\alb[1] {\mbox{$\bar\alpha^{(#1)}$}}
\def\alg           {algebra}
\def\Alg           {Algebra}
\def\alpa          {\gamma}

\def\ao            {{\alpha^{(0)}}}
\def\aS            {\mbox{${\rm S}_X$}}
\def\asi           {apparent singularity}
\def\asis          {apparent singularities}
\newcommand\B[6]   {{\sf B}_{\sss#1#2}\Barray{#3#4}{#5#6}}

\newcommand\Barray[2] {[\!\!{\scriptstyle\begin{array}{c}{}\\[-1.88em]
                   {\scriptstyle #1}\\[-.57em]{\scriptstyle #2}\\[-.4em]
                   \end{array}}\!\!]}
\def\be            {\begin{equation}}
\def\bearl         {\begin{array}{l}}
\def\bearll        {\begin{array}{ll}}
\def\bearlll       {\begin{array}{lll}}
\def\bfe           {{\bf1}}
\newcommand\Bigoplus[1]{\mbox{{\large$\bigoplus$}\hsp{-.18}\raisebox{-.35em}
                   {$\scs#1$}\,}}
\newcommand\blec[1]{\\[.3em]{}${}$\hsp{.89}{\bf\lec#1}:}
\def\bll           {\mbox{$b_{\ilabel}$}}
\def\bullshit      {not entirely correct}
\def\bz            {{\bar z}}

\def\calf          {{\cal F}}

\def\calg          {{\cal G}}
\def\calgzz        {\mbox{${\cal G}\zbz$}}
\newcommand\calL[3]{\bar{\mbox{\sc n}}_{\sss\!#1#2}^{\sss\;\ #3}}
\def\caln          {{\cal N}}
\def\carm          {Cartan matrix}

\def\CARMS         {Cartan Matrices}
\def\ccft          {chiral conformal field theory}
\newcommand\cd[1]  {{\Delta_{#1}}}
\def\cD            {\Delta}
\newcommand\cdb[1] {{\bar\Delta_{#1}}}

\def\cdim          {conformal dimension}
\def\Cdot          {\!\cdot\!}
\def\cft           {conformal field theory}
\def\Cft           {Conformal field theory}
\def\CFT           {Conformal Field Theory}
\def\cfts          {conformal field theories}
\def\CFTS          {Conformal Field Theories}

\def\chI           {{\raisebox{.18em}{$\chi$}}}

\def\chii          {{\raisebox{.18em}{$\chi$}}}
\def\chil          {{\chI\raisebox{-.2em}{$\scriptstyle\Lambda$}}}
\def\chiv          {{\chI\raisebox{-.2em}{$\scriptstyle V$}}}
\def\chserre       {Cheval\-ley\hy Serre }
\def\chsa          {chiral symmetry algebra}
\def\CHSA          {Chiral Symmetry Algebra}
\def\CHV           {{\cal X}_\Lambda}
\def\cijd          {{\cal C}_{\!AB}^{\;\ D}}
\def\cijk          {{\cal C}_{\!AB}^{\ \ C}}
\def\cijK          {{\cal C}_{\!ABC}^{}}
\def\Circ          {\,{\circ}\,}
\newcommand\Cite[1]{\cite{#1}}
\def\class         {classification}
\def\complex       {{\dl C}}
\def\Complex       {$\complex$}
\def\con           {conformal }
\def\Con           {Conformal }
\newcommand\cont[1]{\vtop{\ialign{##\crcr$\hfil\displaystyle{#1}\hfil$\crcr
                   \noalign{\kern-1pt\nointerlineskip\vskip2pt}
                   \leftrighthookfill\crcr}}}
\newcommand\Cont[1]{\hsp{.4}\vtop{\ialign{##\crcr$\hfil\displaystyle
                   {\hsp{-.4}#1}\hfil\hsp{-.4}$\crcr
                   \noalign{\kern-1.9pt\nointerlineskip\vskip2pt}
                   \leftrighthookfill\crcr}}\hsp{.4}}
\def\corfu         {correlation function}
\def\csa           {Cartan subalgebra}

\def\cwbasis       {Car\-tan\hy$\!$Weyl basis}

\def\ct            {coset theory}
\def\cts           {coset theories}
\def\cvira         {coset Virasoro algebra}
\def\cvo           {chiral vertex operator}
\def\cwei          {conformal weight}
\def\cZ            {{\cal Z}}
\def\DA            {{\mbox{\sc D}_{\!A}}}
\def\DB            {{\mbox{\sc D}_{\!B}}}
\def\DD            {{\rm D}}
\def\deq           {differential equation}
\def\dimgb         {{{\rm dim}\,\gB}}
\def\df            {\,{:=}\,}
\def\dg            {\partial{\cal G}}
\def\dl            {\mathbb }
\def\dop           {{:}\;}
\long\def\drac#1#2 {{\displaystyle\frac{#1}{#2}}}
\def\drei          {{\small({\rm H3})}}
\def\Drei          {\fbx H3}
\def\dsum          {\displaystyle\sum}
\def\dyd           {Dynkin diagram}
\def\eal           {\mbox{$E^{\alpha}$}}
\def\ealb          {\mbox{$E^{\bar\alpha}$}}
\def\eali          {\mbox{$E^{\alpha^{(i)}}$}}
\def\ee            {\end{equation}}
\def\eE            {{\rm e}}
\newcommand\EE[2]  {E^{#1}_{#2}}
\def\eear          {\end{array}}
\def\eins          {{\small({\rm H1})}}
\def\Eins          {\fbx H1}
\def\elambda       {{\rm e}^\lambda_{}}
\def\emal          {\mbox{$E^{-\alpha}$}}
\def\emb           {\hookrightarrow}

\def\emt           {energy-momentum tensor}
\def\enva          {en\-ve\-lo\-ping algebra}

\def\eq            {\,{=}\,}
\def\equ           {equation }
\newcommand\erf[1]{(\ref{#1})}
\newcommand\Erf[2]{(\ref{#1#2})}
\newcommand\F[6]   {{\sf F}_{\!\sss#1#2}\Barray{#3#4}{#5#6}}
\def\fabc          {\mbox{$f^{ab}_{\:\:\:\:c}$}}
\def\fabi          {\mbox{$f^{ab}_{\:\:\:\:i}$}}
\newcommand\fbx[2] {\fbox{\raisebox{.08em}{\tiny(}${\scs\rm#1#2}$%
                   \raisebox{.08em}{\tiny)}}~~}
\newcommand\Fbx[3] {\fbox{\raisebox{.08em}{\tiny(}${\scs\rm#1#2}$%
                   \raisebox{.08em}{\tiny)}}~~#3}
\def\findim        {finite-dimen\-sional}
\def\fline         {{~}\\[1 mm]\noindent ------------------\\[1 mm]}
\def\fmbarz        {\mbox{$\overline{\cal F}^{}_{\!I}(\bz)$}}
\def\fmbari        {\mbox{$\overline{\cal F}^{}_{\!I}$}}
\def\fmh           {\mbox{$\hat{\cal F}^{}_{\!K}$}}
\def\fmi           {\mbox{${\cal F}^{}_{\!I}$}}
\def\fmz           {\mbox{${\cal F}^{}_{\!I}(z)$}}
\def\fmw           {\mbox{$\check{\cal F}_{\!J}$}}
\def\fpc           {four-point correlator}
\def\fpf           {four-point function}
\def\four          {four-point function}

\newcommand\Frac[2]{\mbox{\large$\frac{#1}{#2}$}}
\def\frc           {fusion rule coefficient}
\def\furu          {fusion rule}
\def\FURU          {Fusion Rule}
\def\futnote#1     {\footnote{~#1}\ }

\def\g             {\mbox{$\liefont g$}}

\def\gb            {\mbox{$\bar\gM$}}
\def\gB            {{\bar\gM}}
\def\gbdim         {{{\rm dim}\,\gB}}
\def\gbodual       {\mbox{$\bar{\liefont g}_\circ^\star$}}
\def\gbP           {\mbox{$\bar\gPP$}}

\def\ggp           {$(\gM/\hM)_\kV$}
\def\ghat          {\mbox{$\hat{\gM}$}}
\def\Gid           {\mbox{$G_{\rm id}$}}
\def\GId           {{G_{\rm id}}}

\newcommand\glo[2] {\\#2&#1}
\newcommand\glO[3] {\\#3&#1#2}
\newcommand\Glo[3] {\\#3&#1,\,#2}
\newcommand\GLo[4] {\\#4&#1,\,#2,\,#3}
\newcommand\GlO[5] {\\#5&#1#2,\,#3#4}

\def\gloopb        {\mbox{$\bar\gM_{\rm loop}$}}
\def\gm            {\gM_-}
\def\gM            {{\liefont g}}
\def\gmh           {\mbox{$\hat{\cal E}^{}_K$}}

\def\gmw           {\mbox{$\check{\cal E}^{}_J$}}
\def\gmz           {\mbox{${\cal E}^{}_I(z)$}}

\def\go            {\mbox{${\liefont g}_\circ$}}
\def\godual        {\mbox{${\liefont g}_\circ^\star$}}
\def\gp            {\gM_+}
\def\gP            {${\liefont h}$}
\def\gPP           {{\liefont h}}
\def\grank         {r}
\def\Gstab         {{G_\ilabel}}
\def\Gstabb        {{G_\jlabel}}
\def\gv            {\mbox{$g^\Vee$}}
\def\gV            {{g^\Vee}}

\def\h             {\mbox{$\cal H$}}
\def\ha            {\mbox{${\cal H}_A$}}
\def\hA            {{{\cal H}_A}}
\def\half          {\mbox{$\frac12\,$}}
\def\hamred        {Hamil\-tonian reduction}
\def\hb            {\mbox{$\bar{\liefont h}$}}
\def\hbe           {\mbox{${\cal H}_B$}}
\def\hc            {\mbox{${\cal H}_{\gM/\gPP}$}}

\newcommand\Hh[1]  {H^{#1}_{}}
\def\hill          {\mbox{${\cal H}_\Lambda$}}
\def\hilsp         {Hilbert space}
\def\hl            {\mbox{${\cal H}_\Lambda$}}
\def\hlb           {\mbox{${\cal H}_{\bar\Lambda}$}}
\def\hll           {\mbox{${\cal H}_\ilabel$}}
\def\hLP           {\mbox{${\cal H}_{\LambdaP}$}}
\def\hM            {{\liefont h}}
\def\ho            {\mbox{${\cal H}_\circ$}}
\def\hsa           {horizontal subalgebra}
\newcommand\hsp[1] {\mbox{\hspace{#1 em}}}
\def\hw            {highest weight}
\def\hwm           {highest weight module}

\def\HWM           {Highest Weight Module}

\def\hws           {highest weight state}
\def\hwv           {highest weight vector}
\def\hy            {$\mbox{-\hspace{-.66 mm}-}$}
\def\IA            {{\cal T}_i}

\def\ii            {{\rm i}}
\def\ihwm          {irreducible highest weight module}
\def\ilabel        {{(\Lambda;\Lambda')}}
\def\Ilabel        {{(\Lambda;\!\Lambda')}}
\def\ilabl         {{[\ilabel]}}
\def\Ilabl         {{[\Ilabel]}}
\def\iN            {\!\in\!}
\def\infdim        {infinite-dimen\-si\-o\-nal}
\def\Infdim        {Infinite-dimensional }

\def\intopiw       {\Frac1{2\pi\ii}\oint_0\!{\rm d}w\,}

\def\intopiz       {\Frac1{2\pi\ii}\oint_0\!{\rm d}z\,}

\def\intopizw      {\Frac1{2\pi\ii}\oint_w\!{\rm d}z\,}

\def\intow         {\oint_0\!{\rm d}w\,}
\def\intozw        {\oint_w\!{\rm d}z\,}
\newcommand\ipro[2]{\mbox{$(#1\Mid#2)$}}
\newcommand\Ipro[2]{\mbox{$(#1{\mid}#2)$}}
\def\iproo         {\ipro{\cdot}{\cdot}}
\def\irmod         {irreducible module}
\def\irrep         {irreducible representation}
\def\J             {{\rm J}}
\def\jj            {{(\J;\J')}}
\def\JJ            {{(\J;\!\J')}}
\def\jjj           {{[\J;\J']}}
\def\jlabel        {{(\Mu;\Mu')}}
\def\Jlabel        {{(\Mu;\!\Mu')}}
\def\jlabl         {{[\jlabel]}}
\def\Jlabl         {{[\Jlabel]}}
\def\kappb         {\bar\kappa}
\def\km            {Kac\hy Moo\-dy }
\def\kma           {Kac\hy Moo\-dy algebra}
\def\KMA           {Kac\hy Moo\-dy Algebra}
\def\kpf           {Kac\hy Peterson formula}
\def\kv            {\mbox{$k^\Vee$}}
\def\kV            {{k^\Vee}}
\def\kwf           {Kac\hy$\!$Walton formula}
\def\kze           {Knizh\-nik\hy Za\-mo\-lod\-chi\-kov equation}
\def\KZE           {Knizh\-nik\hy Za\-mo\-lod\-chi\-kov Equation}
\def\L             {\Lambda}
\long\def\labl#1   {\label{#1}\ee\ifnum\draftcontrol=1
                   \mbox{ }\\[-12 mm]\query{#1}\\[5 mm] \fi}
\long\def\Labl#1#2 {\label{#1#2}\ee\ifnum\draftcontrol=1
                   \mbox{ }\\[-12 mm]\query{#1#2}\\[5 mm] \fi}
\def\lama          {\lambda_A}

\def\LambdaP       {{\Lambda'}}
\def\lamo          {\lambda\o}
\def\Lb            {{\bar\Lambda}}

\def\Ldots         {,...\,,}
\def\lec           {Lecture }
\newcommand\lect[4]{\vskip6.3ex \setcounter{equation}{0} \noindent
                   {\fbox{\hsp{#3}\large\bf Lecture #1\,: \ \ #2 \hsp{#4}}}
                   \vskip1.5ex \normalsize\def\thelect{#1} \label{lecture#1}} 
\newcommand\lecT[3]{\vskip6.3ex \setcounter{equation}{0} \noindent
                   {\fbox{\hsp{#2}\large\bf #1 \hsp{#3}}}\vskip1.5ex\normalsize}
\def\lefthook      {{\vrule height5pt width0.4pt depth0pt}}
\def\leftrighthookfill{$\mathsurround=0pt \mathord\lefthook
                   \hrulefill\mathord\righthook$}

\def\li            {\mbox{$\Lambda_{(i)}$}}
\def\libar         {\mbox{$\bar{\Lambda}_{(i)}$}}
\def\lie           {Lie algebra}
\def\LIE           {Lie Algebra}

\def\liefont       {\mathfrak }
\def\liematrixfont {\mathfrak }

\def\llb           {\mbox{\large(}}
\def\LLb           {\mbox{\Large[}}

\def\lrb           {\mbox{\large)}}
\def\LRb           {\mbox{\Large]}}

\def\lo            {\mbox{$\Lambda_{(0)}$}}

\def\lP            {{\Lambda'}}

\def\Mapsto        {\;\mapsto\;}
\def\mapstO        {\!\mapsto\!}
\def\md            {\tilde M}

\def\mi            {\,{-}\,}
\def\Mi            {{-}}
\def\Mid           {\!\mid\!}
\def\Mu            {\Upsilon}
\def\MuP           {\Mu'}
\newcommand\mult[2]{{\rm mult}^{}_{#1}(#2)}
\def\multal        {{{\rm mult}(\alpha)}}
\newcommand\N[3]   {\mbox{\sc n}_{\sss\!#1#2}^{\sss\;\ #3}}
\def\naturals      {{\dl N}}

\def\Ne            {\,{\not=}\,}

\def\nijk          {\N ABC}
\newcommand\NN[1]  {\mbox{\sc N}_{#1}}
\def\NNA           {\NN{\!A}}
\def\nondeg        {non-dege\-ne\-rate}
\def\nonneg        {non-ne\-gative}

\def\nontriv       {non-tri\-vial}

\newcommand\normord[1] {\,\raisebox{.033em}{\large\bf:}#1
                   \raisebox{.033em}{\large\bf:}\,}

\newcommand\Ns[3]  {\mbox{\sc n}_{\sss\!#1#2#3}^{}}
\newcommand\nxt[1] {\\[.07em]\raisebox{.12em}{\rule{.35em}{.35em}}\hsp{.6}#1}
\newcommand\nXt[1] {\\[.07em]\raisebox{.12em}{\rule{.35em}{.35em}}\hsp{.6}#1}
\newcommand\Nxt[1] {\raisebox{.12em}{\rule{.35em}{.35em}}\hsp{.6}#1}
\def\o             {_\circ}
\def\O             {{\scriptstyle\circ}}
\def\oa            {operator algebra}
\def\olie          {orbit Lie algebra}
\def\one           {\mbox{\small $1\!\!$}1}
\def\onedim        {one-di\-men\-si\-o\-nal}
\def\onehalf       {\mbox{$\frac12$}}
\def\op            {operator product}
\def\opa           {operator product algebra}
\def\OPA           {Operator Product Algebra}
\def\opc           {operator product coefficient}
\def\ope           {operator product expansion}
\def\opf           {one-point function}

\def\ooplus        {\,{\oplus}\,}
\def\otim          {\raisebox{.07em}{$\scriptstyle\otimes$}}
\def\otiM          {\,\otim\,}
\def\onetor        {1,2,...\,,\grank}
\def\otor          {0,1,...\,,\grank}
\def\P             {{\phantom|}}
\def\pbw           {Poin\-ca\-r\'e\hy Birk\-hoff\hy $\!$Witt theorem}
\newcommand\phj[1] {\Phi_{\!#1}}
\def\pf            {primary field}
\def\pHi           {X}
\def\phl           {\phi_\L}
\def\pho           {\phj\O}
\def\pii           {\pi{\rm i}}
\def\pslz          {\mbox{PSL$_2(\zet)$}}
\def\poleq         {polynomial equation}
\newcommand\Prod[1]{\mbox{{\large$\prod$}\hsp{-.03}\raisebox{-.33em}
                   {$\scs#1$}\,}}
\def\q             {quantum }
\def\Q             {Quantum }

\def\qft           {quantum field theory}
\def\QFT           {Quantum Field Theory}
\def\qftc          {quantum field theoretic }
\def\qfts          {quantum field theories}
\def\QJ            {{\rm Q}^{}_\J}
\def\qp            {quasi-pri\-ma\-ry}
\def\qpf           {quasi-pri\-ma\-ry field}
\def\Qpf           {Quasi-pri\-ma\-ry field}
\def\qppf          {(qua\-si-)\,pri\-ma\-ry field}
\def\qps           {quasi-pri\-ma\-ries}
\def\qrcft         {qua\-si-rational conformal field theory}
\long\def\query#1{\hskip 0pt{\vadjust{\everypar={}\small\vtop to 0pt{\hbox{}%
     \vskip -13pt\rlap{\hbox to 49.0pc{\hfil{\vtop{\hsize=8pc\tolerance=6000%
     \hfuzz=.5pc\rightskip=0pt plus 3em\noindent#1}}}}\vss}}}}%
\def\qzn           {\rationals(\zeta_n)}

\def\rcft          {rational conformal field theory}
\def\rcfts         {rational conformal field theories}
\def\reals         {{\dl R}}
\def\Reals         {$\reals$}
\newcommand\refl[1]{\pageref{lecture#1}}
\def\Refs          {}

\def\refsC         {\Sect\Refs3}
\def\refsD         {\Sect\Refs4}
\def\refsE         {\Sect\Refs5}
\def\refsG         {\Sect\Refs6}
\def\refsF         {\Sect\Refs7}
\def\RefsF         {\Refs7}

\def\refsI         {\Sect\Refs9}
\def\RefsI         {\Refs9}
\def\refsJ         {\Sect\Refs10}
\def\refsK         {\Sect\Refs11}
\def\refsL         {\Sect\Refs12}

\def\refsN         {\Sect\Refs14}
\def\RefsN         {\Refs14}

\def\refse         {\Sect\Refs19}
\def\refsf         {\Sect\Refs20}
\def\refsh         {\Sect\Refs21}

\def\refsj         {\Sect\Refs24}

\def\refsY         {\Sect\Refs28}

\newcommand\refsec[1]{\pageref{S.#1}}
\def\rep           {representation}
\def\Rep           {Representation}
\def\repsp         {representation space}
\def\resp          {respectively}
\newcommand\restr[1]{|\raisebox{-.44em}{$#1$}}
\def\rhs           {right hand side}
\def\righthook     {{\vrule height5pt width0.4pt depth0pt}}

\def\role          {r\^ole}
\def\rop           {\mbox{$\Re$}}

\def\rosyb         {\mbox{$\overline\Phi$}}
\def\scs           {\scriptstyle}
\def\sec           {section}
\newcommand\sect[2]{\addtocounter{section}1 \vskip1ex%
                   \ifnum\draftcontrol=1\query{\fbox{#2}}\fi%
                   \mbox{}\\[.5ex]{\large\bf\thesection\ \hspace{.18em}#1}
                   \label{S.#2}\\[1.1ex]\normalsize}
\newcommand\secT[2]{\addtocounter{section}1%
                   \ifnum\draftcontrol=1\query{\fbox{#2}}\fi%
                   \noindent{\large\bf\thesection\ \hspace{.18em}#1}
                   \label{S.#2}\\[1.1ex]\normalsize}
\def\Sect          {section$\!$~}
\def\Sects         {sections$\!$~}
\def\Setminus      {\!\setminus\!}
\def\SEtminus      {\,{\setminus}}
\def\shps          {space \h\ of physical states}
\def\sig           {w}

\def\signw         {{\rm sign}(w)}
\def\slthree       {\mbox{$\liematrixfont{sl}(3)$}}
\def\slti          {\gM_{{\sss(}i{\sss)}}}
\def\sltwo         {\mbox{$\liematrixfont{sl}(2)$}}

\def\sltwor        {\mbox{$\liematrixfont{sl}(2,\reals)$}}
\def\slz           {\mbox{SL$_2(\zet)$}}
\newcommand\smallmatrix[1] {\mbox{\footnotesize $\left(\begin{array}#1
                   \end{array}\right)$}}
\def\smat          {S-matrix}
\def\smats         {S-matrices}
\def\sofive        {\mbox{$\liematrixfont{so}(5)$}}
\newcommand\Span[1]{\mbox{span$\{#1\}$}}
\def\sps           {space of physical states}

\def\sss           {\scriptscriptstyle}
\def\suee          {\mbox{$\liematrixfont{su}$(1,1)}}
\newcommand\Sum[1]{\mbox{{\large$\sum$}\hsp{-.03}\raisebox{-.33em}{$\scs#1$}\,}}

\newcommand\SUM[1] {\mbox{\,{\Huge$\sum$}\hsp{-.46}\raisebox{-.65em}{$\scs#1$}}}
\newcommand\sumd[1]{\sum_{#1=1}^\gbdim}

\newcommand\suml[1]{\sum_{#1=1}^\ell}
\def\summ          {\sum_{I=1}^M}
\def\summb         {\sum_{\bar I=1}^{\bar M}}
\newcommand\sumr[1]{\sum_{#1=1}^\grank}

\newcommand\sumro[1]{\sum_{#1=0}^\grank}
\def\sumwb         {\sum_{w\in\overline W}}

\def\sumww         {\sum_{w\in W}}

\def\tam           {\mbox{$\hat T^a_m$}}
\def\tamn          {\mbox{$\hat T^a_{m+n}$}}
\def\tan           {\mbox{$\hat T^a_n$}}
\def\tao           {\mbox{$\hat T^a_0$}}
\def\tbn           {\mbox{$\hat T^b_n$}}
\def\tbnn          {\mbox{$\hat T^b_{-n}$}}
\def\tcmn          {\mbox{$\hat T^c_{m+n}$}}
\def\thetagb       {\bar\theta}
\def\threedim      {three-di\-men\-si\-o\-nal}
\def\threepf       {three-point function}
\def\Times         {\,{\times}\,}
\def\tlbara        {\mbox{${\bar T}^{a}_{\!(\Lb)}$}}
\def\tmat          {T-matrix}
\newcommand\tocen[3]{\addtocounter{toce}1\thetoce~~{#3}\hfill\refsec{#2}\\}
\newcommand\Tocen[3]{\addtocounter{toce}1$\mbox{}$~\thetoce~~{#3}%
                   \hfill\refsec{#2}\\}
\def\tpf           {three-point function}

\def\tta           {\theta}
\def\ttab          {\mbox{$\bar\theta$}}
\def\ttaB          {{\bar\theta}}
\def\ttam          {\mbox{$T^a_m$}}
\def\ttan          {\mbox{$T^a_n$}}
\def\ttao          {\mbox{$T^a_0$}}
\def\ttbn          {\mbox{$T^b_n$}}
\def\ttcmn         {\mbox{$T^c_{m+n}$}}
\def\twpf          {two-point function}
\def\Twtw          {^{\scriptscriptstyle(2)}}
\def\twodim        {two-di\-men\-si\-o\-nal}

\def\Twtwtw        {^{\scriptscriptstyle(3)}}

\def\U             {\mbox{$\liefont U$}}

\def\uenva         {universal enveloping algebra}

\def\uihwm         {unitary irreducible highest weight module}
\def\Untw          {^{\scriptscriptstyle(1)}}
\def\uw            {\mbox{${\liefont U}(\w)$}}
\def\va            {{v^{}_{\!A}}}
\def\Va            {\mbox{${\cal V}^{}_{\!A}$}}

\def\vdd           {v^{}_{\!\cD,\bar\cD}}
\def\vdD           {v_{\!\cD,\bar\cD}}
\def\Vee           {{\scriptscriptstyle\vee}}
\newcommand\version[1] {\ifnum\draftcontrol=1 \typeout{}\typeout{#1}\typeout{}
                   \vskip3mm \centerline{\fbox{{\tt DRAFT -- #1 -- }
                   {\small\draftdate}}} \vskip3mm \fi}
\newcommand\vev[1] {\langle #1 \rangle}
\newcommand\Vev[1] {\langle\, #1 \,\rangle}
\def\vf            {Verlinde formula}
\def\vhw           {\mbox{$v^{}_{\rm h.w.}$}}
\def\vir           {\mbox{${\wfont V}\liefont{ir}$}}
\def\viR           {\vir\,}
\def\virc          {\mbox{${\wfont V}\liefont{ir}_{\gM/\gPP}$}}
\def\virg          {\mbox{${\wfont V}\liefont{ir}_\gM$}}
\def\virgP         {\mbox{${\wfont V}\liefont{ir}_\gPP$}}
\def\vira          {Virasoro algebra}
\def\virb          {\mbox{$\overline{\vir}$}}
\def\virp          {\mbox{$\vir_{0,\pm1}$}}
\def\virP          {\mbox{${\wfont P}$}}
\def\VL            {\mbox{$V_{\!\Lambda}^{}$}}

\def\vLambda       {{v_{\!\Lambda}^{}}}
\def\vml           {\mbox{${\cal V}^{}_{\!\L}$}}
\def\vmle          {\mbox{${\cal V}^{}_{\!\mu_1}$}}
\def\vmlz          {\mbox{${\cal V}^{}_{\!\mu_2}$}}
\def\vnull         {\mbox{$v_{\rm null}^{}$}}
\def\vo            {\mbox{$v_\circ$}}
\def\voa           {vertex operator algebra}
\def\vpdd          {\varphi^{}_{\!\cD,\bar\cD}}
\def\w             {\mbox{$\wfont W$}}

\def\wb            {\mbox{$\overline{\wfont W}$}}
\def\wb            {\mbox{$\overline{\wfont W}$}}
\def\wbp           {\mbox{$\overline{\wfont W}{}'$}}
\def\wbw           {(w{,}\bar w)}
\def\wc            {\mbox{${\wfont W}_{\gM/\gPP}$}}
\def\wfont         {\cal }

\def\Wlog          {without loss of generality}
\def\wm            {\mbox{${\wfont W}_{\!-}$}}
\def\wo            {\mbox{${\wfont W}_{\!\circ}$}}
\def\wp            {\mbox{${\wfont W}_{\!+}$}}
\def\wpm           {\mbox{${\wfont W}_{\!\pm}$}}
\def\wrt           {with re\-spect to }
\def\wrtt          {with re\-spect to the }
\def\wune          {\mbox{$\w_{\rm unext.}$}}
\newcommand\ww[2]  {W^{#1}_{#2}}
\newcommand\Ww[1]  {W^{#1}_{}(w)}
\newcommand\Wwb[1] {\overline W{}_{}^{#1}(\bar w)}
\newcommand\Wz[1]  {W^{#1}_{}(z)}
\def\WZW           {Wess\hy Zu\-mi\-no\hy Wit\-ten}
\def\wzwm          {WZW model}
\def\wzwt          {WZW theory}
\def\wzwts         {WZW theories}
\def\WZWTS         {WZW Theories}
\def\zbz           {(z{,}\bz)}
\def\zet           {{\dl Z}}
\def\Zet           {$\zet$}
\def\zetminuso     {\zet_{\le0}}
\def\zetpluso      {\zet_{\ge0}}

\def\zwei          {{\small({\rm H2})}}
\def\Zwei          {\fbx H2}

\def\draftdate{\number\month/\number\day/\number\year\ \ \ \hourmin }

\global\def\draftcontrol{0}
\catcode`\@=12

 \documentclass[12pt]{article} \usepackage{amssymb,amsfonts}
 \def\theequation{\thelect.\arabic{equation}} 
  \def\wB{\,\,\linebreak[0]}

\begin{document}
 \begin{flushright} {~}\\[-15 mm]{\sf hep-th/9702194} 
 \\[2 mm]{\sf February 1997} \end{flushright} \vskip17mm
 \centerline{\large\bf LECTURES ON CONFORMAL FIELD THEORY}\vspace{.5em} 
 \centerline{\large\bf AND KAC-MOODY ALGEBRAS}
 \begin{center} \vskip 16mm
 {\large J\"urgen Fuchs $^{\sss{\sf X}}$} \\[3mm]
 {DESY\\[1mm] Notkestra\ss e 85 \\[1mm] D -- 22603~~Hamburg}
 \end{center} \vskip 20mm
  \begin{quote}{\bf Abstract}.\\[.5mm]
  This is an introduction to the basic ideas and to
  a few further selected topics in conformal quantum field theory and in the
  theory of Kac-Moody algebras.\\[.5mm]
  These lectures were held at the Graduate Course on Conformal
  Field Theory and Integrable Models (Budapest, August 1996). They will
  appear in a volume of the Sprin\-ger\wB Lecture\wB Notes\wB in\wB Physics
  edited by Z.\ Horvath and L.\ Palla.
  \end{quote}
 \vfill {}\fline{} {\small $^{\sf X}$~~Heisenberg fellow} \newpage

\mbox{$ $}\\[2.5em]
{\bf Contents}\\\small

\noindent {\tt
          {\bf Lecture 1\,: \CFT}             \hfill {\bf\refl1}\\[.2em]
\Tocen1 A {\Con \QFT}
\Tocen2 B {Observables:~The \CHSA}
\Tocen3 C {Physical States:~\HWM s}
\Tocen4 D {Sectors:~The Spectrum}
\Tocen5 E {\Con Fields}
\Tocen6 G {The Operator Product Algebra}
\Tocen7 F {Correlation Functions and Chiral Blocks} [.54em]
          {\bf Lecture 2\,: Fusion Rules, Duality and Modular Transformations}
                                             \hfill {\bf\refl2}\\[.2em]
\Tocen8 H {Fusion Rules}
\Tocen9 I {Duality}
\tocen0 J {Counting States:~Characters}
\tocen1 K {Modularity}
\tocen2 L {Free Bosons}
\tocen3 M {Simple Currents}
\tocen4 N {Operator Product Algebra from \FURU s} [.54em]
          {\bf Lecture 3\,: \KMA s}           \hfill {\bf\refl3}\\[.2em]
\tocen5 a {\CARMS}
\tocen6 b {Symmetrizable Kac-Moody Algebras}
\tocen7 c {Affine \LIE s as Centrally Extended Loop \Alg s}
\tocen8 d {The Triangular Decomposition of Affine Lie \Alg s}
\tocen9 e {\Rep\ Theory}
\tocen0 f {Characters} [.54em]
          {\bf Lecture 4\,: \WZWTS\ and Coset Theories}
                                              \hfill {\bf\refl4}\\[.2em]
\tocen1 h {WZW Theories}
\tocen2 i {WZW Primaries}
\tocen3 T {Modularity, Fusion Rules and WZW Simple Currents}
\tocen4 j {The Knizhnik-Zamolodchikov Equation}
\tocen5 l {Coset \CFTS}
\tocen6 m {Field Identification}
\tocen7 n {Fixed Points}
{}\\{}\\[-1.5em]
\tocen8 Y {Omissions}
\tocen9 Z {Outlook}
\tocen0 X {Glossary}
\\[-1.1em]\mbox{$ $}\hsp{2.03}References\hfill\pageref{thebi}\\
} \newpage\normalsize

\lect1{\CFT}{9.04}{6.27}

\sect{\Con \QFT}A\label{s1}%
Over the years, \qft\ has enjoyed a great number of successes.
Nevertheless, from a conceptual point of view the situation is far from 
satisfactory, and in some respects it even resembles a disaster. Indeed, 
beyond perturbation theory many of the methods used in phenomenologically 
successful models cannot be justified rigorously. For instance,
a thorough understanding of path integrals is essentially only
available for free fields, or for topological `field' theories which have a
\findim\ configuration space. Also, the relation between the elementary 
`point-like' quantum {\em fields\/} (whatever the precise meaning of such 
quantities may be) and the actual {\em particle\/} contents of a theory is 
often 
obscure. (The elementary fields may be regarded as a coordinatization of the 
field space, while the particles have a meaning independent of a choice of
coordinates; typically a distinguished and\,/\,or manageable choice of 
field coordinates which directly correspond to particles is not available.)
On the other hand, frameworks whose foundations are mathematically sound,
like e.g.\ Wightman field theory \Cite{STwi} or $C^*$-\alg ic approaches
\Cite{HAag}, are difficult to relate to concrete models, and indeed have
traditionally been plagued with a scarcity of models to which they apply.
For example, the proper treatment of $U(1)$-charges (which are coulombic 
rather than localized) in the framework of algebras of local observables
is still problematic.

It is worth stressing that `\qft' is not a protected term, 
but is used for a variety of rather different concepts.\,%
\futnote{While a lot of physicists believe to know exactly 
what \qft\ is, you will most probably get two different answers
when you ask any two of them. And not only the methods and results
are under debate, but even the basic questions to be asked.} 
In particular there are two basically different points of view of the
relation between `quantum' versus `classical' field theories. One may either 
try to quantize a classical theory (say by canonical quantization or using
path integrals), or directly start at the quantum level, e.g.\ by describing 
the field and observable \alg s, in which case a classical theory may be
obtained by performing a suitable limit. It is far from obvious
that the resulting quantum, \resp\ classical, theory is uniquely determined.
Another important difference concerns the description of observables. In some
approaches these are characterized quite concretely (e.g.\ as
gauge \resp\ BRS invariant combinations of the elementary field variables of 
the classical theory), in others they appear more 
abstractly as operators on a \hilsp\ which at an initial stage are only known
via their general \alg ic properties.

{\em\Cft\/} admits formulations which directly approach the exact theory at the
quantum level, but nevertheless describe observables quite explicitly and
handle large classes of models simultaneously.
In particular many quantities of interest can be {\em calculated
exactly\/} rather than only in a (typically at best asymptotic) 
perturbation series. In addition, for a large number of models there exists 
(or is at least conjectured) a Lagrangian\,/\,path integral realization, so that
in principle one can study such models in several guises and e.g.\ 
compare with conventional perturbation theory.
\Cft\ can therefore help to bridge the gap between various approaches to \qft.

Here and below, by the term \cft\ I refer to models in
one or two space-time dimensions which are relativistic \qfts\ and possess
conformal symmetry. What precisely this characterization means will be studied 
shortly. Again one must note that the term `\cft' is not protected and can
stand for a variety of different formulations. Here I approach this subject
from the perspective of \qft. However,
in this introductory exposition I will not attempt to build exclusively on 
general field theoretic considerations, but rather I concentrate on certain
basic \alg ic structures and assume that the reader has 
already some faint acquaintance with fundamental field theoretic
notions that will play a \role, such as the \opa, \corfu s and \furu s.
On the other hand, in a sense I start from scratch, in that 
at least I try to {\em indicate\/} each logical step explicitly, even though I 
probably do not provide sufficient explanation for all of them.

The motivations to study \cfts\ are manifold, and I am content 
to just drop a few keywords here:\,%
\futnote{For more details I refer to the references listed in
\refsY, to other lectures delivered at this school, as well as to 
the huge number of works where such motivations are advertised.}
field theories describing systems at the critical point of a
second order phase transition in statistical
mechanics; vacuum configurations of strings and superstrings;
braid group statistics; topological field theory; 
invariants of knots and links, and of three-manifolds;
integrable systems as perturbed \con models; the fractional quantum Hall effect;
high temperature superconductivity. 

In the study of \cfts, the following two issues prove to be most fundamental. 
First, the classification programme; and second,
the complete solution, by reconstruction from certain basic data (possibly
obtained in the course of classification), of specific models.
We will encounter various manifestations of these two aspects of \cft\
in the sequel. They indicate that these theories are very special indeed: while 
(contrary to the impression that is occasionally given)
a classification of {\em all\/} \cfts\
or a complete solution of any {\em arbitrary\/} given \cft\ model is totally out
of reach, one {\em can\/} classify large classes of \cfts, and one 
{\em can\/} solve very many models to a large extent.

\sect{Observables: The \CHSA}B\label{s2}%
One of the ultimate goals of a physical theory is to predict the outcome of all
possible measurements. Accordingly, a basic step in the investigation
of a \cft\ consists of characterizing its {\em observables}. In \q physics,
one thinks of observables as operators acting on a \hilsp\ of physical states.
Moreover, in relativistic \qft\ it is usually necessary to consider
operators which are bounded (only these can be continuous) and are
localized in some compact space-time region, i.e.\ act as the identity outside
\Cite{HAag}. Observables localized in space-like separated regions commute
({\em locality\/} principle).
In contrast, below I will work with non-localized unbounded operators; 
that this does not lead to any severe problems has its origin in the fact that
these observables enjoy an underlying {\em\lie\/} structure.
To obtain localized bounded operators one would, roughly speaking,
have to smear these operators
with test functions of compact support and form bounded functions thereof.

The most important subset of these observables of a \cft\ is provided by the
conformal symmetry itself. Conformal transformations of space-time -- i.e.\ of 
a (pseu\-do-)Riemannian manifold -- are by definition those general coordinate
transformations which preserve the angles between any two vectors, or what 
is the same, which scale the metric locally by an over-all factor
(Weyl transformations). In a flat space-time of $D\!>\!2$ dimensions the 
infinitesimal \con transformations consist of translations, rotations (\resp\ 
Lorentz boosts), a dilation and so-called special conformal transformations;
these generate a $(D{+}1)(D{+}2)/2$\,-dimensional \lie, which is a real form of
$\liefont{so}(D{+}2,\complex)$ (e.g.\ $\liefont{so}(D,2)$ in the case of 
Minkowski space).\,%
\futnote{More generally, the real form is $\liefont{so}(p+1,q+1)$ for 
signature $((+)^p,(-)^q)$.}
In contrast, for $D\eq2$ any arbitrary holomorphic mapping 
of the (compactified) complex plane is angle-preserving. Thus in complex 
coordinates $z,\,\bz$, 
infinitesimal conformal transformations are generated by mappings
which transform $z$ as $z \mapstO z + \eta(z) $ by functions which do not 
depend on $\bz$ -- a basis of which is given by $\eta(z)\eq
-z^{n+1} \epsilon$ for $n\iN\zet$ -- and by analogous mappings of $\bz$ with 
functions which do not depend on $z$. On functions of $z,\,\bar z$ 
these mappings are generated by differential operators
  \be  l_n = - z^{n+1} \partial_z \qquad\mbox{and}\qquad
  \bar l_n = - \bz^{n+1} \partial_\bz \,,  \ee
\resp.\ The operators $l_n$ satisfy the commutator relations
  \be  [l_m,l_n] = (m-n)\, l_{m+n} \,, \labl{witt}
and an analogous formula holds for the $\bar l_n$. In mathematical terms, this
means that both the $l_n$ and the $\bar l_n$ span an \infdim\ \lie; moreover, 
these two \alg s are combined as a direct sum, i.e.\ one has $[l_m,\bar l_n]=0$.
The \lie\ defined by \erf{witt} is known as the {\em Witt \alg}.

The formula \erf{witt} was obtained by purely classical considerations.
In quantum theory, there are analogous operators $L_n$ with $n\iN\zet$,
but now they satisfy
  \be  [L_m,L_n]= (m-n)\,L_{m+n} + \Frac1{12}\,(m^3-m)\, \delta_{m+n,0}\,C \,. 
  \labl{vir}
That is, they again give rise to a \lie, but
there appears an additional generator $C$; $C$ is a 
{\em central\/} element, i.e.\ has zero Lie bracket with the whole \alg,
  \be  [C,L_n] = 0  \,. \labl{vir'}
The \infdim\ \lie\,%
\futnote{Let me remind you at this point that, by definition, a 
{\em\lie\/} over \Complex\ (or similarly over \Reals, or more generally
over any field of scalars $\{\xi\}$), is a vector space with a bilinear 
product -- called the Lie bracket and denoted by $[\,\cdot\,,\cdot\,]$ --
which is antisymmetric and satisfies the Jacobi identity. 
In physics, the most common realization of a Lie
bracket is as the commutator \wrt an underlying associative product.}
defined by \erf{vir} and \erf{vir'} is called the {\em \vira\/} and denoted 
by \vir\,. In short, the Lie brackets of \vir\ differ from their
classical counterpart only in the term with $C$; the generator $C$, which in 
mathematics is known as the canonical central element or central charge, 
is therefore in physics also referred to as the \con {\em anomaly}. As in the 
classical situation one deals in fact with the direct sum of two \vira s, the 
second one being generated by operators $\bar L_n$, $n\iN\zet$, and $C$.  

In a purely mathematical context, the relations \erf{vir} arise -- by
solving a cohomology problem \Cite{FUks} -- as the unique non-trivial central 
extension of \erf{witt}. Field-theoretically, \erf{vir} 
is obtained by making a general ansatz for the equal-time commutator of the 
`\emt', whose Fourier\hy Lau\-rent components are the operators $L_n$ and 
$\bar L_n$ (see \equ \Erf Tz below). When one imposes the requirements
that the Wightman axioms are satisfied, that the system is dilation invariant,
and that the \emt\ is a conserved\,%
\futnote{Also note: That the \emt\ is both conserved and traceless implies that
the `dilation currents' $x^\mu T_{\mu\nu}$ are conserved.}
(Noether) current, this equal-time commutator becomes a finite sum of products 
of (derivatives of) the $\delta$-function multiplied with local operators, and
this result -- known as the L\"uscher\hy Mack theorem \Cite{Fust} --
is equivalent to \erf{vir}. 

Let me point out once more that in a \qftc setting the physical meaning of the 
generators of
conformal symmetry is as providing (non-bounded, non-localized) observables,
in the same spirit as e.g.\ momentum and angular momentum are observables 
in ordinary one-particle quantum mechanics. A brief characterization of 
a \cft\ is therefore that the observable algebra
contains (the direct sum of two copies of) the \vira.
In general, the full observable algebra of a \cft\ may be larger, though. But
by considerations based on properties of conserved currents, similarly as in 
the proof of the L\"uscher\hy Mack theorem, one can argue that the 
observables (regarded as Fourier\hy Lau\-rent components of 
conserved currents) still generate a \lie, and the
direct sum structure persists as well. In short, the 
observables of a \cft\ form an \infdim\ \lie\
$\w_{\rm tot}$ over \Complex\ with countable basis, which can be written as 
  \be  \w_{\rm tot}=\w\oplus\wbp \quad{\rm with}\quad \w\!\supset\!\vir 
  \quad{\rm and}\quad \wbp\!\supset\!\virb \,.  \ee
There is a variety of names for the two direct summands \w\ and $\wbp$: the
holomorphic\,/\,anti-holomorphic, or chiral\,/\,antichiral,
or left\,/\,right sub\alg s. When one considers only one of these sub\alg s,
one refers to the observables as the {\em \chsa}. 
As a consequence of the direct sum structure,
for many purposes one can restrict ones attention to
one `chiral half' of the \twodim\ theory, and I will do so for the time
being. (For aspects of the {\em two\/}-di\-men\-si\-o\-nal theory see e.g.\ 
\Sects\,\RefsF,\,\RefsI,\,\RefsN\ below. Also note that one needs not 
necessarily require that the left and right chiral halves have isomorphic 
\chsa s; when $\wbp\not\cong\w$, then one speaks of a `heterotic' theory.)

Unless stated otherwise, from now on it will be assumed that \w\
is {\em maximal\/}, i.e.\ that it already includes {\em all\/} (chiral)
observables. 
\w\ has a basis of the form $\{\ww in\}\cup\{C_\ell\}$, where the index
$n$ of $\ww in$ takes values in \Zet, while the index $i$ takes values in a 
set $I\equiv\{0,1,2,...\}$ which may be finite or possibly infinite, and
where the $C_\ell$ are central elements satisfying $[C_\ell,\cdot\,]=0$.
The Virasoro generators can be identified as $L_n\equiv\ww0n$. At least for 
$m\iN\{0,\pm1\}$ any generator $\ww in$ then satisfies
  \be  [\ww0m,\ww in] \equiv
  [L_m,\ww in] = ((\cd i-1)m-n)\,\ww i{m+n} \,,  \Labl1X
with certain numbers $\cd i$ which must be positive integers (on the 
other hand, for $i\eq0$ and $m\Ne\{0,\pm1\}$, there is 
in addition the central term). In particular, the subscript $n$ already 
determines the Lie bracket with the Virasoro \alg\ generator $L_0$,
  \be  [L_0,\ww in] = -n\,\ww in \,. \labl{l0w}
In fact, this subscript even supplies us with a \Zet-grading of \w, 
i.e.\ the structure constants of \w, defined by
\futnote{In principle, on the \rhs\ infinitely many terms may appear.
In this case strictly speaking one does not deal with a proper basis of 
a \lie, and a completion \wrt a suitable topology is needed. However, 
as it turns out, when applied to any vector in the relevant space \h\ of
physical states, only a finite number of terms is non-zero, owing to the fact 
that the generators $\ww im$ act `locally nilpotently' in \h.
This is also one of the reasons why considering formally infinite series as in
\Erf Wz below does not lead to any serious problems.}
  $  [\ww im,\ww jn] = \sum_{p\in\zet}\,\sum_{k\in I} f^{\;ij\;,p}_{mn,k}\,
  \ww kp  $
are subject to the condition
  \be  f^{\;ij\;,p}_{mn,k}=0 \quad{\rm for}\quad p \ne m+n \,. \labl{grad}
Also, central elements have grade 0. 

For the time being I will not be more specific about the algebra \w.
Later on, we will encounter important examples, the so-called current \alg s. 
Other possibilities are supersymmetric extensions\,%
\futnote{The extension is by normal ordered products of supercurrents; the
supercurrents themselves have half integral $\cD$, i.e.\ are not observables
and hence are not contained in \w.}
of the \vira, and many more
chiral \alg s are described in detail in the lectures by Gerard Watts.

Before proceeding, let me point out that
the restriction to one chiral half requires in particular to
regard $z$ and $\bar z$ as two {\em independent\/} complex variables;
$\bz$ is to be identified with the complex conjugate of $z$ only once the two
chiral halves of a \twodim\ theory
are combined. That this is a sensible way to proceed is
far from obvious, and actually the interpretation of this prescription
depends in part on the context in which the \con symmetry is considered:
\nxt When arising in minkowskian \qft, the variables $z$ 
and $\bar z$ take their values on circles that are
obtained as the two compactified light-cones $x^0\eq\pm x^1$ of the \twodim\ 
theory, from which they are extended to the punctured plane $\complex\SEtminus
\{0\}$. Identification of $\bz$ as the conjugate of $z$ thus amounts to
considering an analytical continuation of the original minkowskian theory to
euclidean signature which does not coincide with the usual Wick rotation.
\nxt In applications to string theory, the punctured plane arises
as the image of the cylindrical world sheet swept out by a closed relativistic
string. In the description of phase transitions, $z$ is the
complex coordinate of a euclidean \twodim\ field theory.
The meaning of separating $z$ and $\bz$ is in these cases less clear.  
\nxt At the level of the chiral \alg, taking $z$ and $\bz$ as
independent amounts to regarding \w\ and \wb\ as \alg s over the
complex numbers. It is tempting to interpret the identification of $\bz$ with
the complex conjugate of $z$ as considering a suitable real form 
of the complex \lie, but this is actually not quite correct.

\sect{Physical States: Highest Weight Modules}C\label{s3}%
Once we know the observables, the next logical step is to investigate how they
`act' on a {\em \shps}. Ideally, this way in the end we describe all 
properties of a \cft\ model in terms of its (maximal) chiral 
symmetry \alg\ \w. Now in the conventional operator approaches to
\qft, \h\ is a separable \hilsp\ and the observables are contained 
in the algebra of bounded linear operators on this \hilsp.
As already pointed out, here we are in a somewhat different situation.
Nevertheless, the following basic requirements are still implicit in 
the concept of a \sps:\\[.12em]
\Eins \h\ is a \repsp\ of the observable \alg, i.e.\ of \w.\\ 
   In the following I will employ the shorter term {\em module\/} common in 
   mathematics as a synonym for `\repsp';
   elements of \h\ are denoted by $v,w,...\,$.\\[.1em] 
\Zwei The \rep\ of \w\ on \h\ is unitary. Correspondingly, \h\ is endowed 
   with a positive hermitian product \iproo, so that it becomes a pre-\hilsp.
\\[.1em]
\Drei The {\em spectrum condition\/}: The energy is bounded from below. 
\vskip.12em

In order that the qualification `unitary' in property \zwei\ makes sense,
\w\ must be endowed with a 
generalized complex conjugation, i.e.\ a mapping $x\mapsto x^*$ satisfying
  \be  (x^*)^*_{} = x \,, \qquad (\xi x)^* = \bar\xi\, x \,, \qquad
  (xy)^* = y^*\,x^*  \labl*
(in mathematical terms: \w\ must be a $^*$-\alg), such that 
  \be  \ipro v{xw} = \ipro{x^*v}w   \ee
holds for the hermitian product on \h. In the last of the relations 
\erf* (and, likewise, often implicitly below), one works in fact within the 
universal enveloping algebra of \w,
i.e.\ with the associative \alg\ \uw\ that (roughly) can be described
as consisting of formal products of elements of \w, with the Lie bracket
of \w\ equal to the commutator in \uw.

In formulating the spectrum condition
\drei, it is in particular assumed that among the 
observables there is an energy operator. In \ccft\ this operator is provided 
by the Virasoro generator $L_0$, as follows from the
\role\ played by $L_0$ in applications. Briefly, $L_0$ acts as $-z\frac\partial
{\partial z}$ and hence measures the mass (or scaling) dimension, i.e.\
$L_0{\cal O}\eq\cD\Cdot{\cal O}$ for operators ${\cal O}\,{\sim}\,({\rm length})
^{-\cD}\,{\sim}\,({\rm mass})^\cD\,{\sim}\,({\rm energy})^\cD$.
Also, it is implicit in formulating property \drei\ that the action of 
$L_0$ on \h\ can be diagonalized, so that $(L_0)^*_{}=L_0$ by \zwei. The 
eigenvalue $\cD$ of $L_0$ on an eigenvector in \h\ 
is called the {\em\cdim\/} or {the }conformal {\em weight\/} of the vector.
Furthermore, together with the relation $[L_0,\ww in]\eq-n\ww in$ \erf{l0w}, 
property \drei\ implies that every vector
in \h\ is annihilated by all $\ww in$ with sufficiently large $n$, and
(because $L_1$ augments the $L_0$-eigenvalue by 1)
that any vector $v\iN\h$ satisfies $(L_1)^Nv\eq0$ for large enough $N$.

\vskip.3em
Let me now analyse the structure of \w\ a bit further. Because of its graded 
structure (see \equ \erf{grad}), as a vector space \w\ is the direct sum
$\w=\w^-\ooplus\w^0$\linebreak[0]$\ooplus\w^+$ of sub\alg s $\w^\pm$ and $\w^0$ 
which are given by
  \be  \w^\pm:=\Span{\ww in \mid i\iN I,\,\pm n\,{>}\,0} \,, 
  \qquad \w^0:=\Span{\ww i0 \mid i\iN I} \,.  \Labl0e
Further, again because of \erf{l0w} (namely, $[L_0,\ww in]\Ne0$ for $n\Ne0$),
the {\em zero mode sub\alg\/} $\w^0$ contains a maximal 
abelian subalgebra $\wo\!\subset\!\w$ with $L_0\iN\wo$, and there are
subalgebras $\wpm\subseteq\w^0\ooplus\w^\pm$ such that
  \be  \w = \wm  \ooplus \wo \ooplus \wp   \labl{tria}
is a {\em triangular decomposition\/} of \w, which means that
  \be  [\wpm,\wo\oplus\wpm]\subseteq\wpm \,, \qquad  
  [\wp ,\wm ]\subseteq \wo \,.  \labl{trib}

\lie s which enjoy a triangular decomposition possess a distinguished class of
modules, the so-called {\em \hwm s}. Such modules $V$ are by definition
generated by a {\em\hwv\/} \vhw. This means the following. First,
  \be  V \;\subseteq\; \U(\wm )\,\vhw \,,  \labl{hau}
i.e.\ all vectors in a \hwm\ $V$ can be obtained from the \hwv\ $\vhw
\iN V$ by acting with the \uenva\ of $\wm $; thus
the elements of $\wm $ play the \role\ of creation operators.
Second, all elements of $\wp $ act on \vhw\ as annihilation operators:
  \be  x_+\vhw = 0 \quad{\rm for\ all}\;\ x_+\iN\wp   \,.  \ee
And third, \vhw\ is an eigenvector of the abelian algebra $\w\o$.
In other words, for any \hwv\ $\va$ there is a linear function 
$\lama\dop\w\o\to\complex$ such that
  \be  x\o\,\va = \lama(x\o)\cdot \va \ee
for all $x\o\iN\w\o$; I will 
call $\lama$ the {\em weight\/} of $\va$ \wrt $\wo$.

Owing to $L_0\iN\w\o$ and relation \erf{l0w}, every \hwm\ of \w\ satisfies
the spectrum condition \drei. However, not any arbitrary \hwm\ of \w\ 
qualifies as a subspace of \h: the unitarity property \zwei\ must be 
satisfied as well.\,%
\futnote{While unitarity is an obvious requirement in \qft, in statistical 
mechanics systems it is not always necessary. Nevertheless the \w-modules 
arising in \cft\ applications to statistical mechanics are 
\hwm s (except in so-called `logarithmic' \cfts, which I disregard here, 
compare \refsY), for a reason which is still mysterious to me.}
Now every \hwm\ $V$ is already endowed with
a natural hermitian product; this is uniquely 
specified by the value on the \hwv, say $\Ipro\vhw\vhw\eq1$, namely by choosing
a basis of \w\ in such a way that
  \be  (\ww in)^* = \ww{\bar i}{-n}  \,, \labl{w*}
where $i\mapstO\bar i$ is some involutive permutation of the index set,
and then implementing the property \erf{hau}. If the so obtained product is 
degenerate, then the subspace of $V$ consisting of all {\em null vectors\/}, 
i.e.\ vectors \vnull\ which 
are orthogonal \wrtt form \iproo\ to all of $V$ (i.e.\ $\ipro\vnull{w}=0$ for 
all $w\iN V$), is a submodule of $V$, and hence
$V$ is reducible (but not fully reducible). On the other hand, every unitary
\hwm\ $V$ is in fact also irreducible, i.e.\ \w\ 
acts on $V$ by an \irrep\ $R$.

Now for any vector $\va$, a simple way to construct a
\hwm\ is to act {\em freely\/} with the enveloping \alg\ $\U(\wm )$; the module
  \be  \Va = \U(\wm )\,\va  \,.  \Labl1v
obtained by this construction is known as the {\em Verma module\/}
generated by $\va$. Verma modules are typically neither irreducible
nor unitary; however, in the cases of interest in \cft\ the
submodule of \Va\ consisting of null vectors is a maximal submodule,
and by `setting the null vectors to zero' (or, in more mathematical terms, by
taking the quotient of \Va\ by this submodule) one arrives at a module
\ha\ which is unitary and irreducible.
Also, any \hwm\ with \hwv\ $\va$ can be understood as some quotient of \Va.
It follows that each irreducible sub-\w-module of the space \h\ can be 
obtained in this fashion, with $A$ some index labelling the module.

The \cwei\ of the \hws\ $\va$ of \ha\ will be denoted by $\cd A$.
When $v\iN\ha$ is an eigenvector of $R_A(L_0)$\,%
\futnote{Henceforth, the symbol $R$ will almost always be suppressed.}
of eigenvalue $\cD$, then $m\,{:=}\,\cd A\Mi\cD$ is a non-negative integer, 
which for $v\eq\ww{i_1}{-m_1}{\cdots}\,\ww{i_l}{-m_l}\va$ 
($m_p\,{\ge}\,0$ for $p\eq1,2\Ldots l$) can be computed as 
$m\eq\sum_{p=1}^l\! m_p$; $m$ is called the {\em grade\/} of $v$.

\sect{Sectors: The Spectrum}D\label{s4}%
One can summarize some of the results reviewed above by stating that the
properties \eins\ to \drei\ imply that \h\ has the structure of a 
(possibly infinite) direct sum 
  \be  \h = \Bigoplus A \ha   \labl{hha}
of \uihwm s \ha\ of \w.
However, these requirements do {\em not\/} specify which of the various
\uihwm s of \w\ appear in the sum \erf{hha}. The set of \irmod s which do
appear constitutes the {\em spectrum} of the \cft; these spaces are also called 
the {\em physical\/} \irmod s of \w, or the superselection sectors
of the theory, or briefly the {\em sectors}.

Any observable which appears as a central element of \w\ not only acts as a 
constant in each sector \ha\ (this is just a corollary of Schur's lemma), but 
in fact it must act by one and the {\em same\/} constant on all sectors.
This condition constitutes a strong restriction on the spectrum. Its origin 
is that central charges cannot be localized; in the context of local observables
which are required to form a simple associative \alg, any central charge must
therefore be a scalar multiple of the unit operator. Frequently this requirement
also arises, at a more practical level, from the fact that one must work with
expressions which only make sense when these charges are regarded as numbers;
cf.\ e.g.\ the Sugawara formula \erf{sug} below.

It is an important result, known as {\em naturality}, that
once the eigenvalues of all central elements of \w\ are fixed (and provided 
that \w\ is maximal), in the spectrum of a chiral half of a \cft\
each (isomorphism class of) unitary \ihwm\ appears precisely once.
(The derivation of this assertion \Cite{mose2} is beyond the scope of these
lectures.) When the number of sectors is finite, then the \cft\ is called a
{\em rational\/} theory. This name originates from the observation that
in a rational \cft\ both the eigenvalues of (canonically normalized)
central elements and all \cdim s $\cd A$ are rational numbers.

Among the spectrum of a \cft\ there is a distinguished sector \ho\ with
$\cd\circ\eq0$. Its presence is implied by
another basic property of \qft, namely the existence of a vacuum state.
This is a vector \vo\ in \ho\ (unique up to scalar multiplication)
which is invariant under all unbroken symmetries of the theory. In the
present context this means that \vo\ is in particular $\w\o$-neutral, i.e.
$x\o\vo = 0$ for all $x\o\iN\w\o$, or in short, $\lamo\equiv0$.
Naively one might expect that \vo\ also has all other symmetries of the theory,
in particular that it is conformally invariant, i.e.\ is annihilated by 
{\em all\/} $L_n$. But for $c\ne0$ this would be incompatible with the 
relations \erf{vir} of \vir. Rather, one can only require that \vo\ be
annihilated by $L_n$ with $n\,{\ge}\,{-}1$.
Similar remarks apply to the action of other elements $\ww in\iN\w$ on
the vacuum; in short, \vo\ respects the maximally possible number of 
symmetries of the theory. As a consequence, \vo\ is a \hwv\ of \w.
The irreducible submodule \ho\ of \h\ which contains the special vector 
\vo\ is called the {\em vacuum sector\/} of the theory.

That \vo\ is annihilated by $L_n$ with $n\,{\ge}\,{-}1$ means that it
is both a \hwv\ \wrt \vir\ and invariant under the subalgebra 
  \be  \virP:=\virp\equiv \Span{L_1,L_0,L_{-1}} \;\cong\; \sltwo  \labl{su11}
of \vir\ (the {\em projective}, or {\em M\"obius}, transformations); this
subalgebra is isomorphic to \sltwo.\,%
\futnote{When the field of scalars is restricted from
\Complex\ to \Reals, one obtains the real form $\suee\cong\sltwor$ of \sltwo.
Together with the anti-holomorphic counterparts $\bar L_0,\,\bar L_{\pm1}$ one
arrives at a real form of $\liefont{so}(4)$; this is precisely the
conformal algebra that one would get when `naively' extrapolating from
arbitrary $D>2$ to $D=2$.}
Note that the projective transformations are precisely those whose classical
counterpart is well-defined on the whole Riemann sphere so that they can be
exponentiated so as to provide finite conformal transformations. Namely, the
action of $-z^{n+1}\partial_z$ is
non-singular at $z=0$ only for $n\ge-1$, and non-singular at $z=\infty$ 
(where it acts as $w^{1-n}\partial_w$ with $w=1/z$) only for $n\le1$.

Let me also remark that in some applications one may well need the full 
conformal invariance, so that according to the remarks above the
conformal symmetry must not have an anomaly, i.e.\ one needs $c=0$. This 
happens e.g.\ in string theory. In that case there is an
extra contribution to $c$ from the ghost fields which are employed to 
gauge-fix \twodim\ gravity, so that the requirement is that $c_{\rm matter}=
-c_{\rm ghost}$, e.g.\ $c_{\rm matter}\eq26$ for the bosonic string.\,%
\futnote{In this context it is less obvious that (the matter part of) the
\cft\ should be unitary.
(What one has to achieve is unitarity on a certain cohomology of \vir\,; for 
this it is certainly helpful to start from a unitary \cft, but I do not know
whether this is really necessary.)
Also, typically one `compactifies' the string theory by stipulating that 
$c_{\rm matter}$ is the sum of a contribution from space-time and one from
some `internal' theory. When the space-time is flat, each
space-time dimension corresponds to one free boson and hence (see 
\refsL) contributes 1 to $c$\,; but it is not clear to me why in 
our world where space-time is only asymptotically flat its contribution to
$c$ should have precisely the integral value $c\eq4$.}

For unitarity, the numbers $\cd A$, and hence the \cwei s of all 
$L_0$-eigenvectors in \h, must in fact be non-negative. Namely, consider first 
vectors $v$ which are {\em quasi-primary}, which means that they are annihilated
by $L_1$. Then due to $[L_1,L_{-1}]\eq2L_0$ and $(L_{-1})^*_{}\eq L_1$ one 
has $\|L_{-1}\,v\|^2 = 2\cd v\, \|v\|^2$, which by unitarity
implies that $\cd v\,{\ge}\,0$. (Also, $\cd v\eq0$ iff 
$L_{-1}v\eq0$, in which case together with $L_1 v\eq0$
it follows that $v$ is invariant under the projective subalgebra
\virP\ \erf{su11} of \vir.) Further, for any arbitrary vector $v\iN\h$
let $N$ be the smallest natural number such that $(L_1)^N v=0$ 
(which exists, see before \equ \Erf0e); then the vector
$\tilde v:=(L_1)^{N-1} v$ is non-zero and also quasi-primary. Since the 
\cdim s of $v$ and $\tilde v$ are related by $\cd{\tilde v}=\cd v\mi(N\Mi1)$,
$\cd v\,{<}\,0$ would imply that $\cd{\tilde v}\,{<}\,0$, which as already 
discussed cannot happen.

A similar argument shows that $\|L_{-n}\vo\|^2=(n^3\mi n)\Cdot c/12$ for all $n
\iN\naturals$, so that (taking $n>1$) for unitarity it is required that $c>0$.
Further constraints on the spectrum are obtained by studying the positivity
of $\|\ww{i_1}{\!-n_1}\!\cdots\ww{i_p}{\!-n_p}\va\|$ systematically. That is, 
at any fixed grade one considers the matrix of 
inner products of all basis states; a necessary condition for unitarity is that
each such matrix has \nonneg\ determinant (one needs not impose strict
positivity, because one can `decouple' null vectors, which corresponds to
quotienting the Verma module \Va). To analyze this condition one writes the 
inner  products as expectation values of \w-generators \wrt the \hwv, which can
be calculated by using the bracket relations of \w. In practice this can 
become very hard; for $\w\eq\vir$ the result is that for $c\,{\ge}\,1$ there 
is no restriction on the \cdim s, while below $c\eq1$ only the discrete set
  \be  c=c_m:=1-\Frac6{m(m+1)} \qquad{\rm with}\;\  m\ge3 \labl{mimo}
of $c$-values is allowed, and for each of these specific values only a finite 
set
  \be  \cD=\cD_{m;p,q}:=\Frac{((m+1)p-mq)^2-1}{4m(m+1)} \qquad{\rm with}\;\
  1\le p,q\le m\mi1  \ee
of \cwei s is possible; the theories with these values of $c$ and $\cD$ are 
known as the minimal unitary models of the \vira. 
Thus all unitary theories with $c\,{<}\,1$ are rational. But even when the 
unitarity requirement is dropped, one still needs $c\,{<}\,1$ for rationality. 
In other words, for rational theories with central charge
$c\ge1$, the chiral symmetry \alg\ \w\ is always larger than \vir.

\sect{\Con Fields}E\label{s5}%
Even though one is ultimately interested in observable quantities only,
i.e.\ in matrix elements of observables, in \qft\ it is often 
convenient to employ non-observable objects in intermediate steps of 
calculations. When such objects are operators on the space \h, they
are referred to as {\em fields}. Generic fields are distinguished from 
the observables by the fact that in terms of the sector decomposition 
\erf{hha} of \h, they act as $\ha\,{\to}\,\hbe$ with $B\Ne A$, while
observables act within each individual subspace \ha.\,%
\futnote{Such a distinction only makes sense if one demands that the
superposition principle is not universally valid. In other words,
only those vectors in \h\ which lie in one of the subspaces \ha\ are 
regarded as corresponding to proper physical states.}
In \cft, one realizes field operators $\varphi$ via 
the so-called {\em state-field correspondence\/}; the basic idea is that 
one generates all vectors in \h\ from the vacuum \vo\ by 
applying suitable point-like {\em \con fields\/} $\varphi\zbz$, according to
  \be  \vdd = \lim_{z,\bar z\to0} \vpdd(z,\bar z)\,\vo  \,.  \labl{vdd}
Here the labels indicate the \cwei s of the fields, \resp\ of the vectors 
$\vdD$ (usually more labels are needed to distinguish the fields, 
especially when $\w\!\supset\!\vir$). The formula \erf{vdd} is often read 
as the definition of the vector $\vdD\iN\h$; in the present
spirit, it rather specifies the \role\ played by the fields $\varphi$.

The precise meaning of the prescription \erf{vdd} and of the formal 
objects $\varphi$ introduced there is not at all obvious, and depending on the
framework adopted to interpret these quantities it can be a delicate
issue to verify that the limit in \erf{vdd} makes sense. To investigate this
one best starts with the case where the vector $\vdD$ lies in the 
vacuum sector \ho, so that the fields are actually observables. 
Now the maximality requirement imposed on the \chsa\ \w\ means in particular
that any operator on \h\ which does not belong to \w\ changes the sector;
thus in order for $\vdD$ to lie in \ho, the fields must be suitable 
combinations of the generators $\ww in$. The correct interpretation of the 
fields is then as generating functions
  \be  \Wz i = \sum_{n\in\zet} z^{-n-\cd i}\, \ww in \,,  \Labl Wz
or in other words, the generators of \w\ are regarded as moments or 
{\em modes\/} of $\Wz i$.  In the case of the Virasoro generators, this 
generating function 
  \be  T(z) = \sum_{n\in\zet} z^{-n-2}\, L_n \,,  \Labl Tz
plays the \role\ of the {\em\emt\/}; that is, in models which can also be
formulated in a Lagrangian setting, it describes the response of the Lagrangian 
density to a variation of the space-time metric.\,%
\futnote{This manipulation makes sense even for theories on flat 
space-times, because the operations of variation and of taking the limit that 
the metric becomes flat do not commute.}
One must note that in the formula \erf{Wz} the variable $z$ is introduced
just as a formal indeterminate. When $z$ is interpreted in terms of a world 
sheet, and hence is regarded as a complex variable, then the $\ww in$ are just 
the Fourier\hy Laurent modes of $\Wz i$, and one must worry about the
convergence of expressions like \erf{Wz}. One way to attack this problem
is to think of fields as operator valued distributions, and to work with
localized bounded operators (by smearing with test functions and 
exponentiating), which can e.g.\ be analyzed in the framework of
Wightman field theory \Cite{STwi} or of the
algebraic theory of superselection sectors \Cite{HAag}. Here I rather
describe the (much less rigorous) formulation that is based on the work
of Belavin, Polyakov and Zamolodchikov \Cite{bepz} and is sometimes called
the {\em bootstrap approach}. Another
alternative, at least as far as the mathematical aspects of the theory, rather
than its physical interpretation, are concerned, is to think of $z$ as a 
{\em formal\/} (in the well-defined mathematical sense) variable; this 
leads to the concept of a {\em \voa\/} \Cite{FRlm}.

Interpreting $z$ as a complex variable, one can invert 
the relation \erf{Wz}, so as to write the
Laurent modes as contour integrals of fields multiplied with powers of $z$,
  \be  \ww im = \intopiz z^{m+\cd i-1}\,\Wz i  \labl{int0}
(recall that $\cd i\iN\zet$), where integration is over some curve encircling
zero.

When it comes to genuine fields $\varphi\zbz$ rather than observables,
it is worth to return to the issue of which properties characterize a `\cft'.
What one has to demand is that the `space of all fields' carries a \rep\
of the conformal group, \resp\ of the conformal algebra -- analogously
to what one is used to from Lorentz covariant theories in which the fields
carry an action of the Lorentz group and in which, as a consequence, the 
excitations can be classified by \rep s of that group.
But not all fields one can think of are on the same footing. In particular, 
the following ones are distinguished:\\[.15em]
\Nxt The (\vir-)\,{\em\pf s\/}. These are defined by the requirement that 
the state $\vdD$ \erf{vdd} is a \viR-\hws, i.e.\ is annihilated by all $L_n$ 
with $n>0$.\\[.1em]
\Nxt The (\vir-)\,{\em\qpf s\/}. For these, $\vdD$ is annihilated by $L_1$, and 
hence is a \hws\ of the M\"obius subalgebra \virP\ \erf{su11} of \viR. 
\vskip.1em

In the relations of the 
\sltwo-algebra \virP\ the central term is absent, so that one can identify the 
generators $L_0,\,L_{\pm1}$ with the corresponding generators of the Witt \alg\
\erf{witt}. \Qpf s are therefore also called `non-derivative' fields.
In terms of the Laurent modes, the distinguished behavior of these types of
fields is demonstrated by \Erf1X, which in terms of fields can be rewritten
as $[L_m,\Wz i]\eq z^{m+1}\partial\Wz i+(m{+}1)\cd i z^m\Wz i$: when 
\Erf1X holds for $m\eq0$ and $\pm1$, then
$\Wz i$ is \qp, while if it holds for all $m\iN\zet$, then $\Wz i$ is primary.
Analogous relations hold for every \qppf\ $\phi$:\,
$[L_m,\phi(z)]\eq z^{m+1}\partial\phi(z)+(m{+}1)\cd\phi z^m\phi(z)$;
in particular, in the limit $z\,{\to}\,0$ one obtains
  \be  [L_n,\phi(0)] = \left\{ \bearll 0 & {\rm for}\ n>0 \,, \\[.1em]
  \cD\cdot\phi(0) & {\rm for}\ n=0 \,, \\[.1em] \partial\phi(0)& {\rm for}\
  n=-1 \,. \eear\right.  \Labl Lp

Non-primary fields are called {\em secondary\/} fields or
{\em descendants\/} (and also the term `ancestor' for primary field is 
used). Among the fields that correspond to the vectors in an \ihwm\ \ha\ of \w\
there is precisely one primary field $\phi_A$; it corresponds to the
\hwv. The collection of these fields, consisting of the primary $\phi_A$
and all its descendants, is called the {\em family\/} of $\phi_A$ and is denoted
by $[\phi_A]$. By the {\em grade\/} of a descendant of $\phi_A$ one means the
grade of the corresponding vector in $\ha$.

The simplest example of a \pf, present in every \cft,
is the {\em identity field\/} \bfe. It satisfies $\partial\bfe=0$, 
and hence corresponds to the vacuum vector \vo; its first non-trivial 
descendant is the \emt\ $T(z)$:
  \be  \lim_{z\to0} T(z)\,\vo = \Sum{n\in\zet}z^{-n-2}L_n\vo=L_{-2}\vo \,. \ee
$T(z)$ is the prime example of a \qpf.

The Lie brackets of the Laurent components of all \qpf s in the chiral
symmetry \alg\ \w\ read
  $  [\ww im,\ww jn] \eq d_{ij} p^i\, \delta_{m+n,0}
  + \sum_k C^{ij}_k q^{ijk}_{mn}\, \ww k{m+n}  $,
where the structure constants $p^i$ (which are combinations of eigenvalues of
central charges) and $q^{ijk}_{mn}$ are rational numbers for which explicit
expressions in terms of $\cd i$, \resp, $m,\,n$ and $\cd i,\,\cd j,\,\cd k$
are known \Cite{bfknrv}. Here I use the notations
  $d^{ij}\df\vev{\ww i{\!\cd i}\ww j{\!-\cd j}}$ and $C^{ijk}\df
  \vev{\ww i{\!\cd i}\ww j{\!-\cd j}$\linebreak[0]$\ww k{\!\cd j-\cd i}}$
as well as $\sum_j\!d^{ij}d_{jk}\eq\delta^i_{\,k}$ and 
$C^{ij}_k\df\sum_l\! C^{ijl} d_{lk}$,
where $\vev{\cdots}$ denotes the {\em vacuum expectation value\/}, defined as
  \be  \vev X := \ipro\vo{X\,\vo} \,.  \labl{cf}

Bounded operators on a Hilbert space can be multiplied. Analogously one would 
like to multiply in a suitable way the fields $\varphi\zbz$ that occur in the
present setting. Just like in other approaches to \qft\ there
arises the problem that such products tend to be singular,
at least in the limit of `coinciding points'. A way out is to consider
{\em normal ordered\/} expressions, i.e.\ set 
  \be  \normord{\ww im \ww jn} = \left\{ \bearlll
  \ww im \ww jn &\;& {\rm for}\ n>0\,,  \\[.3em]
  \ww jn \ww im &\;& {\rm for}\ n\le0  \eear\right. \labl{normord}
(say) for bilinears in the modes $\ww in$. (The normal ordering
prescription adopted here is not at all the only possible one. Whenever one
changes the treatment of only a finite number of terms (in a quite arbitrary 
manner), one obtains another sensible normal ordering.)
The normal ordered products \erf{normord} share the property of the generators
$\ww in$ to act locally nilpotently on \h. 
Also, the normal ordering of fields $\Wz i$ is obtained from the normal
ordering of their modes $\ww in$ via the series expansion \erf{Wz}. It follows 
e.g.\ that all commutators of normal ordered products of fields in \w\ are fully
determined by those of the \emt\ and of the Virasoro-primary fields in \w,
and that normal ordered products of \qpf s are generically not quasi-primary.

\sect{The \OPA}G\label{s6}%
In a chiral \cft\ -- a chiral half of a \twodim\ theory -- the fields depend
on a single complex variable $z$. As long as one deals with a single field,
say $\Wz i$, it suffices to describe it for values of $z$ which lie on a closed 
curve encircling zero, say the unit circle, since then one recovers all its 
modes $\ww in$ by the formula \erf{int0}. However, as soon as one intends
to analyze products of fields, one must at least `smear' the circle. 
Namely, when expressed in terms of fields, the abstract Lie bracket of modes
should correspond to the commutator \wrt a suitable associative product; now
inserting \erf{int0} into a Lie bracket yields
(with a suitable labelling of integration variables) the difference
  \be  [\ww im,\ww jn] = \LLb \intopiz\,\intopiw \mi \intopiw\,\intopiz\!\LRb \,
  z^{m+\cd i-1}w^{n+\cd j-1}\,\Wz i\,\Ww j  \Labl zw
of two double contour integrals, where in each term the integrations must be 
performed in the indicated order. One would like to rewrite this expression
in a form where there is a single contour integration over fields
$\Wz k$ multiplied with suitable powers of the position, so as to
recover the result for the Lie bracket $[\ww im,\ww jn]$ as a linear
combination of modes $\ww kp$. To achieve this, one  
interprets the first pair of integrations as to be performed for
$|z|\!>\!|w|$ and the second for $|w|\!>\!|z|$. Thus the formal product of
fields appearing in \Erf zw is to be understood as
the {\em radially ordered product\/} \rop, defined for arbitrary fields
$\varphi_1$ and $\varphi_2$ by
  \be  \rop\llb \varphi_1(z)\,\varphi_2(w) \lrb := \left\{ \bearlll
  \varphi_1(z)\,\varphi_2(w) &\;& {\rm for}\ |z|>|w|\,, \\[.3em] 
  \varphi_2(w)\,\varphi_1(z) &\;& {\rm for}\ |w|>|z| \,. \eear\right. \labl{rop}
For fixed value of $w$ (say) one can now deform the two $z$-contours in 
the formula \Erf zw and merge them
to a single one encircling the point $w$. One then arrives at the desired
result for the commutator by postulating that the radially ordered product
of two fields whose positions are sufficiently close can be written as
a linear combination of fields evaluated at (say\,%
\futnote{One may also choose the positions of these fields at $z$, or also
at $(z+w)/2$ or $\sqrt{wz}$ or any other function which smoothly approaches 
$w$ in the limit $z\to w$.}\mbox{$ $} \hsp{-.4})
$w$, multiplied with suitable powers of $z\!-\!w$. The so obtained 
decomposition is called the {\em\ope\/} of the fields $\Wz i$ and $\Ww j$.

When one knows the structure constants of \w, one can determine the singular 
part of \ope s by requiring that the contour integrations $\intow\!\intozw$ 
yield the correct result. E.g.\ the \vira\ amounts to the product
  \be  \rop\llb T(z)\,T(w) \lrb = \Frac{1/2}{(z-w)^4}\,C + \Frac2{(z-w)^2}\,T(w)
  + \Frac1{z-w}\,\partial_wT(w) + {\cal O}_{\rm reg}  \,.  \Labl33
Here ${\cal O}_{\rm reg}$ stands for (an infinite power series of) terms
which in the limit $z\to w$ are regular and hence do not affect the result for
$[L_m,L_n]$. (More generally, only singular terms in expansions like \Erf33
are relevant to the observables; the regular terms in fact do not have any 
meaning independent of the sector, e.g.\ their precise form depends on the 
choice of normal ordering prescription in the 
different sectors.) It is common to refer to radially ordered products 
\erf{rop} just as {\em operator products\/} and to omit the symbol \rop; 
I will usually follow this habit.

In the \twodim\ theory, we have both the chiral and the antichiral algebras 
\w\ and 
\wb; they commute, and correspondingly the operator product of fields $\Wz i$
and $\Wwb j$ just coincides with their normal ordered product. But we may
also consider operator products of the $\Wz i$ with arbitrary fields $\varphi$.
For instance, for the operator product of a \pf\ with $T$ one finds 
  \be  {T(z)\, \phi}(w,\bar w) = \Frac\cD{(z-w)^2}\,\phi(w,\bar w)
  + \Frac1{z-w}\,\partial_w\phi(w,\bar w) + {\cal O}_{\rm reg} \,, \labl{tf}
which upon Laurent expansion correctly reproduces the formula \Erf Lp
for the commutator $[L_n,\phi]$.

The singular terms in an operator product $\varphi_1(z)\varphi_2(w)$ of the 
type above are called the contraction of $\varphi_1$ and $\varphi_2$ and 
are denoted by $\Cont{\varphi_1(z)\varphi_2}(w)$.
Also, the term of order $(z-w)^0$ is essentially the normal ordered product 
$\normord{\varphi_1\,\varphi_2}$ -- when the particular normal ordering 
prescription 
  \be  \normord{\varphi_1(z)\,\varphi_2(z)}:= \intopizw (w-z)^{-1}\,\rop\llb 
  \varphi_1(z)\,\varphi_2(w) \lrb \ee
(which is slightly different from the one presented in \erf{normord}) is 
adopted, then it is indeed precisely the normal ordered product, i.e.\ one has
  \be  \varphi_1(z)\varphi_2(w) = \Cont{\varphi_1(z)\varphi_2}(w)
  + \normord{\varphi_1(z)\varphi_2(z)} + O(z\Mi w)  \,.  \ee

Besides the conformal invariance, a second basic property of \twodim\ \cfts\
is that operator products cannot only be defined when one of the fields
belongs to the chiral algebra, but for arbitrary pairs of fields. That is,
there is a {\em closed \opa\/} among {\em all\/} fields, and the operation
involved is indeed a 
product in the sense that it is associative. It is expected that this structure
is a direct consequence of fundamental properties of \qft. But when trying 
to construct it this way one faces quite a few subtleties, and to the best of 
my knowledge a complete derivation from first principles has never been 
given. Instead, one usually postulates the existence of a closed associative
\opa\ as a {\em separate input\/}, called the {\em bootstrap\/} hypothesis or 
requirement. Note, however, that \ope s do {\em not\/} converge in the operator 
norm, but only weakly, i.e.\ when
applied to a vector in the state space \h. Thus while one may regard the
collection of all fields as a vector space (over some function field), 
the operator product does not really make this space into an associative
\alg\ in the mathematical sense.\,%
\futnote{The theory of \voa s \Cite{FRlm} provides a
means for formulating operator products in a
purely \alg ic setting, without having to resort to arguments based on
complex analysis, e.g.\ contour deformation or analytic continuation.
The basic structures of a vertex operator algebra are an
infinite set of products and the action of the derivative $\partial$;
the commutator \wrtt first of these products yields the \lie\ structure of
the space $\wo$ of zero modes. The variables $z$ etc.\ are treated as formal 
variables, which implies e.g.\ that $(z\mi w)^{-1}$ is interpreted as 
$z^{-1}(1\Mi w/z)^{-1}=z^{-1}\sum_{n\ge0}(w/z)^n$. 
Thus in particular $1/(z\mi w)+1/(w\mi z)=z^{-1}\sum_{n\in\zet}(w/z)^n$,
which in terms of complex analysis corresponds to the delta function on the 
unit circle. The relation with the radial ordering prescription is that upon
re-interpreting the variables as complex numbers, the formal power series 
occurring in the vertex operator formulation become convergent series precisely 
if the relevant numbers are radially ordered in the appropriate manner.}

Operator products of primary fields $\phi$ can be written in the form
  \be  \phi_{\!A}^{}\zbz\,\phi_{\!B}^{}\wbw \eq \sum_C\cijk\,
  (z\Mi w)^{\cd C-\cd A-\cd B}
  (\bz\Mi\bar w)^{\cdb C-\cdb A-\cdb B}\phi_{\!C}^{}\wbw+\ldots, \Labl oa 
where the ellipsis stands for contributions of descendant fields 
$\varphi_C\iN[\phi_C]$. The numbers $\cijk$ introduced in \Erf oa are called 
the {\em\opc s\/} of the theory. The corresponding coefficients
which involve descendants are completely fixed by the \w-symmetry
in terms of those of the primaries (and of their \cwei s and the
values of central charges). For instance, the coefficient with which the
descendant $\partial^n\varphi$ of a \qpf\ $\varphi$ appears equals the
coefficient for $\varphi$ multiplied by $(n!)^{-1}\prod_{l=0}^{n-1}
(l+\cd A-\cd B+\cd\varphi)/(l+2\cd\varphi)$. On the other hand, the relation 
between the \opc s for \qps\ and those for primaries has been worked out in
some detail only for $\w\eq\vir$ \Cite{bepz}.
Nevertheless, at least in principle it is possible to solve the theory 
completely, i.e.\ to compute all correlation functions, by expressing them 
in terms of the coefficients $\cijk$ involving only primaries and of
the structure constants of \w. The determination of the numbers $\cijk$ 
is therefore one of the main goals in \cft.

\sect{Correlation Functions and Chiral Blocks}F\label{s7}%
Operator product expansions like \Erf tf or \Erf oa can only be valid 
when applied to vectors in \h. In fact, because the presence of additional 
fields typically affects manipulations with integration contours,
strictly speaking they even only hold when matrix elements \wrt vectors
in \h\ are taken, e.g.\ for vacuum matrix elements
  \be \bearll  \calg(\{z_j\},\!\{\bz_j\})\!&\, \equiv\vev{\varphi_1(z_1,\bz_1)\,
  \varphi_2(z_2,\bz_2)\cdots\varphi_p(z_p,\bz_p)} \\[.5em] &\!
  :=\ipro\vo{\varphi_1(z_1,\bz_1)\,\varphi_2(z_2,\bz_2)\cdots
  \varphi_p(z_p,\bz_p) \,\vo} \,. \eear \labl{Cf}
Such a vacuum expectation value is called the ($p$\,-point) {\em correlation 
function\/} of the fields $\varphi_1,\,\varphi_2 \Ldots\varphi_p$.

Given the \opa, one can in principle evaluate any \corfu\ by expressing
successively all products as linear combinations of fields until one ends up
with a linear combination of fields acting on $\vo$,
and then using the fact (see below) that the only field with non-vanishing
one-point function is the identity primary field \bfe. In practice, this 
procedure only works in simple cases such as free field theories.
However, some general properties of correlators can be derived without a
detailed knowledge of the operator products. They are implied by the
so-called {\em Ward identities\/} for \w, which are obtained as follows.
Consider a \corfu\ of the form $\vev{\Wz i\,$\linebreak[0]$\varphi_1
(z_1,\!\bz_1)\,$\linebreak[0]$\varphi_2(z_2,\!\bz_2)\cdots}$, multiplied with 
some power $z^n$, and perform a $z$-integration over a contour 
$\cal C$ that encircles all of the
`insertion points' $z_i$. By deforming $\cal C$ to infinity one learns
that for $n\,{>}\,{-}\cd i$ this integrated correlator vanishes; on the other
hand, after deforming the contour into a union of contours each of which 
encircles precisely one of the insertion points $z_j$, one can insert
the operator product $\Wz i \varphi_1(z_j,\bz_j)$ and perform the integration
for each of these contours separately. These manipulations amount to a linear 
\deq\ for the \corfu\ $\vev{\varphi_1(z_1,\bz_1)\,\varphi_2(z_1,\bz_1)\cdots}$.
In particular, in the case of $\Wz i\eq T(z)$, the operator product \Erf tf 
leads to the {\em projective\/} Ward identities
  \be  \sum_{i=1}^p z_i^n\,\llb z_i\Frac\partial{\partial z_i}+(n+1)\cd i\lrb
  \, \calg(\{z_j\},\!\{\bz_j\})=0 \qquad{\rm for}\;\ n\iN\{0,\pm1 \}  \labl{pwi}
for \corfu s of \pf s. When $\w\eq\vir$\,%
\futnote{When \w\ is strictly larger than \vir, a more detailed analysis 
is required.}
and the number $p$ of insertions is small, the projective 
Ward identities are particularly effective:
\nxt The relation \erf{pwi} with $n=-1$ tells us that \opf s are 
constant. Thus they satisfy
$\vev{\phi_A}=\!\raisebox{-.43em}{${\displaystyle\lim}\atop\scs z\to0$} \vev
{\phi_A(z)}=\ipro\vo\va$, which is non-zero only if $\phi_A=\phi\o=\bfe$.
\nxt 
In the case of \twpf s $\calg\,{\equiv}\,\calg_{\!AB}\eq\vev{\phi_A(z_1)
\,\phi_B(z_2)}$, the general solution of the $n\eq{-}1$ identity
is $\calg\eq\calg(z_1\mi z_2)$; the $n\eq0$ identity then gives 
$\calg_{\!AB}\,{\propto}\,(z_1\mi z_2)^{-\cd A-\cd B}$, and finally $n\eq1$ shows 
that $\cd A\eq\cd B$.
When the chiral symmetry algebra \w\ is maximally extended, the constant of
proportionality (which is not fixed by the Ward identities, since they are 
linear), is zero unless $\phi_2$ is the {\em conjugate field\/} of $\phi_1$,
which carries the complex conjugate \rep\ of \w.
Thus $\vev{\phi_A\,\phi_B}\propto\delta_{\!A,B^+}$, where $\phi_{A^+}^{}=
(\phi_A)^+$.
\nxt Similarly, \tpf s\ are completely fixed up to a multiplicative constant:
  \be  \calg_{\!ABC} \propto (z_1-z_2)^{\cd C-\cd A-\cd B} (z_2-z_3)
  ^{\cd A-\cd B-\cd C} (z_3-z_1)^{\cd B-\cd C-\cd A}  \,. \ee
\Nxt For \fpf s the situation is already more involved (they
will be studied in more detail in \refsN\ and \refsj). The projective 
Ward identities only imply that every \fpf\ can be written as an arbitrary
function of a certain variable $w$, multiplied with definite powers (which
are linear combinations of the \cwei s) of the coordinate differences
$z_i\mi z_j$. This variable $w$ is a cross-ratio of the insertion points;
e.g.\ one may 
choose $w=(z_2\Mi z_1)(z_3\Mi z_4)/$\linebreak[0]$(z_3\Mi z_1)(z_2\Mi z_4)$,
which is obtained by applying to all insertion points the (finite) projective 
transformation $z\mapsto(z_2\Mi z_1)(z\Mi z_4)/(z\Mi z_1)(z_2\Mi z_4)$, upon 
which $z_1,\,z_2,\,z_3,\,z_4$ are mapped to 
$\infty,\,1,\,w,\,0$. In other words, \Wlog\ we can set three of the insertion 
points to $\infty,\,1,\,0$, i.e.\ consider correlators of the form
  \be  \calgzz\equiv\calg_{\!ABCD}\zbz =
  \langle\phi_A(\infty,\!\infty)\,\phi_B(1,\!1)\,\phi_C\zbz\,\phi_D
  (0,\!0)\rangle \,.  \Labl11  
(In the case of \tpf s one can analogously
map the insertion points to the fixed values $\infty,\,1,\,0$, which explains
why in that case the dependence on the $z_i$ is completely fixed. More 
generally, any $p$-point function with $p\ge3$ can be written as a function 
of $p-3$ cross-ratios, multiplied by powers of the coordinate 
differences $z_i\mi z_j$.)  

The Ward identities involve only one chiral half of the theory,
so that when studying them one can ignore the presence of the antichiral half.
Of course, there are analogous identities involving the antichiral \alg, too. 
As a consequence, the \corfu\ of a {\em two\/}-di\-men\-si\-o\-nal theory that
is obtained by combining its two chiral halves can be expressed as 
a linear combination of the products of independent solutions \fmz\ and
\fmbarz\ of the chiral and antichiral Ward identities,
  \be  \calgzz=\summ \summb a_{I\bar I}\, \fmz\,\fmbarz \,.   \Labl14
The individual solutions \fmi\ and \fmbari\ are called the 
{\em chiral blocks\/} (or also the conformal blocks) of $\calg$.
 
For $p\,{>}\,3$ the sum \Erf14 contains, in general, more then one term (see 
also \refsN), and hence does {\em not\/} factorize into a chiral and an 
antichiral part. This means in particular that (except for the observables $\Wz 
i$) such a factorization is not possible for the fields $\varphi\eq\varphi
\zbz$.\,%
\futnote{Note in particular that the notation $\varphi(z)$ that I occasionally
used above was just an abbreviation to indicate the $z$-dependence of
$\varphi\zbz$.}
But still, as also indicated by the form \Erf14 of correlators, one can regard
primary fields $\phi$ as combinations
  \be  \phi_A(z,\bar z) = \Sum q c_q\, \varpi_q(z)\, \bar\varpi_{\bar q}(\bz)
  \Labl vv
of suitable chiral objects $\varpi_q$ and $\bar\varpi_{\bar q}$, which are
called {\em \cvo s}. In order for
this decomposition to make sense, for each primary $\phi_A$ the label $q$ 
must be understood as standing not only for $A$, but also for two additional
primary field labels $B,C$\,%
\futnote{as well as possibly some multiplicity label, compare \refsI.}
such that $\varpi_q\equiv\varpi_{A;B,C}$ constitutes a 
map $\h_B\to\h_C$ which intertwines the action of the chiral \alg, i.e.\
$\varpi_{A;B,C}\circ R_B(\ww in)=R_C(\ww in)\circ\varpi_{A;B,C}$.
In terms of operator products, the restriction to a specific {\em source\/}
$\h_B$ and {\em range\/} $\h_C$ corresponds to considering only the terms
involving the family $[\phi_C]$ in the operator product $\phi_A\phi_B$;
correspondingly, the coefficients $c_q\equiv c_{A;B,C}$ in \Erf vv are 
in fact nothing but the \opc s $\cijk$.

By construction, the chiral blocks can be interpreted as the vacuum
expectation values of suitable \cvo s, e.g.\ $\fmi\eq\vev{\varpi_{\circ;A,A}
\varpi_{A;B,I}\varpi_{I;C,D^+}$\linebreak[0]$\varpi_{D^+;D,\circ}}$ for the 
\fpf\ $\calg_{ABCD}$.
Chiral blocks are generically not functions, but are multi-valued.\,%
\futnote{Sometimes the blocks are also called {\em holomorphic\/} \resp\
anti-holomorphic blocks. Thus in this context the term `holomorphic' is used
in a rather loose sense; even the full \corfu s are in general
only {\em mero\/}morphic (they can have a pole whenever two
insertion points coincide).}
They are single-valued on a suitable multiple covering of the complex plane
(and hence can be regarded as sections of some (projectively flat) 
bundle over the Riemann sphere). Analytic continuation of a chiral 
block connects the different sheets of that covering.
Correspondingly, the exchange of two \cvo s is governed by a \rep\
of the braid group rather than the permutation group.

The requirement that the chiral and antichiral blocks combine to single-valued
\corfu s of the \twodim\ theory yields algebraic equations for the
linear coefficients $a_{I\bar I}$ in \Erf14. In principle, these equations
can be solved to obtain these coefficients up to over-all normalization; 
the latter is left undetermined because the Ward identities are {\em linear\/}
\deq s. For the \tpf s there is often only a single chiral block 
and hence only a single coefficient $a\equiv a_{ABC}$.
In this case comparison with the \opa\ \Erf oa shows that
$a_{ABC}=\cijK:=\sum_D\cijd d_{DC}$, where
  \be  d_{AB}\df(z_1-z_2)^{2\cd A}(\bz_1-\bz_2)^{2\cdb A}
  \vev{\phi_A(z_1)\,\phi_B(z_2)} \propto\delta_{A,B^+}^{}  \ee
plays the \role\ of a metric in the space of fields.
In short, the \tpf s of \pf s are essentially the \opc s.

\lect2{Fusion Rules, Duality and Modular Transformations}{.46}{.08}

\sect{Fusion Rules}H\label{s8}%
In the first lecture we learned that upon forming radially ordered products and 
when considered inside \corfu s, the fields $\varphi$ of a \cft\ realize a 
closed associative \opa. 
Unfortunately, in practice this structure looks extremely complicated.
However, a large amount of information about the \op s is already contained 
in a much more transparent structure, the so-called {\em\furu s\/}. Roughly 
speaking, the \furu s constitute the basis-independent contents of the \opa,
i.e.\ count the number $\nijk$ of times that the family $[\phi_C]$ of the \pf\ 
$\phi_C$ appears in the operator product of primaries $\phi_A$ and $\phi_B$. 
In other words, they tell how many distinct
couplings among primary fields, \resp\ \w-families, are possible.
To encode this information, one associates to each \pf\
$\phi_A$ an abstract object $\phj A$, and introduces 
an abstract multiplication `\,$\star$\,' by writing
  \be  \phj A\star\phj B=\Sum C\nijk\,\phj C \,. \labl{fr}
The integers $\nijk$ are known as the {\em \frc s}. Note that the product
\erf{fr} is {\em not\/} isomorphic to the ordinary tensor product
of modules of the \chsa\ \w.\,%
\futnote{This follows directly from the observation that upon forming tensor 
products all central charges add up (e.g.\ $c_{{\rm tot}} = c_1+c_2$),
whereas the fusion product yields an object which appears in the same theory and
hence has the same values of central 
charges as the original fields. (For a description of fusion products 
which closely resembles a ($z$-dependent) tensor product, see \Cite{gabe}.
Another approach to fusion products \Cite{wass2} is via Connes fusion.)}

Let me comment on how to actually read the \furu s off the \opa.\,%
\futnote{In practice, one rather proceeds the other way round, i.e.\
given the fusion rules, one determines (though not uniquely) \deq s for 
chiral blocks, and thereby the \opc s,
from general principles of \cft. For some details see \refsN.}
The \w-family $[\phi_C]$ occurs in the 
product of $\phi_A$ with $\phi_B$ iff $\nijk$ is non-zero. This
might suggest that $\nijk$ can take only the values 0 or 1,
corresponding to the alternative whether $[\phi_C]$
appears in $\phi_A(z) \phi_B(w)$ or not. But one and the same family
$[\phi_C]$ may couple in several distinct ways to $\phi_A$ and $\phi_B$ so that 
values $\nijk\,{>}\,1$ are possible as well. Accordingly, the fields 
$\varphi_C$ appearing in the expansion \Erf oa are not necessarily all
distinct, i.e.\ it may happen that $\varphi_{C_1}^{}\eq\varphi_{C_2}^{}$
for $C_1\Ne C_2$; nevertheless it is not allowed just to add up the
corresponding \opc s, because the relative values of coefficients
involving different members of a family $[\phi_k]$ are fixed by the
Ward identities of \w. If the full \opa\ is known,
the presence of a coefficient $\nijk$ larger than unity can therefore 
be inferred as follows. One subtracts the leading contribution to the
operator product $\phi_A\phi_B$, corresponding to the descendant
$\varphi$ of $\phi_C$ with the lowest allowed grade, and likewise 
all contributions involving descendants at higher grades whose
presence is dictated by the Ward identities; then $\nijk>1$ iff afterwards
there is still some contribution to $\phi_A\phi_B$ left
over which involves members of $[\phi_C].$ It follows e.g.\ that
for $\nijk>1$ the minimal grades of the fields that 
contribute to the $\nijk$~different couplings between $\phi_A,\,\phi_B$ and
$[\phi_C]$ must all be distinct. (E.g.\ there is at most one non-vanishing 
coupling among the {\em primaries\/} $\phi_A,\,\phi_B,\,\phi_C$;
this is actually a necessary
condition for excluding chiral blocks with logarithmic singularities.)

It turns out to be convenient to
formalize the general properties of \furu s which are implied by the 
principles of \cft\ that I outlined in lecture 1. To this end one regards the
objects $\phj A$ and numbers $\nijk$ as the basis elements and structure
constants of a ring over the integers \Zet\ or of an algebra over the
complex numbers \Complex\,. These structures are called the {\em fusion
ring\/}, \resp\ the {\em fusion \alg\/}, of the \con field theory; 
their defining properties are the following: \\[.15em]
\Fbx F1 they are commutative and associative, and they have a unit element
(namely, $\pho$, the abstract object associated to the identity \pf);\\[.1em]
\Fbx F2 there is a distinguished basis (namely the one consisting
of the $\phj A$) in which the structure constants are non-negative 
and which contains the unit element;\\[.1em]
\Fbx F3 the evaluation at the unit element provides an involutive automorphism,
called the {\em conjugation\/} and denoted by\,
$\phj A\mapsto(\phj A)^+_{}$.\\[.15em]
In terms of the structure constants $\nijk$, these properties read as follows.
\\[.1em]
\Fbx N1{}$\N BAC\eq\N ABC,\ \sum_{\sss D}\!\N ABD\N DCE\eq\sum_{\sss D}\!\N BCD
   \N DAE$\, and\, $\N\O BC\eq\delta_{\sss\!B}^{\sss\;C}$;\\[.18em]
\Fbx N2{}$\N BAC\ge0$\,;\\[.18em]
\fbx N3\mbox{$C_{\sss AB}\equiv\N AB{\,\O}=\delta_{\!\sss A,B^+}$} for some 
   order-two permutation $A\mapsto A^+$
   of the index set (such that $(\phj A)^+_{}\eq\Phi^{}_{\!A^+}$),
   and ${(A^+)}^+_{}\eq A$ as well as $\N{A^+}{B^+}{\ \ C^+}\eq\N ABC$.\\[.1em]
It then also follows that $\Ns ABC:=\N AB{C^+}$ is totally symmetric.

Note that in the identification of the $\phj A$ as a basis it is implicit that
for all $A,B$ the number $\sum_C\nijk$ is finite; \cfts\ which satisfy this
requirement are called {\em quasi-rational\/}.
A {\em rational\/} \cft, i.e.\ one with only a finite number of sectors, is 
also quasi-rational; while this is not manifest in the definition, it follows 
easily from the very existence of a fusion ring. It is worth
stressing that (quasi-)\,rationality is {\em not\/} a fundamental property of 
\cfts. However, in practice it is often indispensable, since it allows one
to perform many calculations explicitly.
This manifests itself for the first time when one studies the \rep\ theory
of the fusion ring. (Whether the concept of a fusion ring is still applicable in
non-quasi-rational theories is not known.) 
Accordingly, I will from now on restrict my attention to rational \cfts\ only,
unless stated otherwise.

The fusion ring of a rational \cft\ is a \findim\ commutative associative
ring. As a consequence, each of its \irrep s
is \onedim, and every \findim\ \rep\
is isomorphic to the direct sum of such \irrep s. In particular, the adjoint
\rep\ $\pi_{\rm ad}$, defined by $\pi_{\rm ad}(\phj A)\df\NNA$, where $\NNA$ 
denotes the matrix with entries $(\NNA)_{\sss B}^{\sss \ C}\eq\N ABC$, must be 
isomorphic to the direct sum of \onedim\ \irrep s (in fact, each inequivalent 
\onedim\ \rep\ appears in this sum precisely once). In other words, there 
exists a unitary matrix $S$ which `diagonalizes the \furu s' in the sense that 
-- simultaneously for all values of the label $A$ -- the matrix
  \be  \DA := S^{-1}\,\NNA\,S  \labl{D A}
is diagonal, and also the relation $CS=S^*$ is valid. 

It should be noted that the row and column labels of the matrix $S$ that is
introduced this way are a priori on a rather distinct footing: the row index
labels the elements $\phj A$ of the distinguished basis, while the 
column index counts the inequivalent \onedim\ \irrep s $\pi_B$.
While the two sets of labels have the same order (so that, as I already did
above, one can use the same symbols for either type of labels), in general 
there does not exist a canonical bijection between them.
As it turns out, however, for those fusion rings which occur in \cft\
such a canonical bijection does exist, and moreover, when implementing this
bijection the diagonalizing matrix $S$ possesses the highly non-trivial 
property of being symmetric, and also satisfies $S_{A\O}\,{>}\,0$ (in 
particular $S_{A\O}\iN\reals$) for all values of the label $A$. It then 
follows that the \onedim\ \irrep s $\pi_A$ obey
  \be  \pi_A(\phj B) \equiv (\DB)_A^{\;A} = {S_{BA}}/{S_{\O A}} \,. \ee
As a consequence, the relation \erf{D A} amounts to an expression\,%
\futnote{This is not yet the Verlinde \Cite{verl2} formula, because 
at this point there is no connection to modular transformations yet 
-- this interpretation must be postponed until \refsK.}
  \be  \nijk = \sum_D \frac{S_{AD}S_{BD}(S^{-1})_{CD}}{S_{\O D}} \labl{-verl}
of the \frc s in terms of the matrix $S$,
and also that $S^2\eq C$, so that \erf{-verl} can be rewritten more 
symmetrically as $\Ns ABC = \sum_D {S_{AD}S_{BD}S_{CD}}/{S_{\O D}}$. 

\sect{Duality}I\label{s9}%
The associativity of the \opa\ implies that the \furu s are associative as 
well. I will now use the same information to deduce properties of the \corfu s. 
Let us study the \fpf s $\calg=\Vev{\phi_A\phi_B\phi_C\phi_D}$ of \pf s. 
Applying the \opa\ on the first two and on the last two fields, one
has $\Vev{\phi_A\phi_B\phi_C\phi_D}=\Vev{(\phi_A\phi_B)\,(\phi_C\phi_D)}
\equiv\Vev{\rop(\rop(\phi_A\phi_B)\,\rop(\phi_C\phi_D))}$, which
pictorially amounts to
  \begin{eqnarray} &&{}\nonumber\\[-5.5em] &
\begin{picture}(85,80)(8,47) \put(50,50){\circle*{5}}
\put(50,50){\line(-1,1){22}} \put(50,50){\line(-1,-1){22}}
\put(50,50){\line(1,1){22}} \put(50,50){\line(1,-1){22}}
\put(19.6,75){$\scs B$}\put(76,75){$\scs C$} \put(19.6,20){$\scs A$}
\put(76,20){$\scs D$} \end{picture} 
=\ \dsum_I \begin{picture}(115,80)(8,47) \put(50,50){\line(1,0){30}}
\put(50,50){\line(-1,1){22}} \put(50,50){\line(-1,-1){22}}
\put(80,50){\line(1,1){22}} \put(80,50){\line(1,-1){22}}
\put(19.6,75){$\scs B$}\put(106,75){$\scs C$} \put(19.6,20){$\scs A$}
\put(106,20){$\scs D$} \put(63,41){$\scs I$} \end{picture}
  \label{p1} &\\[1.1em] &&\nonumber \end{eqnarray}
But using associativity of the \opa, one could also form products in a
different order (in fact this is implicit in the very notation: usually 
I do not write any brackets to indicate the order in which multiple products
are performed), resulting in pictures like
  \begin{eqnarray} &&{}\nonumber\\[-6.2em]
\dsum_K \begin{picture}(97,100)(-9,65) \put(30,50){\line(0,1){30}}
\put(32,63){$\scs K$} \put(0.5,104){$\scs B$} \put(54.5,104){$\scs C$}
\put(30,80){\line(1,1){22}} \put(30,80){\line(-1,1){22}}
\put(30,50){\line(1,-1){23}} \put(30,50){\line(-1,-1){23}}
\put(0.5,19){$\scs A$} \put(54.5,19){$\scs D$}  \end{picture}
  & \ {\rm or}\quad\quad &  
\dsum_J \begin{picture}(115,80)(8,47) \put(50,50){\line(1,0){30}}
\put(50,50){\line(5,2){51.5}} \put(50,50){\line(-1,-1){22}}
\put(80,50){\line(-5,2){12.5}} \put(28.5,70.6){\line(5,-2){33.5}}
\put(80,50){\line(1,-1){22}} \put(19.6,74.2){$\scs B$}\put(106,74.2){$\scs C$}
\put(19.6,20.6){$\scs A$}\put(106,20.6){$\scs D$} \put(63,41){$\scs J$}
\end{picture}
  \label{p2} \\[2.3em] &&\nonumber \end{eqnarray}
Of course, because of the radial ordering prescription each of these choices
is only valid for a definite order of the absolute values of the insertion 
points $z_i$ of the
fields $\phi_A\Ldots\phi_D$, and hence the different pictures describe
functions which coincide only upon an appropriate analytic continuation.
Now when the two chiral halves of a \twodim\ theory are combined, so that
$\bz_i$ is identified as the complex conjugate of $z_i$, one must require 
that every \corfu\ is single-valued on the Riemann sphere when the dependence 
on both $z_j$ and $\bz_j$ is accounted for; thus the three descriptions above 
for the \fpc s $\calg$ must yield one and the same
function. Together with \Erf14 this yields
the {\em crossing symmetry\/} relations for the \fpf s, which when 
choosing the insertion points as in formula \Erf11 read
  \be  \calg_{\!ABCD}\zbz = \calg_{\!BCDA}(1\mi z,1\mi \bz)=
  z^{-2\cd C}\bz^{-2\cdb C}\, \calg_{\!ACBD}(z^{-1},\bz^{-1}) \,. \Labl19
On the other hand, each of the
pictures describes the chiral blocks in a different `channel', and hence
combining the two chiral halves amounts to three different decompositions
of the \fpf\ as a sum over products of chiral blocks
(for more details, see \refsN\ below).
In a quasi-rational theory the sums are finite, and as a consequence
\Cite{lewe} the different systems of chiral blocks are linearly related:
  \be  \bearl 
  \hat{\cal F}_{\!ABCD,K}(z_1,z_2,z_3,z_4) =\Sum{\sss I}\,\F KIBCAD\, 
  {\cal F}_{\!ABCD,I}(z_2,z_3,z_4,z_1) \,,\\ {}\\[-.4em] 
  \check{\cal F}_{\!ABCD,J}(z_1,z_2,z_3,z_4)
  \,=\Sum{\sss I}\,\B JIBCAD\, {\cal F}_{\!ABCD,I}(z_1,z_3,z_2,z_4)
  \,. \eear  \labl{BF}
The coefficients $\F{}{}BCAD$ and $\B{}{}BCAD$ that appear in these linear
relations are called the {\em fusing\/} and {\em braiding matrices}, and
the corresponding manipulations with chiral blocks as visualized in the
pictures above are referred to as fusing and braiding transformations, or
generically as {\em duality\/} transformations. Strictly speaking, the indices 
$I$ and $J$ that I used to label the various blocks in a given channel
generically do not just count the intermediate primary fields, but for the case 
of \frc s larger than 1 also the various possible couplings among their 
families, i.e.\ they stand for multi-indices, e.g.\ $I\equiv(\alpha,I,
\beta)$ with $\alpha\iN\{1,2,\Ldots\N AB{\,I}\},\,\beta\iN\{1,2,\Ldots\N ICD
\}$. For notational simplicity, in \erf{BF} and also
below I suppress all multiplicity indices.\,%
\futnote{Those readers who want to see them are referred to e.g.\ \Cite{fugv3}. 
\\The relation with the notation used there is 
$\F{\alpha I\beta}{,\gamma J\delta}ABCD
=F^{(ABC)_{\scs D}}_{\alpha I\beta,\gamma J\delta}$ \,and\\
$\B{\alpha I\beta}{,\gamma J\delta}ABCD = \sum_K\sum_{\lambda,\mu,\nu}
((F^{(BAC)_{\scs D}}_{})^{-1})^{}_{\alpha I\beta,\lambda K\nu}
R^{(AB)_{\scs K}}_{\lambda,\mu}F^{(ABC)_{\scs D}}_{\mu K\nu,\gamma J\delta}$.}

The fusing and braiding matrices obey a number of compatibility relations, 
which can be derived
by considering suitable duality transformations of higher $p$-point functions 
and demanding that the result does not depend on the individual set of
transformations, but only on the initial and final configurations of 
(external and intermediate) fields. These relations can all be reduced to those 
obtained for
five-point functions and which involve five and six different configurations,
\resp\ \Cite{Mose}; the latter are called the {\em polynomial equations}, or 
more specifically, the {\em pentagon\/} and {\em hexagon\/} equations and read
  \be  \bearl 
  \F HIABEF\, \F JFCDEG = \Sum{\sss K}\F KFBDIG\,\F JIAKEG\,\F HKABJD \,,\\{}
  \\[-.5em] \Sum{\sss K}\B KIABDE\, \B GEACKF\, \B HKBCDG
  = \Sum{\sss L}\B LEBCIF\, \B HIACDL\, \B GLABHF   \eear \labl{poleq}
(note that the required summations do {\em not\/} amount to matrix 
multiplication of the {\sf F} and {\sf B} matrices).
When also correlation functions on higher genus Riemann surfaces are
considered, then there arises one additional independent relation, obtained from
manipulating one-point blocks on the torus.
In principle one may start from the \furu s to set up the system \erf{poleq} of
polynomial equations and then solve this system for the {\sf F} and {\sf B} 
matrices, but already for a small number of sectors this becomes
very hard (see \Cite{fugv3} for examples).

\sect{Counting States: Characters}J\label{s10}%
The structure of an \ihwm\ \ha\ of the chiral algebra \w\ is rather 
complicated. But a lot of information is already contained in its
{\em weight system\/}, i.e.\ in the collection of weights $\lambda$
of the vectors in a basis of \ha.
To keep track of the weight system of a module, one sums over 
formal exponentials $\eE^\lambda$ of its weights, counting their 
multiplicities, or what is equivalent (writing $\eE^\lambda\eq\ipro{v_\lambda}
{\eE^{W\o}_{}v_\lambda}$ with $v_\lambda$ the orthonormal basis vectors of
\ha), takes the trace over the module of formal exponentials of elements $W\o$ 
of $\w\o$. The so obtained quantity\,%
\futnote{That this construction makes sense has its origin in the
fact that for any set of numbers $a_n$ indexed by \Zet, the formal Laurent 
series $f(q):= \sum_n a_n q^n$ in some indeterminate $q$
contains the same information as the $a_n$.
Here this recipe is generalized to the case where
multiplicities are labelled by weights rather than by
integers; accordingly one associates to each weight $\lambda$ a formal variable 
$\elambda$ on which one imposes the usual properties of exponentials, i.e.\
$\elambda\eE^\mu\eq\eE^{\lambda+\mu}$ and $\eE^0\eq1$.
In more mathematical terms, the basic observation is that the weights
form an abelian group $L$ under vector addition. The formal exponentials 
constitute a basis for the group algebra $\complex L$ of $L$.
When considered as elements of $\complex L$, one can add up these
exponentials, and hence can consider $\chI_A$ as an element
of this group algebra, \resp, in the case of \infdim\ modules, of some
completion of the group \alg.} 
  \be  \chI_A(W\o) :={\rm tr}_\hA\eE^{W\o}  \ee
is called the {\em character\/} of the irreducible \w-module \ha.
(Compare the characters of \irmod s of \lie s, see \refsf).

It is often convenient to regard $\chI_A$ as
a function of complex variables, which are the coefficients 
in an expansion of a generic element $W\o$ of $\w\o$ in some suitable basis,
and for some purposes it is sufficient to specialize to characters at vanishing
values of some of these variables. 
A very specific, but nevertheless most important, case is provided by the 
{\em Virasoro-specialized\/} character, for which $W\o$ is taken to have only
a component in the direction of $L_0\mi\frac1{24}\,C$ (the inclusion of 
the central term is for simpler modular transformation properties):
  \be  \chI_A(\tau):={\rm tr}^{}_\hA\,\eE^{2\pi\ii\tau(L_0-C/24)} \,. \labl{chd}
It is a function of one complex variable $\tau$ and (provided that the
chiral \alg\ \w\ is of `at most polynomial growth', which seems to be the
case in all \cfts) converges in the upper half-plane. 
By construction, $\chI_A(\tau)\cdot\exp(-2\pi\ii\tau(\cd A{-}c/24))$ is a power
series in $q=\exp(2\pi\ii\tau)$, with the coefficient of $q^n$ the
dimension of the subspace of vectors of grade $n$. 

The \vir-specialized character $\chI_A(\tau)$ 
can be interpreted as a chiral block for the `0-point \corfu' (or
vacuum-to-vacuum amplitude) on a torus with modular parameter $\tau$, which in 
the diagrammatic description used in \Erf p1 corresponds to a one-loop graph 
without any external lines. By identifying $L_0{-}C/24$ as a 
Hamiltonian, $\chI_A(\tau)$ can also be regarded as a partition function,
  \be  \chI_A(\tau) = {\rm tr}^{}_\hA \eE^{-\beta H} \,,  \ee
in a thermal state of complex inverse temperature $\beta\eq-2\pii\tau$;
in particular, $\tau\!\to\!\ii0$ and $\tau\!\to\!\ii\infty$ correspond
to the high and low temperature limit, \resp.

Once the characters are known, the computation of the 0-point function on the 
torus for the full \twodim\ \con field theory with symmetry algebra 
$\w\ooplus\wb$ 
amounts to specifying non-negative integers $Z_{AB}$ which tell how often the 
chiral block corresponding to the irreducible \w-\rep\ labelled by $A$ gets 
combined with the antichiral block for the irreducible $\overline{\w}$-\rep\ 
labelled by $B$.
The total {\em partition function\/} of the \cft\ is then given by
  \be \cZ\equiv \cZ(\tau,\bar\tau)\restr{\scriptstyle{\bar\tau^*=\tau}} := 
  \sum_{A,B} \chI_A(\tau)
  Z_{AB}\chI_B(\bar\tau) \restr{\scriptstyle{\bar\tau^*=\tau}} ,\labl{partfunct}
where we have to identify $\bar\tau$ with the complex conjugate of $\tau$.
To qualify as a partition function of a physical theory, the matrix $Z$ 
clearly must fulfill a number of consistency requirements. 
By construction, its entries $Z_{AB}$ are non-negative integers; moreover,
the vacuum sector \ho\ must be unique, and hence one needs $Z_{00}=1$.
Further properties arise from modular invariance, to be discussed in the
next \sec.

\sect{Modularity}K\label{s11}%
The space of characters of the \uihwm s
of the chiral algebra \w\ of a \rcft\ carries a unitary \rep\ of the group
\slz. (The elements of \slz\ are $2\Times2$\,-matrices with integral entries
and determinant one.) To show this crucial fact of life 
would go much beyond the scope of this 
review, and in fact I am not aware of any rigorous proof.\,%
\futnote{Actually in all known theories the \slz-\rep\ factorizes through
some finite-index subgroup.}
A matrix $M\iN\slz$ acts on the characters by a change of 
parameters which depends in a specific way on the entries of $M$; in 
the particular case of Virasoro-specialized characters, the action is given by
  \be  \tau \mapsto M\tau := \frac{a\tau+b}{c\tau+d} \qquad {\rm for}\;\
  M=\smallmatrix{{cc}a&\!b\\c&\!d}  \labl{mm}
($a,b,c,d\iN\zet,\;ad\mi bc=1$), i.e.\ $M{\cdot}\chii(\tau)\,{:=}\,
\chii(M\tau)$. Note that here the elements $M$ and $-M$ of \slz\ act in the 
same way, so that the Virasoro-specialized characters actually carry a \rep\
of the quotient group $\pslz=\slz/\{\one,\!-\one\}$. This group is known as the
{\em modular group\/} of the torus, and correspondingly the mapping \erf{mm} 
is called a {\em modular transformation}.

The parameter $\tau$ which (for convergence of $\chii(\tau)$) takes values in
the upper complex half-plane can indeed be interpreted geometrically
as parametrizing a torus, namely the one obtained by identifying the 
opposite edges of the quadrangle
with corners $0,1,\tau$ and $1+\tau$ in the complex plane. The complex
structure on this torus depends on $\tau$ only up to modular
transformations.

The group \slz\ is freely generated by
two elements $S$ and $T$ modulo the relations $S^2\eq(ST)^3$ and $S^4\eq\one$.
For \pslz\ these get supplemented by the relation $S^2\eq\one$; in the general
case, $S^2\eq C$ is the matrix for the conjugation $A\mapsto A^+$, i.e.\
$C_{AB}\eq\delta_{\!A,B^+}$; thus the
action of modular transformations on the space of characters of primary 
fields is a genuine \rep\ of the twofold cover \slz\ of the modular group 
\pslz\ precisely when the conjugation is non-trivial. The elements $S$ 
and $T$ are represented on the modular parameter $\tau$ as
  \be T:\,\ \tau \mapsto \tau +1 \qquad\mbox{and}\qquad S:\,\ \tau\mapsto-1/\tau
  \,. \ee
(In terms of the temperature $\beta^{-1}\eq(-2\pii\tau)^{-1}\!$, $S$ 
exchanges the low and high temperature regimes -- or, thinking of $\beta$ as
a coupling constant, the strong and weak coupling regions.)
The corresponding matrices that act on the space of characters as
  $   \chI_A(\tau+1) = \sum_B T_{AB}\, \chI_B(\tau)$ and
  $\chI_A(-1/\tau)= \sum_B S_{AB}\, \chI_B(\tau)$
are, correspondingly, referred to as the (modular) \smat\ and \tmat\ of 
the \cft. {}From the definition \erf{chd} of Virasoro-specialized characters it 
follows immediately that the \tmat\ is diagonal and unitary, with entries 
  \be  T_{AA}^{}=\exp(2\pii\cd A-c/24) \,.  \Labl Te
In contrast, the explicit form 
of the \smat, which turns out to be a symmetric and unitary matrix, is
much more difficult to obtain; in fact it is by no means obvious that the
modular inversion $S$ closes on the space of characters.
In particular, knowing only the Virasoro-specialized characters
is typically not sufficient to determine $S$ (e.g., conjugate fields
$\phi_A$ and $\phi_{A^+}^{}$ have identical Virasoro-specialized characters); 
rather one must use the full characters. 

It is by no means accidental that in the definition \erf{D A} I used the same 
letter $S$ for the matrix that diagonalizes the \furu s as for the matrix that
implements $\tau\mapstO-1/\tau$. Namely, for a modular invariant \rcft\
this diagonalizing matrix (when normalizations are chosen
in such a way that it is symmetric) coincides with the modular inversion $S$ as
introduced here. With this identification,
the identity \erf{-verl} expressing the \frc s through $S$ is known
as the {\em Verlinde formula} \Cite{verl2}.\,%
\futnote{The existing derivations of the Verlinde formula range from formal 
manipulations with chiral blocks (see e.g.\ exercise 3.6 in \Cite{Mose}), 
localization of path integrals \Cite{blth} and surgery manipulations on
three-manifolds \Cite{axdw} to highly non-trivial arguments in \alg ic 
geometry (\Cite{beLa,falt}; for reviews and further references, see 
\Cite{beau2} and \Cite{Scho}). Typically the more rigorous the arguments are, 
the more restricted is the range of theories to which they apply; e.g.\ the 
\alg ic geometry derivation applies only to the case of (most) \wzwts.}

In order to construct a \cft\ consistently on a Riemann surface of
genus larger than zero, the partition function \erf{partfunct} must be
invariant under the modular transformations of that surface, in particular
under \erf{mm} in the case of the torus which has genus 1. I should point out
that, as a consequence, modular invariance is a basic property of any \cft\ 
model that is relevant to string theory, but that it is {\em not\/} a 
fundamental property of \cft\ as such, and there are situations where 
it is absent. (For instance, in the case of the critical Ising model
modular invariance restricts
the number of sectors to three, which have the same highest weight \wrtt chiral
and the antichiral \vira. These correspond to the identity field, the
energy operator and the order parameter of the Ising model, as listed in table
\Erf IM below. But for many purposes one must also consider other sectors, which
correspond to the disorder parameter and to a chiral or antichiral free 
fermion.) Also, in general more requirements than just modular invariance 
are necessary for the existence and consistency of a \cft. It is therefore
not surprising that there exist modular invariant partition functions
which cannot correspond to any \cft\ (for some examples, namely so-called
extensions of \wzwts\ by spin-1 currents which do not describe a
conformal embedding, see \Cite{fusS}).

The requirement that \erf{partfunct} is invariant under the modular
transformations \erf{mm} is equivalent to demanding that the matrix
$Z=(Z_{AB})$ satisfies
  \be  [Z,S] = [Z,T] = 0 \,,  \labl{zszt}
i.e.\ commutes with both the S- and the \tmat. 
An immediate consequence of $[Z,T]\eq0$ is that the `conformal spin', 
i.e.\ the combination $\cD\mi\bar\cD$ of \cdim s, must be an integer, 
so that each field is `local' \wrt all other fields. As a consequence, 
at any genus all \corfu s of the \twodim\ theory are single-valued functions.

There always exists a straightforward solution to the constraints \erf{zszt}, 
namely $Z\eq\one$. This is called the {\em diagonal\/} or $A$-{\em type\/} 
modular invariant. When \w\ is maximal, then this is in fact \Cite{mose2}
the only solution, up
to possibly a permutation which is an automorphism of the \furu s, i.e.\
$Z_{AB}=\delta_{A,\pi B}$ with $\N{\pi A}{,\pi B}{\;\ \ \pi C}\eq\N ABC$.
However, often one does not know the maximal chiral \alg, so that it makes
sense to analyze the constraint
\erf{zszt} also in the case of non-maximal \w\ (but with
the theory still being rational). As it turns out, \erf{zszt} constitutes a
powerful restriction, and its implementation
is a highly non-trivial task. So far the solutions have been classified
only for very specific types of theories, e.g.\ for the \wzwts\ (see 
\refsh) based on $A_1\Untw$, the
Virasoro minimal models \Cite{caiz2}, for \wzwts\ based on $A_2\Untw$ 
\Cite{gann4}, and for $N\eq2$ superconformal minimal models \Cite{gann10}.

\sect{Free Bosons}L\label{s12}%
The simplest examples of \cfts\ are those which describe massless free bosons  
or fermions. Here I will sketch the bosonic case. The classical action
for a massless free boson $\pHi$ living
on a $D$-dimensional space-time manifold of metric 
$g_{\mu\nu}$ reads $\aS\,{\propto}\int\!{\rm d}^D\!x\,\sqrt{{\rm det}\,g}\,
g^{\mu\nu}\partial_\mu\pHi\,\partial_\nu\pHi$. Variation of $\aS$ \wrt 
$g_{\mu\nu}$ yields the \emt\
  \be  T_{\mu\nu} = -\partial_\mu\pHi\partial_\nu\pHi+\onehalf\,g_{\mu\nu}\,
  \Sum{\sigma,\tau}g^{\sigma\tau}\,\partial_\sigma\pHi\,\partial_\tau\pHi
  \,, \ee
which is conserved ($\sum_\mu\partial^\mu T_{\mu\nu}\eq0$)
and has trace $\sum_\mu T^\mu_{\;\mu}\,{\propto}\,1\mi D/2$. Thus in $D\eq2$
dimensions $T_{\mu\nu}$ is traceless; in complex coordinates, this
means $T_{z\bar z}\eq0$, so that conservation reduces to
$\partial_\bz T_{zz} \eq0\eq \partial_z T_{\bz\bz}$, or in other
words, $T\,{\equiv}\,T_{zz}\eq T(z)$ and $\bar T\,{\equiv}\,T_{\bz\bz}\eq
\bar T(\bz)$, as is needed for a \con field theory.

As the \emt\ $T$ has conformal weight $\cD\eq2$, at the classical level 
the boson $\pHi$ has scaling dimension zero. Correspondingly, $\pHi$ is {\em 
not\/} a proper conformal field in the sense of \refsE; it can be written as
$\pHi\zbz=\pHi(z)+\overline\pHi(\bz)$ where $\pHi(z)$ has an expansion
  \be  \pHi(z) = X_0 - \ii\,P\,\ln(z)+\ii\,\sum_{n\in\zet\setminus\{0\}}
  \Frac1n\,J_n\, z^{-n} \,.  \ee
Thus $\pHi$ is not single-valued on the complex plane. But the derivatives
$\partial\pHi$ and $\bar\partial\overline\pHi$ (which appear in the action \aS) 
are, and so are (suitably normal ordered) exponentials $\normord{\eE^
{\ii q\pHi}}$. The canonical commutation relations for the free boson $\pHi$ 
yield Heisenberg relations for its modes:\, $[X_0,J_n]\eq0$ for $n\Ne0$ and
  \be  [X_0,P]=\ii \,, \qquad [J_m,J_n] = m\,\delta_{m+n,0}  \,. \Labl u1

The chiral \alg\ \w\ of this theory is the semi-direct sum of the \vira\
generated by $T(z)$ and a {\em Heisenberg \alg\/} which is generated by 
the Virasoro-\pf\ $J\eq\sum_{n\in\zet}\!J_n z^{-n-1}$ ($J_0\,{\equiv}\,P$). 
In terms of $\pHi$ these fields read
  \be  J(z)=\ii\,\partial\pHi(z) \,,\qquad T(z)=-\half\normord{\partial\pHi(z)\,
  \partial\pHi(z)}=\half\normord{J^2(z)} \,.  \Labl JT
Using the two-point function $\vev{\pHi(z)\,\pHi(w)}\,{\simeq}\,-\ln(z\Mi w)$ 
and Wick's rule for calculating correlators of normal ordered products of free
fields, one checks that the conformal central charge has the value $c\eq1$ 
and that $J$ has conformal weight $\cd J\eq1$. The zero mode sub\alg\
$\wo$ is spanned by $L_0$ and the zero mode $J_0\,{\equiv}\,P$ of the field 
$J$. The primary fields of this theory are $\phi_q\eq\normord{\eE^{\ii q\pHi}}$;
they are labelled by charges, i.e.\ $J_0$-eigenvalues, $q\iN\reals$, and their 
fusion rules read $\Phi^{}_p\star\Phi^{}_q=\Phi^{}_{p+q}$. 

To be precise, these statements refer to the situation that the modes of
the classical boson $\pHi$, in particular the zero mode $X_0$, are allowed
to take values on the whole real line \Reals. In contrast, when the
boson is {\em compactified\/} on a circle of radius $R$, in the sense that
one identifies the values $X_0$ and $X_0+2\pi R$ of the zero mode, and
when in addition $R^2\eq2{\cal N}$ is an even
integer, then there are two additional 
(Virasoro- and $J$-primary) fields in \w\ which have conformal dimension 
$\cD={\cal N}$. Moreover, in this case the theory is rational, and with 
a suitable normalization of the current $J$ the possible charge values 
are $q\in\{0,1,...\,,2{\cal N}\Mi1\}$.
The conformal dimension of a \pf\ of charge $q$ is then $\cd q=q^2/4{\cal N}$.
The fusion rules read $\Phi^{}_p\star\Phi^{}_q=
\Phi^{}_{p+q\,{\rm mod}\,2{\cal N}}$, and the modular \smat\ has entries
  \be  S_{pq} = \exp(-\pi\ii\,pq/{\cal N})/\sqrt{2{\cal N}} \,.  \labl{su}
Similarly, when $R^2\eq2r/s$ ($r,s$ coprime) is any other rational number, the 
theory is still rational and the chiral \alg\ \w\ contains additional 
fields of \con weight $\cD\eq{\cal N}\df rs$ in \w.
But in the general case \w\ contains fields of still higher weight, and one
deals with a non-diagonal extension of the theory that has squared radius 
$2{\cal N}$. However, the theories with radius $R$ and radius $2/R$, which look
most different when formulated in terms of a classical action, are actually
one and the same \cft; in particular for $R^2\eq2/{\cal N}$ one recovers a
diagonal theory.

All the theories just described have $c\eq1$. In fact, any other
unitary rational $c\eq1$ \cft\ can be formulated in terms of a free boson 
as well. Namely, each of those theories can be obtained as a so-called
{\em orbifold\/} of the theory of a free boson on a circle. In general,
forming an orbifold of a \cft\ means that one restricts the observables to
those invariant under some (discrete) group Z of automorphisms. In the case 
at hand, Z must be a discrete subgroup of O(3), which leads to 
three isolated rational theories for which Z is non-abelian, and to a 
continuous family of theories for which ${\rm Z}\eq\zet_2$, with the non-trivial
element corresponding to the identification of 
$\pHi$ with $-\pHi$. For rational square radius, the latter $\zet_2$-orbifolds
are again rational theories, but for non-rational values of $R^2$ they
are not even quasi-rational.

More generally,
free bosons are important not only as theories in their own right, but can 
serve as a starting point for the description of many other theories, too.
A particularly interesting possibility is to modify the theory in such a
way that $\pHi$ still has the correlation functions of a free boson, but 
that the form \Erf JT of $T(z)$ is changed the by addition of a term 
proportional to $\partial^2\pHi$, which
corresponds to the presence of a background charge in a Coulomb gas.
This way one can in particular obtain a free field realization of
the minimal series \erf{mimo} of unitary models with $c\,{<}\,1$.

\sect{Simple Currents}M\label{s13}%
Each fusion ring contains a distinguished basis element, the unit element 
$\pho$. This element clearly has the property to be invertible
within the ring, i.e.\ in mathematical terms, it is a {\em unit\/} of the ring.
But a fusion ring may contain also further
basis elements which are units in this sense.
These are called {\em simple currents\/} of the ring, or of the \cft, and
they turn out to be of considerable importance. In the theory of a 
free boson at rational square radius that was just described, in fact each 
basis element $\Phi_p$ is a simple current, with inverse $(\Phi_p)^{-1}=
\Phi_{2{\cal N}-p}$. Another important example for a simple current is the 
field which implements the GSO projection in superstring theory.
In this \sec\ I sketch the most intriguing properties
of simple currents; for more details see  e.g.\ \Cite{scya6} and \Cite{jf24}.

A simple current \J\ can be equivalently characterized as a primary field for 
which the fusion product with the conjugate field just yields the unit element,
  \be  \J \star  \J^+ = \pho \,,  \ee
or for which the fusion rules are {\em simple\/} in the sense that for 
each $A$ the fusion product $\J\star \phj A$ belongs
again to the distinguished basis (it is then simply written as 
$\phj{\J A}\equiv\phj{\J\star A}$). In terms of the \frc s this means that
$\sum_{\sss C} \N\J BC\eq1$ for all $B$. Yet another characterization is 
that $S_{\O\J}\eq S_{\O\O}$ (while $S_{\O A}\,{>}\,S_{\O\O}$ for every $\phj A$ 
which is not a simple current).\,%
\futnote{The term `current' is chosen \Cite{scya} in anticipation of 
the fact that these fields can be used for an extension of the 
chiral symmetry \alg, and (also when the {\em conformal\/} dimension 
is not equal to 1) such a field is often called a current.
The qualification `simple' refers to the behavior \wrt fusion rules
(and to \corfu s as well, see \Cite{fuge}), but also fits with the fact that
in \alg ic \qft\ a superselection sector with 
{\em statistical\/} dimension equal to 1 is called a simple sector.}
Due to the associativity of the fusion product, the product of two simple 
currents is again a simple current. Simple currents thus form an abelian group
under the fusion product, i.e.\ a sub{\em group\/} of the fusion ring (this
group has been termed the {\em center\/} of the theory). A rational theory can 
of course accommodate only finitely many simple currents. 
This implies that simple currents are unipotent; the smallest positive integer
$N\equiv N_\J$ such that $\J^N = \pho$ is called the {\em order\/} of $\J$
(here $\J^2\,{\equiv}\,\J\star\J$, etc.).
Furthermore, any simple current organizes the primary fields into 
orbits $[\phj A]:=\{\phj A,\phj{\J A},\phj{\J^2 A}\Ldots\phj{\J^{N-1}\!A}\}$;
the size $N_A:=|[\phj A]|$ of any orbit $[\phj A]$ is a divisor of the order 
$N\eq|[\pho]|$ of $\J$.

A crucial result about simple currents \Cite{scya6,intr} is that
in a unitary theory, \smat\ elements involving fields on the same simple 
current orbit differ only by a phase:
  \be  S_{\J_\P^p\! A,\J_\P^q B} = \eE^{2\pi\ii p\QJ(B)}\eE^{2\pi\ii q\QJ(A)}
  \eE^{2\pi\ii pq \QJ(\J)}\cdot S_{AB} \,.  \labl J
Here $\QJ(A)$ is the {\em monodromy charge\/} of $\phi_A$ \wrt \J, defined
as the combination
  \be  \QJ(A) := \cd\J + \cd A - \cd{\J A}\, \bmod \zet  \labl{moch}
of conformal weights (in a rational theory, $\QJ(A)$ is a rational number). 
The monodromy charge is additive under the operator product, i.e.\ for any 
$C$ with $\nijk\ne0$ one has $\QJ(A)=\QJ(B)+\QJ(C)\bmod\zet$.
Based on \erf J, one can show that for any simple current which 
satisfies $N_\J\cd\J\iN\zet$, the matrix $Z$ with entries 
  \be  Z_{AB}^{}= \Frac N{N_A}\sum_{n=0}^{N_A-1}\delta^{}_{B,\J_\P^n\!A}\, 
  \delta^{(\zet)}(Q_\J(A)+\Frac n2\, Q_\J(\J))  \labl{sina}
provides a modular invariant.
Here $\delta^{(\zet)}$ denotes the function with $\delta^{(\zet)}(x)=1$ 
for $x\iN\zet$ and $\delta^{(\zet)}(x)=0$ otherwise.
Note that all non-zero entries of $Z$ are between fields on 
the same simple current orbit. If the center is the cyclic group  
$\{\pho,\J,\J^2\Ldots\J^{N-1}\}\cong\zet_N$ generated
by a simple current \J, then \erf{sina} is the only modular invariant of the
theory with this property. When the center is not cyclic, then 
there are further possibilities for simple current extensions which are
parametrized by the so-called discrete torsion \Cite{krSc}. 

When $\cd\J$ is integral, the presence of $\delta^{(\zet)}$ in \erf{sina} 
means that only fields with vanishing monodromy charge occur, and the modular 
invariant reduces to 
  \be  {\cal Z}= \sum_{A:\,\QJ(A)=0} \Frac{N}{N_A} \, \mbox{\Large$|$}
  \sum_{n=0}^{N_A-1} \chii^{}_{\J_\P^n\!A} \mbox{\Large$|$}^2 . \labl{sinv}
Invariants of this type are called {\em integer spin simple current extensions}.
They are interpreted as diagonal invariants of a theory with an extended 
chiral algebra \w, with the additional fields in \w\ being the simple 
currents $\J^n_\P$. Note that typically several \irmod s of the original 
chiral algebra \wune\ combine to a single \irmod\ of $\w$; according to 
\erf{sinv}, the characters of the irreducible \w-modules are proportional to
the orbit sum $\sum_n\!\chii_{\J^{n_\P}\!\!\!A}$.
Moreover, not every irreducible \wune-module will be contained
in a module of \w, and this is precisely encoded in the requirement of
vanishing monodromy charge. 

The orbit sums of characters of the original `unextended' theory appear in 
\erf{sinv} with multiplicities $N/N_A$. Orbits $A$ with $N/N_A>1$ are called 
{\em fixed points\/} of the simple current $\J$. To interpret these
multiplicities, one must recall (see \refsK) that for maximally extended
\w\ each (equivalence class of) irreducible \w-module appears precisely once. 
Thus a multiplicity in front of the complete square indicates that
there exist several non-isomorphic \w-modules which reduce to one and the 
same module of \wune, or in other words, that such a term corresponds to 
several distinct primary fields in the \cft\ with the enlarged chiral algebra.
However, this prescription is unambiguous only for 
$N/N_A=2$ or $3$. For larger values it can also happen that one must include 
(part of) the prefactor $N/N_A$ into the extended character; as this can 
only be done for complete squares, the number of distinct possible 
interpretations is then equal to the number of ways that $N/N_A$
can be written as a sum of squares. For example, when $N/N_A\eq5$ one either 
deals with five inequivalent \w-modules which have (\wune-specialized) character
$\sum_{n=0}^{N_A-1}\!\chii_{\J^{n_\P}\!\!\!A}$, or else with one such module 
plus another \w-module whose character equals twice that sum.

Given an invariant of the type \erf{sinv}, it is in general a highly
non-trivial task to investigate whether a fully consistent conformal field
theory can be associated to that modular invariant.
In particular one would like to compute the relevant \slz-\rep.
While the \tmat\ for the extended theory is just the restriction
of the original \tmat\ to allowed orbits, for the \smat\ one only gets
some constraints involving the \smat\ of the original theory, and determining 
those elements which involve two fixed points proves to be an intricate
problem. However, recently a general class of solutions to this {\em fixed 
point resolution\/} problem was constructed \Cite{fusS6,bant6}; this
construction yields in particular a general closed expression for the \smat. 

\sect{\OPA\ from \FURU s}N\label{s14}%
The results of \refsF\ show that the \opc s of \pf s are 
the normalizations of the corresponding \tpf s, but these normalizations
cannot be determined from the Ward identities. With the help of \four s,
on the other hand, one {\em can\/} compute the \opc s. Indeed, for maximal \w\
the coefficients $a_{I\bar I}$ that appear in the decomposition \erf{14} of
\fpc s into \threepf s are of the form $a_{I\bar I}=
a_I^{}\delta_{\bar I,\pi I}$ for some permutation $\pi$.
Moreover, provided that the chiral
blocks are correctly normalized, the $a_I$ are just the products 
$C_{\!AB}^{\ \ I}C_{ICD}$ of the relevant \opc s. As a consequence, one can 
regard a \cft\ as completely solved once all \fpf s of its
primaries are known. These correlators, in turn, can be deduced from the
\furu s of the theory by using general analytic properties 
of the correlators to construct linear \deq s which they must satisfy.\,%
\futnote{Often these differential equations also follow 
from the presence of null vectors in the Verma modules of \w. An example is 
provided by the \kze s which will be discussed in \refsj.}
The independent solutions of these
\deq s are the chiral blocks, which transform into each other under
duality transformations (see \erf{BF}). Having obtained explicit 
formul\ae\ for the blocks, one can therefore determine the duality matrices 
{\sf F} and {\sf B} and express the coefficients $a_I$, and hence the
\opc s, through the entries of these matrices \Cite{blya2,jf14,jf21}.

To achieve this result in practice, one just exploits the following basic
properties of (rational or quasi-rational, genus zero) \cfts:
\nxt invariance of the vacuum vector \vo\
   under the projective \sltwo-\alg\ \erf{su11} (see \refsD);
\nxt existence of a closed associative operator \alg\ \Erf oa whose
   structure `constants' are of a specific form that is compatible with 
   the \w-symmetry (\refsG); 
\nxt factorization into chiral and antichiral blocks (\refsF);
\nxt independence of the chiral blocks.

First, one can invoke projective invariance to work with
four-point correlators \calgzz\ which are of the special form \Erf11.
Next, the closure of the \opa\ implies that $\calg$ can be written as a sum 
of products of \threepf s (compare the picture \erf{p1}), and
because of the \w-symmetry the contributions from each
`intermediate' family $[\phi_I]$ combine to a block of 
definite analytic behavior. Furthermore, by the fact that the
symmetry algebra is the direct sum $\w\ooplus\wb$ of two chiral
halves, the $z$- and $\bz$-dependence factorize, which amounts to the
decomposition \Erf14. The number $M$ ($\eq\bar M$ for maximally extended \w) 
of blocks in \Erf14 is determined as
  \be  M = \sum_{\sss I} \N AB{\;I} \Ns ICD \,,  \Labl MM
with $I$ labelling the intermediate families $[\phi_I]$,
through the fusion rule coefficients $\nijk$.

Now recall that the chiral blocks are, generically, not ordinary functions
but multi-valued. When the chiral blocks \fmz\ are known, then requiring that 
\calgzz\ becomes single-valued when $\bz$ is taken to be
the complex conjugate of $z$ -- which 
can be encoded in the crossing symmetry relations \Erf19 -- determines
the complex constants $a_{I\bar I}$ up to an over-all factor.
By considering the collection of {\em all\/} \fpf s of primary fields
(and fixing a definite normalization of each \pf),
one can compute these over-all factors, too.

To determine the chiral blocks, one observes that according to the \ope\ 
\erf{oa}, $\,\phi_i\zbz\phi_j\wbw$ depends on $z$ and $w$ as $(z\Mi w)^\alpha$, 
with $\alpha$ determined by the \cdim s of the fields involved. More precisely,
  \be  \fmz \,\propto\, z^{-\cd C-\cd D+\tilde\cD_I^{\sss(CD)}}
  \ \ \ {\rm for}\ z\to0 \,.  \labl{16}
Here $\cd C$ is the \cdim\ of the \pf\ $\phi_C$, while
$\tilde\cD_I^{\sss(CD)}=\cD_I^{\sss(CD)} + \mu_I^{\sss(CD)}$
is the \cdim\ of that field $\varphi_I\iN[\phi_I]$, of grade $\mu_I^{\sss(CD)}$,
in the family of the primary $\phi_I^{}\equiv\phi_I^{\sss(CD)}$
which gives the leading contribution to the coupling between
$\phi_C$, $\phi_D$ and $[\phi_I]$. The result \erf{16} corresponds to
the decomposition of $\calg$ into three-point blocks as indicated in figure 
\erf{p1}. Analogously, decomposing as in picture \erf{p2} yields
different chiral blocks \fmh\ and \fmw\ which behave as
  \be   \bearll \fmh\,{\propto}\, (1\Mi z)^{-\cd C-\cd B+
  \tilde\cD_K^{\sss(CB)}} &  {\rm for}\ z\,{\rightarrow}\,1 , \\{}\\[-.5em]
  \fmw\,{\propto}\, (z^{-1})^{\cd C-\cd A+\tilde\cD_J^{\sss(CA)}}
  & {\rm for}\ z\,{\rightarrow}\,\infty  \eear \labl{18}
with $\tilde\cD_J^{\sss(CB)}$ and $\tilde\cD_K^{\sss(CA)}$ defined 
analogously as $\tilde\cD_I^{\sss(CD)}$. 
(Thus the systems $\{\fmi\},\,\{\fmh\}$ and $\{\fmw\}$
have diagonal monodromy around $z=0,\,1$ and $\infty,$ respectively.)
As already described in \refsI,
the systems $\{\fmi\},\,\{\fmh\}$ and $\{\fmw\}$ of chiral blocks
transform into each other upon analytic continuation.
On the other hand, since they correspond to distinct couplings they must be 
\alg ically independent; in technical terms, this amounts to the property
that their Wronskian determinant must not vanish identically. According to 
elementary results from the theory of ordinary linear \deq s (see e.g.\
\Cite{YOsh}) it then follows that the chiral blocks $\fmi$ constitute the $M$ 
independent solutions of an $M$th order \deq.
More precisely, the blocks obey
  \be  \partial^M\calf(z)+\sum_{m=0}^{M-1}h_m(z)\,\partial^m{\cal F}(z)
  =0  \labl{115}
with certain rational coefficient functions $h_m(z)$.
As prescribed by the singular terms in the relevant operator products, 
the solutions of \erf{115} have branch cut singularities at the 
points $z=0,\,1$ and $\infty$.
It can happen that the coefficient functions $h_m(z)$ in \erf{115} also 
possesses additional singular points, at which all of the solutions are regular;
such singular points are known as {\em\asis\/}.

The construction of the \deq\ \erf{115} is essentially equivalent to 
solving a {\em Riemann monodromy problem\/}, i.e.\ determine a collection
of functions with prescribed monodromy. The Riemann monodromy problem
always possesses a solution, provided that one allows for the possible 
presence of \asis, around which the monodromy is trivial, in the \deq. 
The general solution of \erf{115} can be summarized in a so-called
{\em Riemann scheme\/} {\sf P},
which specifies the positions of the singularities $z_i$ of the \deq\
together with the {\em exponents\/} \aim\ at $z_i$, i.e.\ the roots of the
$M$th order algebraic equation for $\alpa^{(i)}$ that one obtains
in lowest order in $z\mi z_i$ 
by making the power series ansatz $\calf(z)=(z\Mi z_i)^{\alpa^{(i)}}\!
\sum_{p=0}^\infty a_p(z\Mi z_i)^p\,$ ($a_0\Ne0$;
for $z_i\eq\infty$, $z\Mi z_i$ is to be interpreted as $z^{-1}$).
The Riemann scheme {\sf P} does not, in general, determine uniquely the 
functions \fmz. But one finds that as long as there are no \asis, all of the 
parameters not specified by {\sf P} (called {\em accessory\/} parameters) 
can be fixed by imposing the \poleq s \erf{poleq}. When \asis\ are present, 
the situation is more involved because
only the positions and exponents of the real singularities
of correlation functions are dictated by the operator products.
In other words, the singularity structure of the \four s is
fully known if and only if both the conformal weights of the \pf s
(including the integer part) and -- in the case of \frc s larger than
1 -- also the grades of the exchanged fields which are responsible for
the leading singular behavior of a chiral block, are given.
Once this is the case, the results for the blocks
are unique provided that the \deq s satisfied by the
four-point blocks do not possess any \asis\ \Cite{blya2}. On the other hand, 
the presence of \asis\ spoils the uniqueness of the solution. 

The position of an \asi\ is almost arbitrary. There is only a 
mild restriction which results from the crossing symmetry \Cite{jf21}.
The polynomial equations \erf{poleq} do {\em not\/} impose any further
restriction, as follows 
from the theory of isomonodromic deformations of \deq s. Namely, for 
arbitrarily chosen rational coefficient functions $f_n(z)$ the combinations
$\gmz\,{:=}\,\sum_{n=0}^{M-1}f_n(z)\,\partial^n\fmz$ and similarly defined 
functions $\gmh,\,\gmw$ not only possess the same monodromy as the blocks
$\fmi,\,\fmh\,,\fmw$, but also the same fusing and braiding matrices
(roughly, duality can be regarded as the `square root' of monodromy). 
By varying the rational functions $f_n$, the position (and, generically, even 
the number) of \asis\ can be varied freely. The positions of the \asis\ of 
various distinct \four s are related, but only by the constraints that are 
implied by the crossing symmetry.

\lect3{\KMA s}{9.73}{6.8}

\mbox{$ $}\\[1ex]
This lecture is plain mathematics. In the final lecture 4
the mathematical structures studied here will be applied to \cft.

\sect{\CARMS}a\label{s15}%
\kma s constitute a certain class of (generically \infdim) \lie s. Roughly, 
they are those \lie s which possess both a Cartan matrix (implying 
the existence of a triangular decomposition) and a Killing form.
They include all \findim\ simple \lie s, and I will start 
by summarizing some pertinent features of those well-known \alg s.
A \findim\ simple \lie\ $\g\eq\g(A)$ is characterized by its {\em\carm\/} $A$; 
the basic Lie bracket relations of \g\ can be expressed through $A$ as\,%
\futnote{This definition is asymmetric in $i$ and $j$. In the literature the 
\carm\ is sometimes defined as the transpose of the matrix used here.}
  \be  \begin{tabular}{l|l|c|l|}  \multicolumn3c{} \\[-.7em]
       \cline{2-2}\cline{4-4} &&&\\[-.8em] \fbox{\footnotesize A}\hsp{.55}
  & $\,[H^\alpha,H^\beta] = 0$ & \hsp{1.9}\fbox{\footnotesize B}\hsp{.55}& 
    $\,[H^i, H^j] = 0$  \\[.58em]
  & $\,[H^\alpha,E^\beta] = (\alpha^\Vee,\beta)\, E^\beta$ &&
    $\,[H^i,E^j_\pm] = \pm \A ji\,E^j_\pm \ $  \\[.58em]
  & $\,[E^\alpha,E^{-\alpha}] = H^\alpha$ &&
    $\,[E^i_+,E^j_-] = \delta_{ij}\, H^j$  \\[.58em]
  & $\,[E^\alpha,E^\beta] = e_{\alpha,\beta}\,E^{\alpha+\beta} \ \ \mbox{for}\
    \alpha\Ne-\beta\,$  && $\,({\rm ad}_{E^i_\pm} )^{1-\A ji}\! E^j_\pm = 0$
        \\[-.8em]&&&\\ \cline{2-2}\cline{4-4}
  \multicolumn4c{} \\[-.7em] \end{tabular} \labl T
in a so-called \cwbasis\ and \chserre basis, \resp.
Let me explain some of the notation that is used here.
When \g\ has rank $r$, $A$ is an $r\times r$-matrix, and
\g\ has a basis $\{E^{\alpha}\} \cup \{H^i\,{|}\,i=1,2,...\,,r\}$ 
obeying the relations (\ref T\,A) (in addition to antisymmetry and
the Jacobi identity, the two defining properties of a Lie bracket). The $r$
generators $H^i$ span an abelian \lie, called the {\em \csa\/} \go\ of \g; 
the maps $\ad{H^i}{:}\; x\mapstO[H^i,x]$ are diagonalizable, and to each 
{\em root\/}, i.e.\ vector $\alpha=(\alpha^i)$ of eigenvalues of these maps, 
there corresponds a generator $E^{\alpha}$ of \g, and one writes
$H^\alpha\df(\alpha^\Vee,H)$ with $\alpha^\Vee\df2\alpha/(\alpha,\alpha)$. 
Among the roots there is a
special subset $\{\aI\,{|}\,i\eq1,2,...\,,r\}$, called the {\em simple roots\/},
through which all roots can be expressed as integral linear
combinations with either only positive or only negative coefficients; the
corresponding roots are called positive and negative roots, \resp.
(Actually there are several distinct subsets of roots which can serve as
systems of simple roots; but all of them are equivalent.)
Writing $E^{\pm\aI}\,{=:}\,E^i_\pm$,\, $\g\eq\g(A)$ is generated \alg ically\,%
\futnote{By `(algebraically) generated' -- as opposed to `(linearly) spanned' --
one means that the elements of the \lie\ \g\ are obtained
as arbitrary linear combinations of arbitrary (multiple) Lie brackets of
the basic generators. Such a characterization of the \lie\ \g, 
not to be confused with the description of \g\ as the linear span of a
basis, is called a {\em presentation\/} of \g\ by generators modulo relations.}
by $3r$ generators $\{E^{i}_{\pm},H^i\,{|}\,i\eq1,2,...\,,r\}$ (called the
Chevalley generators of \g) modulo the relations (\ref T\,B). 
The relations in the last line of (\ref T\,B) are called the Serre relations
(the symbol $(\ad x)^n$ is a shorthand 
for $\ad x\Circ\ad x\Circ\cdots\,\Circ\ad x$ ($n$ factors),
e.g.\ $(\ad x)^3(y)\equiv[x,[x,[x,y]]]$\,),
and the full set (\ref T\,B) is known as the Chevalley\hy Serre relations.
Cartan matrices are conveniently encoded into {\em\dyd s\/}: to each possible
value of the label $i$ one associates a node or vertex, and for $i\Ne j$
the nodes $i$ and $j$ are connected by $A^{ij}\!A^{ji}$ lines.

Of course, not any arbitrary square matrix qualifies as a Cartan matrix. In 
order that $\g(A)$ is \findim\ and simple, the matrix $A$ must satisfy
  \be  A^{ij}\iN\zet\,,\qquad  A^{ii}=2\,,\qquad
  A^{ij}\leq 0\;\;{\rm for}\;\; i\ne j\,, \qquad 
  A^{ij}=0 \,\Leftrightarrow\, A^{ji}=0 \,, \Labl1a
as well as\\[-1.7em]{}
  \be  {\rm det}\,A>0 \,, \hsp4  \labl{detA}
and must be indecomposable (i.e.\ must not be rearrangeable to block-matrix 
form by any simultaneous permutation of its rows and columns;
decomposable matrices obeying \Erf1a and \erf{detA}
describe direct sums of simple \lie s). The analysis of these conditions 
leads to the following classification of all \findim\ simple \lie s:
  \be  A_{\grank}\, {\scs(\grank\geq 1)}\,,\;\ B_{\grank}\,
  {\scs(\grank\geq 2)}\,,\;\ C_{\grank} \,{\scs(\grank\geq 3)}\,,\;\
  D_{\grank}\, {\scs(\grank\geq 4)}\,,  \;\ 
  E_6\,,\;\ E_7\,,\;\ E_8\,,\;\ F_4\,,\;\ G_2\,.  \Labl5b
Thus there are four infinite series, the {\em classical\/} \lie s 
$A_{\grank}\Ldots D_{\grank}$, as well as the five {\em exceptional\/} simple 
\lie s $E_6\Ldots G_2$. In particular there is a unique simple \lie\ of
rank one, namely $A_1\cong\sltwo$, with Cartan matrix $A(A_1)=2$, and 
there are three simple \lie s of rank two, namely $A_2\cong\slthree$,
$B_2\,{\equiv}\,C_2\cong\sofive$ and $G_2$, with Cartan matrices 
  \be A(A_2)=\smallmatrix{{rr} 2&-1\\-1&2 }\,,  \quad
  A(B_2)=\smallmatrix{{rr} 2&-2\\-1&2 }\,,  \quad
  A(G_2)=\smallmatrix{{rr} 2&-3\\-1&2 }\,.  \Labl5t

By construction, the space spanned by the \g-roots $\alpha$ is the space 
\godual\ dual to the \csa\ \go. \godual, called the {\em weight space\/} of \g,
is an $r$-dimensional vector space with euclidean scalar product 
$(\cdot\,,\cdot)$; the simple roots, and likewise the simple {\em coroots\/}
$\alpha^{(i)\Vee}\df2\alpha^{(i)}/(\alpha^{(i)},\alpha^{(i)})$, form a basis 
of \godual. The entries of the \carm\ encode the non-orthonormality
of these bases; more precisely,
  \be  A^{ij} = (\alpha^{(i)\Vee}\!,\aj) \equiv 2 \,(\ai,\aj)/(\aj,\aj) \,.
  \Labl4G
Further properties of the root system, i.e.\ the collection of all roots of \g,
are the following. The only multiple of a root $\alpha$ that is again
a root is $-\alpha$. Also, for any pair $\alpha,\beta$ of roots,
the linear combination $\alpha\,{+}\,m\beta$ is a root for a set of consecutive
integers $m$, but not for any other value of $m$; for pairs of simple roots,
this {\em root string\/} can be directly read off the Serre relations.
For the lengths of roots at most two different values are possible; the
relative length squared of a longer and a shorter root is given by
$\max\{|A^{ij}|\;|\,i\Ne j\}$. In particular, when $A^{ij}\iN\{0,\!-1\}$ for
$i\Ne j$ (in which case \g\ is said to be {\em simply laced\/}), then all
roots have the same length. Also, when different root lengths do occur, then 
this already happens among the simple roots, and each of the two sets of simple 
roots of equal length corresponds to a connected sub-diagram of the \dyd. (Also,
in order that the \dyd\ specifies the \lie\ \g\ uniquely, one supplements 
the multiple link which connects these two sub-diagrams by an arrow pointing 
from the longer to the shorter simple roots.)

The {\em Killing form\/} of \g\ associates to any two elements $x$ and $y$
of \g\ the number
  \be  \kappa(x,y) \equiv (x|y) := {\rm tr}(\ad x\Circ\ad y)  \,.  \Labl kf
It is bilinear, symmetric, and satisfies $\kappa(x,[y,z])=\kappa([x,y],z)$.
Conversely, any map with these properties equals the Killing form up to a
multiplicative constant. When restricted to the \csa\ \go, the Killing form
induces by duality an inner product on the weight space \godual, which,
when suitably normalized, coincides with the euclidean product on \godual\ that
was already used above.

Another important basis of the weight space \godual\ consists of the 
{\em fundamental weights\/} \li, which are defined by the requirement that
  \be  (\li,\ajv)=\delta_i^{\;j}   \Labl li
for $i,j\eq\onetor$. This basis plays a special \role\ in the \rep\ theory of 
\g,
or more precisely, for its \hwm s (the concept of \hwm s was already discussed
in the context of chiral \alg s in \refsC, and will be dealt with in more detail
in \refse\ below). Namely, among the \ihwm s \hl\ of \g, precisely those are
\findim\ whose \hw\ $\L$ is {\em dominant integral\/}, i.e.\ is a non-negative
integral linear combination of the fundamental weights \li. Moreover, each
such module can be obtained by forming tensor products of just a few
special \hwm s, for all of which the \hw\ is a fundamental weight.

The {\em Weyl group\/} $W$ of the \lie\ \g\ is a finite group which is 
generated by the reflections $r_i$, $i\iN\{\onetor\}$, in the weight space 
\godual\ of \g\ about 
the hyperplanes through the origin which are perpendicular to the simple roots 
$\al i$. On the fundamental weights $\Lambda_{(j)}$ these reflections act as
  \be  r_i:\quad \Lambda_{(j)} \,\mapsto\, \Lambda_{(j)} - \delta_{i,j}^{}\,
  \Sum{k} A^{ik}_{}\Lambda_{(k)}  \,.  \ee
Each element $w\in W$ can be written (non-uniquely) as a product of the
reflections $r_i$. The minimal number of $r_i$ that is needed to form such a 
{\em Weyl word\/} for $w$ is called the {\em length\/} $\ell(w)$ of $w$,
and $\,(-1)^{\ell(w)}{=:}\,\signw$ is called the {\em sign\/} of $w$.

\secT{Symmetrizable \KMA s}b\label{s16}%
A considerable portion of the description above directly generalizes to a 
much larger class of \lie s. Recall that the \findim\ simple Lie algebras 
are obtained when imposing indecomposability as well as
the properties \erf{1a} and \erf{detA} of $A$;
in particular the rank of $A$ is equal to $r$. One arrives at more 
general \lie s \g\ when one stipulates again that $\g\eq\g(A)$ 
is algebraically generated by Chevalley generators $\EE i\pm,\, \Hh i$ modulo
the relations (\ref T\,B), with the matrix $A$ still satisfying 
\erf{1a}, but relaxes the condition \erf{detA} on the determinant of $A$.
The \alg s obtained this way are known as {\em Kac\hy Moody\/} Lie algebras,
and $A$ as a {\em generalized\/} Cartan matrix.
All \kma s, except for the simple ones described above, are \infdim\
(for the case of affine \kma s, this follows from their realization 
that I will present in the next \sec).

Remarkably, even removing the condition \erf{detA} altogether
does not destroy much of the power and beauty of these \lie s and
their \rep\ theory. But to keep some specific nice properties,
it is necessary to restrict oneself to a (still comprehensive)
subclass, the so-called {\em symmetrizable\/} \kma s. These are
characterized by the existence of a \nondeg\ diagonal matrix
$\DD$ such that the matrix $\DD A$ is symmetric. A symmetrizable \kma\ 
is simple (that is, it does not possess any \nontriv\ ideal, i.e.\ sub\alg\
\gP\ such that $[\gPP,\gM]\,{\subseteq}\,\gPP$,
and has dimension larger than 1), if an only if ${\rm det}\,A\Ne0$. 

Every symmetrizable \kma\ possesses a bilinear symmetric invariant form
$(\cdot\,|\,\cdot)$ -- in the \findim\ case this is the Killing form \Erf kf.
However, when \g\ is strictly defined as above, this bilinear form is 
degenerate as soon as the symmetrized \carm\ $\DD A$ has a zero 
eigenvalue. To make the form
\nondeg, one needs to enlarge the \csa\ \go\ of \g\ by so-called outer 
{\em derivations\/} $D$. More precisely, to every zero eigenvalue of $A$ 
there is associated an independent central
element $K^a\iN\g$ which is a suitable linear combination of the 
generators $H^i$. For each $K^a$ one must introduce an independent derivation
$D^a$; this can be done in such a way that $(K^a|D^b)\eq\delta^{ab}$ and 
$(K^a|K^b)\eq0\eq(D^a|D^b)$, and then $(\cdot\,|\,\cdot)$ is non-degenerate.
It is usually this enlarged \lie\ that is referred to as the \kma\ $\g\eq\g(A)$
associated to $A$.

When \erf{detA} is not removed completely, but only relaxed to the requirement
  \be  {\rm det}\,A_{\{i\}}>0 \quad\mbox{for all }\, i=\otor  \Labl5s
for the matrices $A_{\{i\}}$ that are obtained from $A$ by deleting
the $i$th row and $i$th column (and when $A$ is indecomposable),
then $A$ is called an {\em affine\/} Cartan matrix, and $\g\eq\g(A)$ an
{\em affine Lie algebra\/}. In \Erf5s I changed the labelling convention for 
the entries of $A$: they now take values in $\{\otor\}$, so that
$A$ is an $(r+1)\Times(r+1)$\,-matrix; note that
the validity of \Erf5s implies that the rank of $A$ is at least $\grank$.
One can also characterize an affine Cartan matrix by the requirement that 
the symmetrized \carm\ $\DD A$ is positive
semidefinite, but not positive definite. (If $\DD A$ is positive definite, then
$\g(A)$ is a \findim\ simple Lie algebra.) The center of an \aff\ is \onedim,
and correspondingly one includes into the definition of \g\ an outer
derivation $D$.

Once the classification of simple Lie algebras up to some rank
is known, classifying the affine Lie algebras with that
rank is straightforward. For instance, just like for $\grank=1$ 
requiring $\,{\rm det}\,A>0$ immediately leads to
the three simple rank-two Cartan matrices \Erf5t, the requirement that 
$A$ has one positive and one zero eigenvalue yields two rank-one
affine \alg s, denoted by $A_1\Untw$ and $A_1\Twtw$:\,%
\futnote{The value $A^{12}A^{21}=4$ is an exception:
for every other \aff\ one has $A^{ij}A^{ji}\leq 3$ for all $i,j$.}
  \be  A(A_1\Untw)= \smallmatrix{{rr} 2&-2\\-2&2 }\,, \qquad
  A(A_1\Twtw)= \smallmatrix{{rr} 2&-4\\-1&2 }\,.  \Labl5u
For $\grank>1$, the crucial observation is that by
deleting the $i$th row and $i$th column ($i\in\{\otor\}$ arbitrary) 
from an affine Cartan matrix, one must produce the Cartan matrix of a direct sum
of \findim\ simple \lie s. This leads to the following classification. There 
are seven infinite series of affine Lie algebras,
and in addition nine isolated cases; they are denoted by
  \begin{eqnarray} &&\hsp{-.4}
  A_{\grank}\Untw\, {\scs(\grank\geq 2)}\,,\,\ B_{\grank}\Untw\,
  {\scs(\grank\geq 3)}\,,\,\ C_{\grank}\Untw \,{\scs(\grank\geq 2)}\,,\,\
  D_{\grank}\Untw\, {\scs(\grank\geq 4)}\,,\,\ B_{\grank}\Twtw\,
  {\scs(\grank\geq 3)}\,,\,\ C_{\grank}\Twtw\, {\scs(\grank\geq 2)}\,,\,\
  \tilde{B}_{\grank}\Twtw {\scs(\grank\geq 2)} \,, \nonumber\\[.6em]&&\hsp{-.2}
  A_1\Untw\,,\;\ E_6\Untw\,,\;\ E_7\Untw\,,\;\ E_8\Untw\,,\;\ F_4\Untw
  \,,\;\ G_2\Untw\,,\;\ A_1\Twtw\,,\;\ F_4\Twtw\,,\;\ G_2\Twtwtw \,. \label{5W}
  \end{eqnarray}
The affine algebras are thus all denoted by symbols $X_{\grank}^{\sss(\ell)}$,
with $X_\grank$ the symbol for one of the simple Lie \alg s
\Erf5b and $\ell\iN\{1,2,3\}$.\,%
\futnote{For $\ell\eq2,3$ several different conventions are in use.
The notation adopted here is taken from \Cite{FUch}; 
a rather different one is used in \Cite{KAc3}.}
The algebras with $\ell\eq1$ are called {\em untwisted\/} affine algebras, while
those with $\ell=2,3$ are the {\em twisted\/} ones.
For the time being, the distinction between these two types of algebras
is only a matter of nomenclature, but the reason for this choice of 
terminology will become clear later on.

\sect{Affine \LIE s as Centrally Extended Loop \Alg s}c\label{s17}%
The description of a \kma\ \g\ in terms of generators $\{E^i_\pm,\,H^i\}$
and relations (\ref T\,B) encodes the structure of \g\ in a very compact 
form,\,%
\futnote{The most elegant formulation \Cite{KAc3} is to start with a \csa\ of
sufficiently large dimension and describe \g\ as
the \lie\ generated modulo (\ref T\,B), divided by its maximal ideal that
intersecting \nontriv ly the \csa.}
but it is in fact not too transparent. (For instance it is difficult to deduce 
the root system, and it even requires some effort to decide 
whether \g\ is \findim.) But {\em affine\/} \kma s 
possess a specific realization which for many purposes is much more convenient
and which arises naturally in \cft.

Consider the space \gloopb\ of analytic mappings from the circle $S^1$ to some 
\lie\ \gb. (For future convenience I denote the Lie \alg\ to start with
by \gb\ rather than \g.) Fourier analysis shows that when
$S^1$ is regarded as the unit circle in the complex plane with coordinate
$z\eq\eE^{2\pi\ii t}$, then a (topological\,%
\futnote{A topological or analytic
basis, sometimes also called Hilbert space basis, of a vector space
$V$ is characterized by the property that the closure in the topology of $V$
of the set of all finite linear combinations of basis vectors is the
whole space $V$. In contrast, an ordinary vector space basis of $V$ 
has the property that every vector in $V$ is a finite linear combination
of basis vectors.} \hsp{-.31})
basis of the function space \gloopb\ is given by
$\{\ttan\,|\,a\eq1,2,...\,,\gbdim;\; n\iN\zet\}$, where
  \be  \ttan := \bar T^a\otiM z^n \equiv \bar T^a\otiM \eE^{2\pi\ii nt}  \ee
with $z\iN S^1$ and $\{\bar T^a\,|\,a\eq1,2,...\,,\dimgb\}$ a basis of \gb.
Moreover, \gloopb\ inherits a natural bracket operation 
from \gb\ \wrt which it becomes a Lie algebra itself, namely
\def\theequation{\mbox{$\simeq\pi$}}
  \be  [\ttam,\ttbn]:=[\bar T^a,\bar T^b] \otiM(z^m z^n)
  = \sumd c\fabc\, \bar T^c\otiM z^{m+n} = \sumd c\fabc\,\ttcmn \,. \ee
Accordingly, \gloopb\ is called the {\em loop algebra\/} over \gb.
The subset of the loop \alg\ that is spanned by the generators $\ttao$ is a 
Lie subalgebra, called the zero mode subalgebra
of \gloopb; it is isomorphic to \gb.
\def\theequation{\thelect.\arabic{equation}}

Now let $\gb\eq X_\grank$ be a \findim\ simple \lie. Its loop \alg\ \gloopb, 
which is \infdim, turns out to be closely related to the affine \alg\
$\gM\eq X_\grank\Untw$.\,%
\futnote{The notation in the lists \Erf5b and \Erf5W is chosen in such a way 
that $\gM\eq X_\grank\Untw$ with $X\iN\{A,B,C,D,E\}$ precisely corresponds to
the simple \lie\ $\gb\eq X_\grank$.}
However, unlike \aff s, \gloopb\ has a trivial
center (and also does not possess any interesting \rep s).  
But starting from \gloopb\ one can obtain a \lie\ which does possess a center,
by a canonical construction known as {\em central extension\/}. Namely,
by simply adjoining $\ell$ additional generators $K^j$,
$j\eq1,2,...\,,\ell$, to a basis $\{T^a\}$ of 
an arbitrary Lie algebra \g, one constructs an algebra \ghat\
whose dimension exceeds the dimension of \g\ by $\ell$ and for which
the $K^j$ are central, as follows. One 
keeps the original values \fabc\ of those structure constants which
involve only the generators $\hat T^a$ and imposes the relations $[K^i,K^j]=0$ 
and $[\hat T^a,K^j]=0$ for all $i,j$ and all $a$. Here by $\hat T^a$ I denote 
the image of the generator $T^a$ of \g\ in the extended \alg\ \ghat. The 
general form of the brackets among these generators $\hat T^a$ reads
$[\hat T^a,\hat T^b] = \sum_c\fabc\,\hat T^c + \suml i\fabi\,K^i$.
The new additional structure constants $\fabi$ are not 
arbitrary, but are restricted by the Jacobi identity.
Clearly, one should count only those extensions 
as genuine central extensions for which \ghat\ is not 
just the direct sum of \g\ and an abelian \lie;
such a direct sum is certainly obtained when $\fabi\equiv 0$, but
this is typically not the only choice for which this happens.

It is in general a difficult cohomological question whether a given \lie\ 
allows for \nontriv\ central extensions or not. But it is quite straightforward 
to see that \findim\ simple Lie algebras $\gb\eq X_\grank$
do not possess \nontriv\ central extensions, whereas their loop \alg s \gloopb\
possess indeed a unique \nontriv\ extension by a single central generator
$K$, with brackets 
  \be  [K,\tan]=0 \,, \qquad
  [\tam,\tbn]=\sumd c \fabc\,\tcmn + m\,\delta_{m+n,0}\kappb^{ab}\, K\,; \Labl9r
here $\kappb^{ab}\eq\kappb(\bar T^a,\bar T^b)$, with $\kappb$ the Killing form 
\Erf kf of \gb. Moreover, the untwisted \aff\ $\gM=X_\grank\Untw$ is obtained 
from this centrally extended loop algebra $\widehat{\gloopb}$ by including a 
derivation $D$, i.e.\ a generator with Lie brackets
  \be   [D,\tam]=m\,\tam \,, \qquad [D,K]=0 \,.  \Labl9w
(Thus $D$ measures the mode number $m$ and hence provides a \Zet-grading, 
which is of precisely the same form as that of the Virasoro 
operator $-L_0$ in the case of chiral \alg s \w, cf.\ equation \erf{l0w}.)
In short, the vector space structure of \g\ is
  \be  \gM=\ghat\oplus\complex D= \complex\,[z,z^{-1}] \otimes^{}_\complex
  \gB\,\oplus\complex K\oplus\complex D \,.  \ee
Just like for loop \alg s, the zero modes $\hat T^a_0$ of \g\
span a subalgebra which is isomorphic to the simple Lie \alg\ \gb; 
this is called the {\em horizontal\/} subalgebra of \g. 

Upon choosing a \cwbasis\ $\{\bar T^a\}=\{H^i\Mid i\eq\onetor\}\cup\{\ealb\}$ 
of $\gb\,{\equiv}\,\gb(\bar A)$, the defining Lie bracket relations (\ref T\,A) 
of \g\ take the form
  \be \bearl
  [H^i_m,H^j_n]=m\,(\bar\DD\bar A)^{ij}\delta_{m+n,0}\,K \,, \qquad
  [H^i_m,E^{\bar\alpha}_n]=\abar^iE^{\bar\alpha}_{m+n} \,,  \\[.55em]
  [E^{\bar\alpha}_m,E^{\bar\beta}_n]=e_{\bar\alpha,\bar\beta}^{}
    E^{\bar\alpha+\bar\beta}_{m+n} \qquad {\rm for}\
    \bar\alpha+\bar\beta\ \mbox{a \gb-root}, \\ \mbox{ }\\[-.6em]
  [E^{\bar\alpha}_n,E^{-\bar\alpha}_{-n}]=\sumr i \abar_i^{} H^i_0+nK \,.
  \eear\Labl9X
Here $m,n\iN\zet$, $i,j\iN\{\onetor\}$,
and $\abar,\,\bar\beta$ are arbitrary \gb-roots.

So far only the untwisted affine algebras have been obtained. But along
similar lines one can realize the twisted ones as well.
The only modification is to give up the single-valuedness of the maps $f$
from $S^1$ to \gb\ that were used in the loop construction above.
More precisely, the {\em twisted\/} affine Lie \alg s are obtained when 
instead of $f(\eE^{2\pii}z)\eq f(z)$
one imposes the {\em twisted\/} boundary conditions
  \be  x\otiM f(\eE^{2\pii}z)=\omega (x)\otiM f(z)  \ee
for all $x\iN\gB$, where $\omega$ is an automorphism of the \hsa\ \gb\ of
finite order $N$. In other words, $f$ is now no longer a function on
the circle $S^1$, but rather on an $N$-fold covering of $S^1$. However, when
performing this construction for two automorphisms $\omega$ and $\omega'$
for which $\omega^{-1}\omega'$ is an inner automorphism of \gb, one obtains
two realizations of one and the same abstract \lie; in particular, when
$\omega$ is inner, then one gets a realization of an untwisted \alg\ which
differs from the realization \Erf9X. As a consequence, the twisted \aff s
correspond in fact to equivalence classes of outer modulo inner
automorphisms of \gb; in each such class there is a distinguished
representative which is induced by a symmetry of the \dyd\ of \gb.

\sect{The Triangular Decomposition of Affine \LIE s}d\label{s18}%
For any \aff\ \g\ one
can determine a triangular decomposition of \g\ that is analogous to \erf{tria}
for chiral \alg s. One first observes that a Cartan subalgebra \go\ of \g\ is
spanned by $\{K, D, H^i_0 \Mid i\eq\onetor\}$. The roots (i.e., vectors of
eigenvalues) with respect to $(H_0,K,D)$ are given by
  \be  \alpha=(\abar,0,n) \quad{\rm for}\;\ \abar\iN\rosyb\,,\;n\iN\zet 
  \qquad\,{\rm and}\qquad\,
  \alpha=(0,0,n) \quad{\rm for}\;\ n\iN\zet\Setminus\!\{0\}\,,  \Labl9B
where $\rosyb\eq\{\abar\}$ denotes the root system of \gb. These roots 
correspond to the generators $E^{\bar\alpha}_{n}$ with $n\iN\zet$ and
to $H^j_n$ with $n\ne 0$, \resp. While the roots $(\abar,0,n)$ are \nondeg, 
i.e.\ appear with multiplicity one, the roots $(0,0,n)$ do not depend on the 
label $j$ of $H^j_n$ and hence have multiplicity $\grank$. The degenerate 
roots are all integral multiples of $\delta\df(0,0,1)$. Also note that the 
roots of the derived algebra $[\gM,\gM]\,{=:}\,\ghat$ are all infinitely
degenerate, because in $\ghat$ there is no Cartan subalgebra generator
that is able to distinguish between different labels $n$. Thus another 
\role\ of the derivation $D$ is to avoid such infinite multiplicities. 

Having made a choice for distinguishing between positive and negative roots
of \gb, a system of positive roots ($\alpha\,{>}\,0$) of \g\
can be chosen to consist of those roots $(\abar,0,n)$ for which either 
$n\,{>}\,0$ and $\abar\iN\rosyb$ is an arbitrary \gb-root or zero, or else 
$n\eq0$ and $\abar$ is a positive \gb-root. The remaining roots are then
the negative ones ($\alpha\,{<}\,0$). Together with \go, the subalgebras
$\gp\df{\rm span}\{\eal|\,\alpha\,{>}\,0\}$ and $\gm\df{\rm span}\{\emal|\,
\alpha\,{>}\,0\}$ then provide a triangular decomposition 
  \be \gM = \gp\oplus\go\oplus\gm  \Labl9W
of \g, i.e.\ satisfy $[\gM_\pm,\go\oplus\gM_\pm]\subseteq\gM_\pm$ 
and $[\gp,\gm]\subseteq \go$, analogous to \erf{trib}.

The simple roots $\ai$ of \g\ provide a basis of the root space such that
in the expansion $\alpha\eq\sumro ib_i\,\alpha^{(i)}$ one has, for all
$i\eq\otor$, $\,b_i\iN\zetpluso$ if $\alpha\,{>}\,0$ and $b_i\iN\zetminuso$ if 
$\alpha\,{<}\,0$.  With the above choice of positive roots, the simple roots are
  \be  \ai \eq (\aibar,0,0)\,{\equiv}\,\aibar\,\ \ {\rm for}\;\ i \eq
  \onetor \qquad {\rm and}\qquad
  \ao \eq (-\bar{\tta},0,1) \eq \delta\mi\bar{\tta}\,,  \Labl9Y
where $\aibar$ are the simple roots of the \hsa\
\gb\ and \ttab\ is the highest root of \gb. For each $i\iN\{\otor\}$
the generator \eali\ corresponding to a simple root is precisely the
generator that in the \chserre formulation is denoted by
$E^i_+$. The relation to the $\EE{\bar\alpha}n$-notation is thus
  \be  E^i_+=E^{\bar\alpha^{(i)}}_0 \quad{\rm for}\;\ i=\onetor \,, \qquad\ 
  E^0_+=E^{-\bar\theta}_1.  \ee

\sect{\Rep\ Theory}e\label{s19}%
Every \kma\ \g\ has a triangular decomposition 
$\gM \eq \gp\ooplus\go\ooplus\gm$ similar to the one
described above. Correspondingly a particularly interesting class of modules
(\rep\ spaces) $V$ of \g\ are those which have 
a basis on which the whole \csa\ \go\ acts diagonally, and hence possess
a decomposition $V\eq\bigoplus_\lambda{} V_{(\lambda)}$ into subspaces such that
  \be  R(H^i)\, v_\lambda=\lambda^i \cdot v_{\lambda}   \Labl Hi
for all $v_\lambda\iN V_{(\lambda)}$. The vectors
$\lambda\equiv(\lambda^i)_i$ are called the {\em weights\/} of the module $V$;
they are elements of the weight space \godual, and the numbers $\lambda^i$ are
their components in the basis furnished by the fundamental weights \li.

A {\em maximal\/} weight $\L$ satisfies by definition $R(\eal)\, \vLambda\eq0$ 
for all positive roots $\alpha$ and all $\vLambda\iN V_{(\Lambda)}$.
A module which has exactly one weight $\Lambda$ with this property is called a
{\em\hwm\/} and is denoted by \VL.
All other elements of \VL\ can be obtained by applying step operators for
negative roots to $\vLambda$, i.e.\ every $v\iN\VL$ is contained in 
$\U(\gm)\,\vLambda$, where $\U$ denotes the universal enveloping \alg.

A \kma\ \g\ is \alg ically generated by its sub\alg s $\slti$ that are 
associated to the simple \g-roots, i.e.\ $\slti\eq{\rm span}\{E^i_\pm,H^i\}$ 
(together with the derivations, but these do not concern us here). {}From 
(\ref T\,B) one reads off that $\slti\cong\sltwo$ for each $i$;
one can therefore reduce many issues in the \rep\ theory of \g\ to that of the
simple \lie\ \sltwo. A particularly important application is the following.
All \nontriv\ \g-modules are \infdim, but in favorable circumstances
the subspaces into which an {\em irreducible\/} \hwm\ \hl\ decomposes with 
respect to the subalgebras $\slti$ are all \findim. 
(Loosely speaking, such \g-modules are `less \infdim' than others.) This happens
precisely when the \hw\ $\Lambda=\sum_i\Lambda^i\li$ of \hl\ is {\em dominant 
integral\/}, that is, a positive integral linear combination
of fundamental weights, i.e.
  \be  \Lambda^i\equiv(\Lambda,\aiv)\in\zetpluso \quad \mbox{for all}\ i
  \Labl xz
(i.e.\ for $i\eq\otor$ in the case of \aff s).

For affine \g\ it is convenient to write \g-weights (cf.\  
\equ \Erf9B) as triples $\lambda\eq(\bar\lambda,k,m)$
with $k$ the eigenvalue of the central element $K$ and $m$ the 
{\em grade\/}, i.e.\ the ei\-gen\-value of the derivation $D$. Then
for untwisted \g, the fundamental weights are\,%
\futnote{$\thetagb$ denotes again the highest root of \gb. 
The presence of the length squared of $\thetagb$ implements the dependence on 
the normalization of the inner product on the weight space \gbodual\ of \gb.}  
  \be  \li=(\libar,\onehalf(\ttab,\ttab) \Aiv,0) \;\ {\rm for}\ i=\onetor 
  \quad{\rm and}\quad \lo = (0,\onehalf(\ttab,\ttab),0) \,.  \Labl1A
Here $\Aiv$ are the dual Coxeter labels, defined as the coefficients of 
$\thetagb^\Vee\equiv2\thetagb/(\thetagb,\thetagb)$ (the highest coroot)
in the basis of simple coroots of \gb. Also, the property \Erf xz means, first,
that the horizontal part $\Lb$ of $\L$ is a dominant integral 
\gb-weight, and second, that 
  \be  \kv:=2k /(\thetagb,\thetagb) \,,  \ee
called the {\em level} of \hl,
is a non-negative integer; and third, there is the bound
  \be  0\,\le\; \sum_{i=1}^r
  \Aiv\Lb^i\equiv(\Lb,\ttab^\Vee) \;\le\, \kv \,. \Labl2x

The precise structure of an \ihwm\ \hl\
strongly depends on whether the \hw\ $\L$ is dominant integral or not.
For generic $\L$ already the Verma module \vml, which is isomorphic to
$\U(\gm)\,\vLambda$, without any additional relations (compare equation \Erf1v),
is irreducible. In contrast, for dominant integral $\L$ the Verma module 
contains null vectors. The entire information on the null vector structure of
\vml\ is encoded in the weight $\L$: all null vectors can be obtained
by acting with $\U(\gm)$ on the so-called {\em primitive\/} null vectors
  \be  v^{(i)}_{}:= (E^{-\alpha^{(i)}})^{\Lambda^i+1}_\P\vLambda \,, \Labl1q
where $i$ can take all possible values, i.e.\ $i\iN\{\otor\}$ for affine 
\alg s. The \irmod\ \hl\ is obtained from the Verma module \vml\ by taking 
the quotient \wrtt submodule spanned by all null vectors. For affine \g,
the quotienting by the null vectors $v^{(i)}_{}$
with $i\eq\onetor$ can be implemented by just restricting to irreducible 
modules rather than Verma modules with respect to the \hsa\ \gb\ of \g; 
also, one has $v^{(0)}_{}\eq(E^{-\alpha^{(0)}})^{\Lambda^0+1}_{} \vLambda
\eq$\linebreak[0]$(E^\ttaB_{-1})^{k^\vee-(\ttaB^\vee\!,\,\Lb)+1} \vLambda$.

\sect{Characters}f\label{s20}%
Precisely as in the case of the chiral \alg s \w\ that arise in \cft\
(see \refsJ), one can encode the information contained in the
weight system of a module $V$ of a \kma\ \g\ in its {\em character}
  \be  \chiv:=\sum_\lambda \mult V\lambda\,\elambda \,. \ee
Since the weights $\lambda$ are linear functions on the \csa\ \go\ of \g,
the formal exponential $\elambda$ -- and hence also the character 
$\chiv$ -- can be regarded as a function on \go, i.e.\ $\elambda{:}\;\go\,{\to}
\,\complex\,$, $h\,{\mapsto}\,\elambda(h){:=}\exp(\lambda(h))$, where 
`exp' is the ordinary exponential function. 
Moreover, via the invariant bilinear form the \csa\ \go\ can be identified 
with its dual space, i.e.\ with the weight space \godual. Correspondingly, 
by setting
  \be  \elambda:\quad \mu\Mapsto\elambda(\mu):=\exp[(\lambda,\mu)] 
  \quad\mbox{for all } \mu\iN\godual\,,  \ee
where $(\cdot\,,\,\cdot)$ is the inner product in weight space, one can 
interpret the exponentials $\elambda$ and the character $\chiv$ also 
as functions from the weight space to the complex numbers. 

One of the most intriguing results in the \rep\ theory of symmetrizable \kma s 
is the {\em Weyl\hy Kac character formula\/} which states that the character
$\chil$ of an irreducible \g-module \hl\ with \hw\ $\Lambda$ satisfies
  \be  \chil(\mu) = \frac { \sumww \signw\,\exp[(\sig(\L+\rho),\mu)] }
  {\sumww \signw\,\exp[(\sig(\rho),\mu)] } \,. \Labl1V
Here $W$ is the Weyl group of \g\ which, just like in the \findim\ case, is
the group generated by reflections about the
hyperplanes perpendicular to the simple roots,
$\signw\iN\{\pm1\}$ is the sign of a Weyl group element $w$, and
$\rho=\sum_i\li$ is the {\em Weyl vector\/} of \g.
The formula \Erf1V is a consequence of a deep interplay between the structure
of the Weyl group and the relation between irreducible and Verma modules.
The character $\CHV$ of a Verma module \vml\ is easily obtained with the help 
of the \pbw, which tells how to construct bases of \uenva s; from the form of
such a basis it follows that each generator \eal\ associated to a positive 
\g-root $\alpha$ contributes a geometric series in $\eE^{-\alpha}$ to $\CHV$, 
and by summing these series one finds that
  \be \CHV = \eE^\Lambda\cdot \Prod{\alpha>0} ( 1 - \eE^{-\alpha})_{}
  ^{- \multal} \,. \ee
Moreover, irreducible characters are invariant under the Weyl group $W$,
$\chil(\sig(\mu))\eq\chil(\mu)$, while Verma characters acquire a factor of
$\,\signw$. When combining these results with a linear relation that relates
irreducible and Verma characters and with basic properties of the Weyl 
group $W$, one obtains the character formula in the form
  \be \chil = \frac { \sumww \signw\, \eE^{w(\Lambda+\rho) - \rho} }
  { \prod_{\alpha>0} (1-\eE^{-\alpha})^\multal_{} } \,. \labl W
Furthermore, for $\Lambda\eq0$, \hl\ is just the trivial \onedim\ module so 
that it has character $\chii_0\equiv1$, and hence \erf W implies the 
{\em denominator identity}
  \be \prod_{\alpha>0} (1-\eE^{-\alpha})^\multal_{} =
  \sumww \signw\, \eE^{w(\rho) - \rho}  \ee
(which is the source of many most non-trivial combinatorial identities).
Upon re-inserting this result into \erf W one arrives at the character
formula in the form \Erf1V.

With the help of Poisson resummation, the character formula allows one to
analyze the behavior of characters under modular transformations. One finds
that the set of characters of \ihwm s at fixed level of every untwisted 
(and of some twisted)\,%
\futnote{Among the twisted \aff s, precisely those which do have nice
modular transformation properties have appeared \cite{fusS4} in \cft.}
\aff\ spans a module for a unitary \rep\ of \slz\ -- compare the \kpf\ 
\erf{kpf} below. As observed after equation \Erf9w, the derivation $D\iN\go$ 
of an affine \kma s \g\ behaves just like $-L_0$. Thus the restriction
of the character to the span $\complex D$ of $D$ yields precisely the
Virasoro-specialized character that I already introduced in \erf{chd};
however, the Virasoro-specialized characters are usually not sufficient to
determine the \rep\ matrices of \slz. 

The character formula is valid for arbitrary symmetrizable \kma s. 
Untwisted \aff s are nevertheless distinguished, as the formula
is then often easy to evaluate. It becomes
particularly simple when \g\ is simply laced and at level 1. In this case 
the Virasoro-specialized character $\chil(\tau)$ is, up to an over-all power
of $q\eq\exp(2\pi\ii\tau)$, just the $-r$\,th power of Euler's 
product function $\varphi(q)=\prod_{n=1}^\infty(1-q^n)$ multiplied 
with a theta function for the \gb-root lattice shifted by the \hw. E.g.\ 
for $A_{N-1}\Untw$ at level one the formula reads (expressing 
$A_{N-1}$-weights in an orthogonal basis of a space of dimension $r+1\eq N$)
  \be  \chii^{}_{\Lambda_{(j)}}(\tau)= (q^{1/24}\varphi(q))^{-(N-1)}\,
  q^{-j^2/2N} \SUM{\scs j_1,j_2,...,j_N\in\zet \atop \scs j_1+j_2+...+j_N=j}
  q_{}^{(j_1^2+j_2^2+\cdots j_N^2)/2}  \,.  \ee
for $j=1,2\Ldots N-1$.

Let me return to the relation between Verma and irreducible modules.
The maximal submodule $V_{\rm max}$ that must be divided out from \vml\ in
order to obtain the \irmod\ \hl\ is {\em not\/} the direct
sum of the Verma modules that are `headed' by the primitive null vectors.
Rather, in $V_{\rm max}$ the latter can overlap (moreover, in general there
also exist proper submodules which are {\em not\/} Verma modules). However, 
when $\Lambda$ is an integrable weight,
the overlap between two Verma submodules \vmle\ and \vmlz\ of \vml\ is
a submodule of both \vmle\ and \vmlz\ and is again
obtained from primitive null vectors (now in \vmle\ \resp\ \vmlz).
Moreover, the precise structure is governed by the Weyl group of \g.
Based on this connection one can construct a semi-infinite complex of 
\g-modules and \g-module homomorphisms, the so-called {\em BGG resolution\/} of
\hl, in which each module except for the one at the right-most place, which is
\hl, is a direct sum of Verma modules. (The structure looks rather simple,
but the proof, which works even for non-symmetrizable 
\kma s \Cite{kuma2}, is most complicated.)

\lect4{\wzwts\ and \cts}{6.18}{4.2}

\sect{WZW Theories}h\label{s21}%
The results of \lec3 show in particular that an untwisted affine \lie\ 
\g\ enjoys all properties that are needed for a chiral \alg\ of \cft, except 
that it does not contain the \vira\ \vir. The latter shortcoming is easily 
remedied by just prescribing the bracket relations $[L_m,K]=0=[C,\tan]$ and 
  \be  [L_m,\tan] = -n\,\tamn  \Labl lj
among the generators of \g\ and \vir. This way one
arrives at the semi-direct sum of \vir\ and the centrally extended loop \alg\
$\ghat=[\g,\g]$ and identifies the Virasoro generator $L_0$ with minus the
derivation $D$ of the affine \alg\ \g.

While this observation does not automatically imply that one can associate a
\cft\ to any untwisted \aff\ \g, indeed there exists a class of \cfts\ with such
a chiral \alg\ $\g\ooplus\vir$, namely the so-called {\em WZW\/} (\WZW) 
{\em theories\/} \Cite{knza}. The\-se theories are characterized by the fact 
that 
their chiral symmetry \alg\ consists of the \emt\ $T(z)$ and the {\em current}
  \be  J(z) = \sum_{m\in\zet} z^{-m-1}\,\tam  \ee
whose modes $L_m$ and $\tan$ satisfy the relations \erf{vir}, \Erf9r and 
\Erf lj, together with the requirement that the Virasoro generators $L_m$
are of the {\em Sugawara\/} form, i.e.
  \be  L_m = \Frac 1{2(\kV+\gV)} \sum_{n\in\zet} \kappb_{ab}
  \normord{\tamn\tbnn} \,.  \labl{sug}
Here the colons $\normord{\,\,}$ denote a normal ordering prescription (compare
equation \erf{normord}), \kv\ is the level of the relevant \g-modules,
and \gv\ is the dual Coxeter number of the \hsa\ \gb\ of \g, i.e., the 
(properly normalized) eigenvalue of the quadratic Casimir operator in the 
adjoint representation of \gb. The Sugawara formula \erf{sug} implies that 
\wzwts\ are governed by the \rep\ theory of \g, and there is no need to 
study the \vira\ independently of \g. Moreover, it follows from the \rep\ 
theory of \g\ that I summarized in \refse\ that
many quantities of interest of a \wzwt\ can be studied entirely in terms of 
the \findim\ simple Lie \alg\ \gb\ and of the level \kv.
E.g., because of the relation \erf{sug} the 
Virasoro central charge $c$ can be expressed as
  \be c(\gM,\kV)=\frac{\kV\,{\rm dim}\, \gB }{\kV+\gV} \,  .  \labl{confc}
Note that in a \qft\ the factor $(\kv+\gv)^{-1}$ makes sense only if \kv\ 
can be regarded as a \Complex\,-number; in other words, it is necessary that
all \g-\rep s appearing in a \wzwt\ have one and the
same value of the level. For unitary theories, \kv\ is a non-negative integer. 

\wzwts\ also possess a Lagrangian realization (from which they in fact received 
their name) as euclidean principal sigma models \Cite{powi}, i.e.\ 
\twodim\ non-linear sigma models for which the target space is a compact simple
Lie group manifold $G$ -- the one whose \lie\ is the compact real form of
\gb\ -- which are supplemented \Cite{witt8} by a Wess\hy Zumino 
\Cite{wezu,novi} term. The action of such a WZW sigma model is
$S_{\rm WZW}\eq\kv(S_{\sigma}+S_{\rm WZ})$ with
  \be  \hsp{-.4} S_\sigma\eq \int\!\Frac{{\rm d}^2z}{16\pi}\,{\rm tr}
  (\partial_{\mu}\gamma\,\partial^{\mu}\gamma^{-1})\,, \quad
  S_{\rm WZ}\eq \int\! \Frac{{\rm d}^3y}{24\pi}\,\epsilon^
  {\mu\nu\tau}{\rm tr}(\dot{\gamma}^{-1}\partial_{\mu}\dot{\gamma}\,
  \dot{\gamma}^{-1}\partial_{\nu}\dot{\gamma}\,\dot{\gamma}^{-1}
  \partial_{\tau}\dot{\gamma}) \,.  \ee
Here $\gamma$ takes values in $G$, and the integral in
$S_{\rm WZ}$ is over a \threedim\ ball whose surface is
to be identified with the compactified
\twodim\ world sheet such that $\dot{\gamma}=\gamma$ on the surface (the WZ 
term is a total derivative, hence can also be written as a surface integral).

Sometimes one uses the term \wzwt\ also to refer to a somewhat more general
class of models, namely those where the affine \alg\ \g\ is replaced by a
{\em current \alg\/}, i.e.\ by a
direct sum of untwisted affine \alg s and of Heisenberg \alg s. (The Lie
bracket relations of the Heisenberg \alg, which generates the observables \w\
for the theory of a free boson, and which is often also called the
$\hat{\liefont u}(1)$ current \alg, were displayed in \Erf u1. Note that this
\lie\ is {\em not\/} a \kma.)
Those models for which \g\ is just an untwisted affine \alg, and hence \gb\ is
simple, are then called `simple' \wzwts.

\sect{WZW Primaries}i\label{s22}%
The primary fields of a unitary \wzwt\ with diagonal modular invariant are in 
one-to-one correspondence with the integrable highest weights $\Lambda$
of an untwisted affine \lie\ \g\ at level \kv, i.e.\ with the
dominant integral \gb-weights $\Lb$ that satisfy \Erf2x.
Only finitely many weights of \gb\ fulfill the condition \Erf2x;
\wzwts\ are therefore rational \cfts.

WZW primary fields $\phi\eq\phl(z)$ can be characterized by the properties that 
(compare also \equ \Erf Lp)
  \be  [\tan,\phl(0)]=0  \quad{\rm for}\ n>0  \qquad{\rm and}\qquad
       [\tao,\phl(0)]=\tlbara\,\phl(0)  \,, \ee
where \tlbara\ is a shorthand for the \rep\ matrix $\bar R_\Lb(\bar T^a)$.
In terms of operator products, these relations correspond to
  \be  \Cont{J^a(z)\,\phl}\wbw=(z-w)^{-1} \,\tlbara\phl\wbw \,.  \Labl Jp
The operator product of two current fields reads
  \be \Cont{J^a(z)\,J^b}(w)=(z-w)^{-2}\bar\kappa^{ab}K -(z-w)^{-1}
  \sum_c\fabc\, J^c(w) \,; \ee
thus while $J^a$ is a primary field of the Virasoro algebra,
it is not primary \wrt \g, but rather a
descendant of the identity field \bfe\ at grade one.

The conformal dimension of a (\g-)\,primary field with highest weight $\L$ is
  \be  \cd\Lambda=\frac{(\Lb,\Lb+2\bar\rho)}{2\,(\kV+\gV)} \,,  \labl{cdim}
with $\bar\rho$ the Weyl vector of \gb.
Note that $(\Lb,\Lb+2\bar\rho)$ is the quadratic Casimir eigenvalue of the
irreducible \gb-module that has \hw\ $\Lb$. Conjugate \pf s have \hw s which 
are `charge-conjugate' \wrt \gb, and hence identical \cdim s, as they should.

The Ward identity corresponding to the zero modes of the current reads
  \be  \sum_{i=1}^p R_i(\bar x)\,\vev{\varphi^{}_{\lambda_1}\,
  \varphi^{}_{\lambda_2} \cdots \varphi^{}_{\lambda_p} } = 0  \,,  \Labl wj
where $\varphi^{}_{\lambda_j}$ are arbitrary conformal fields of 
\gb-weight $\lambda_j$ and $\bar x=\sum_a\xi_a\tao$ is any element of \gb. By 
taking $\bar x$ in the \csa\ of \gb, the formula \Erf wj
tells us that $\llb\sum_{j=1}^p \lambda_j\lrb
\vev{\varphi^{}_{\lambda_1}\!\cdots \varphi^{}_{\lambda_p}}$%
\linebreak[0]$\eq0$. Thus a \corfu\
can be non-vanishing only if it is `\gb-neutral', i.e.\ if $\sum\lambda_j\eq0$.
In particular, $\vev{\phi_{\Lambda_1}\!\cdots\phi_{\Lambda_p}}\equiv0\,$
for each \corfu\ which only involves (non-trivial) fields $\phi_\Lambda$
which are primary not only \wrt \vir, but also \wrt \g.
In order to study non-vanishing \corfu s it is therefore necessary to deal also 
with correlators which involve secondary fields. It is, however, sufficient to
allow just for `horizontal' \g-descendants, which are obtained from the 
primaries $\phi_\Lambda$ by acting only with zero mode currents $\tao$ and 
which are still \vir-primary. (An analogous complication arises whenever 
the zero mode subalgebra $\w^0$ of the chiral symmetry \alg\ is non-abelian.)

\sect{Modularity, Fusion Rules and WZW Simple Currents}T\label{s23}%
As already mentioned in \refsf, the irreducible characters of 
a \wzwt\ span a module for a unitary \rep\ of \slz. 
The explicit form of the modular \tmat\ is obtained by inserting \erf{cdim}
into the general formula \Erf Te.
The modular \smat\ is given by the {\em\kpf\/} \Cite{kape3}
  \be  S_{\Lambda,\Lambda'}^{} =\caln \sumwb \signw \, \exp[-\Frac{2\pi\ii}
  {\kV+\gV}\,(w(\bar\L+\bar\rho),\bar\L'+\bar\rho)]\,. \labl{kpf}
Here the summation is over the Weyl group $\overline W$ of the horizontal 
subalgebra \gb\ of \g; the normalization $\caln$ follows from the 
requirement that $S$ must be unitary.
As an example, consider $\g\eq D_N\Untw$ at level one. Then there are 
four primary fields corresponding to the singlet ($\circ$), vector (v), spinor 
(s), and conjugate spinor (c) representation of $D_N$, with conformal
dimensions $\cd\circ=0$, $\cd{\rm v}=1/2$, $\cd{\rm s}=\cd{\rm c}=N/8$, and
the \smat\ reads
  \be S((D_N\Untw)_1) = \onehalf \smallmatrix{{rrrr} 1&1&1\ &1\ \\[.1em] 
  1&1&-1\ &-1\ \\[.1em] 1&-1&\ii^{-N}&-\ii^{-N}\\[.1em] 1&-1&-\ii^{-N}&\ii^{-N}
  } .  \ee

Using these explicit formul\ae, it has in particular been possible to construct
a large number of modular invariant combinations of characters (by methods
such as simple current extensions, conformal embeddings, automorphisms of 
the \furu s, or (quasi-)Galois symmetries). However, so far the classification 
of physical modular invariants has been completed only for $\gb\eq \sltwo$, 
where an $A$-$D$-$E$ pattern emerges \Cite{caiz2}, and \Cite{gann4} for 
\gb\eq\slthree. In particular, despite much progress (see e.g.\ \Cite{gann8}) 
it is not yet known whether already all `exceptional' modular invariants,
which are similar to the $E_7$-type invariant for \sltwo, have been found.

Via the Verlinde formula \erf{-verl}, the \kpf\ can be employed to compute 
the fusion rules of \wzwts. Moreover, comparison with the 
character formula \Erf1V shows that
$S_{\Lambda,\Lambda'}$ is essentially the character of an \irmod\ of the \hsa\ 
\gb\ evaluated at a specific rational element of the weight space \gbodual.
By combining these observations with the unitarity of the modular \smat\ and
with knowledge about the tensor products of \gb-modules in a clever manner, one
finds the following {\em \kwf\/} \Cite{KAc3,walt3}
  \be  \N{\Lambda}{\Lambda'}{\;\Lambda''}=\sumww
  \signw\,{\rm mult}_{\bar\Lambda'}(\hat w(\bar\Lambda''+\bar\rho)
  \mi\bar\rho\mi\bar\Lambda)    \Labl2Y
for the \frc s $\N{\Lambda}{\Lambda'}{\;\Lambda''}$ of \wzwts.
Here $\bar\rho$ is the Weyl vector of the \hsa\ \gb, $W$ is the Weyl group 
of \g, $\hat w$ denotes the (affine) action of $w\iN W$ that is induced on 
the weight space of \gb,
and ${\rm mult}_{\bar\Lambda}(\bar\mu)$ is the multiplicity with which the 
weight $\bar\mu$ arises for the \ihwm\ \hlb\ of \gb.
Since $W$ is an infinite group, \Erf2Y would not be of practical use if one 
really had to perform the sum over all of $W$, even though only a finite
number of Weyl group elements gives a non-zero contribution. Fortunately, this
is not necessary, because one can express the relevant elements of $W$
through a finite algorithm as products of the (finitely many) reflections
which correspond to simple \g-roots (see e.g.\ Appendix A of \Cite{fusS}).

All simple currents of \wzwts\ are known \Cite{fuge,fhpv,jf15}. Except for an
isolated case appearing for $E_8\Untw$ at level two, they are, besides the
identity field, the \pf s with 
\hw s $\kv\Lambda_{(i)}$, where \kv\ is the level and $\Lambda_{(i)}$ is a 
fundamental weight for which the Coxeter label $a_i$ -- the coefficient of
$\alb i$ in the expansion of $\thetagb$ \wrtt basis of simple \gb-roots -- has
the value $a_i\eq1$.\,%
\futnote{Numbering the fundamental weights as in \Cite{KAc3}, these 
are the following. For $\gb\eq A_r$, all fundamental weights qualify;
their fusion rules are isomorphic to $\zet_{r+1}$. 
For $B$ and $C$ type \alg s, there is a single non-trivial simple current,
with $i=1$ for $B_r$ and $i=r$ for $C_r$. For $D$ type \alg s, there
are three non-trivial simple currents, corresponding to 
$i=1,r,r\mi1$; their fusion products are isomorphic to the multiplication of
the vector, spinor, and conjugate spinor conjugacy classes of $D_r$.
For $E_6$ there is a simple current of order three, with $i\eq1$, and its 
conjugate with $i\eq5$, and for $E_7$ there is a simple current of order 2, 
with $i=6$.}
The action of such a simple current $\J$ on the \hw s of the \pf s
corresponds to a symmetry of the Dynkin diagram of \g, and the monodromy 
charge $\QJ(\Lambda)$ is proportional to the conjugacy class of $\bar\Lambda$. 

\sect{The \KZE}j\label{s24}%
Recall that one obtains \ihwm s of \w\ by `setting to zero' the null vectors in
the corresponding Verma module. In more field theoretical terms, this means 
that application of certain elements of the \uenva\ $\U(\wm)$
to the \hwv\ yields a vector which is not contained in the \sps.
Inserting such an operator into a correlation function must therefore give
zero. Identities for correlators that are obtained this way are called 
{\em null vector equations}. In \wzwts, where $\w\,{\supset}\,\gM$,
one clearly gets null vector equations by acting with
the \enva\ of $\gM_-$, corresponding to the null vectors \Erf1q.
These {\em Gepner\hy$\!$Witten\/} equations are purely \alg ic relations
(and are thus quite different from the null vector differential equations
that arise for $\w\eq\vir$ when $c\,{<}\,1$) and will not be treated here.

But in \wzwts\ the Sugawara relation \erf{sug} provides an
additional source for null vectors. The associated null vector equations
are known as {\em \kze s\/} \Cite{knza}. I will formulate them for the \four s
  \be  \calg\zbz= \langle\phi_{\L_1}(\infty,\!\infty)\,\phi_{\L_2}(1,\!1)\,
  \phi_{\L_3}\zbz\,\phi_{\L_4}(0,\!0)\rangle \,.  \Labl91
As already observed, due to the Ward identity \Erf wj such correlators vanish
identically when the fields $\phi_{\L}$ are primary \wrt \g, and to get 
non-trivial correlators one must also allow for (\vir-primary) horizontal 
descendants which arise from the \g-primaries $\phi_{\L}$
by application of the zero modes $\tao$ of the current;
I will denote such descendants of $\phi_{\L}^{}$ by $\phi_{\L}^\lambda$.
By \gb-symmetry, one can then expand \Erf91 as 
  \be  \calg_{\L_1\L_2\L_3\L_4}^{}\zbz
  \equiv\calg_{\L_1\L_2\L_3\L_4}^{\,\lambda_1\lambda_2\lambda_3\lambda_4}\zbz
  =\sum_{i=1}^{\md} \calg_{\L_1\L_2\L_3\L_4;i}^{}\zbz\cdot
  (\IA)_{}^{\bar\lambda_1\bar\lambda_2\bar\lambda_3\bar\lambda_4}  \,,  \Labl92
where $\{\IA\,|\,i\eq1,2\Ldots\md\}$ is a basis of invariant \gb-tensors $\IA$ 
for the tensor product \rep\ $\bar R_{\Lb_1}{\otimes}\bar R_{\Lb_2}{\otimes}
\bar R_{\Lb_3} {\otimes}\bar R_{\Lb_4}$ of \gb. 
In this description, the specific choice of horizontal descendants is encoded
in the corresponding components $(\IA)^{\bar\lambda_1\bar\lambda_2\bar\lambda_3
\bar\lambda_4}$ of the invariant tensors, while the `reduced' functions 
$\calg_{\L_1\L_2\L_3\L_4;i}\zbz$
only depend on the \pf s. Also, the number $\md$ of independent invariant 
tensors is the \gb-analogue of the number $M\eq\sum_{\sss I}\N AB{\;I} \Ns ICD$ 
(cf.\ \equ \Erf MM) of chiral blocks appearing in \equ \Erf14, i.e.\ one has 
$\md\eq\sum_{\bar\mu}\calL{\Lb_1}{\Lb_2}{\scs\;\ 
\bar\mu}\calL{\scs\bar\mu}{\Lb_3}{\,\Lb_4}$, where $\calL\Lb{\Lb'}{\Lb''}$ are 
the tensor product coefficients which appear in the decomposition 
$\bar R_\Lb{\otimes}\bar R_{\Lb'}\eq$\linebreak[0]$\bigoplus
_{\Lb''}$\linebreak[0]$\calL\Lb{\Lb'}{\Lb''}\bar R_{\Lb''}$ of tensor 
products of \gb-\rep s into irreducible sub\rep s. (For 
any value of the level, the \frc s $\N\L{\L'}{\L''}\eq\N\L{\L'}{\L''}(\kv)$ 
are majorized by the $\calL\Lb{\Lb'}{\Lb''}$, and $\N\L{\L'}{\L''}(\kv)\eq
\calL\Lb{\Lb'}{\Lb''}$ for \kv\ large enough. Hence one has
$\md\,{\ge}\,M(\kv)$ for all \kv, and $\lim_{\kV\to\infty}M(\kv)\eq\md$.)

With the expansion \Erf92, the \kze\ constitutes
an ordinary first order matrix \deq, reading
  \be -\onehalf\,(\kv+\gv)\,\dg_{\L_1\L_2\L_3\L_4;i}\zbz
  = \sum_{j=1}^{\md} \left(\Frac{P_{ij}}{z}+\Frac{Q_{ij}}{z-1}\right)
  \calg_{\L_1\L_2\L_3\L_4;j}\zbz\,, \Labl94
where the matrices $P$ and $Q$ are given by $P\eq\sum_{a,b=1}^{\dim\gB}
\kappb_{ab}\bar R_{\Lb_3}(\bar T^a){\otimes}\bar R_{\Lb_4}(\bar T^b)$ 
and $Q\eq\sum_{a,b=1}^{\dim\gB}\kappb_{ab}$\linebreak[0]%
$ \bar R_{\Lb_3}(\bar T^a){\otimes}\bar R_{\Lb_2}(\bar T^b)$.
When the basis of invariant tensors is chosen in such a way that they correspond
precisely to the various irreducible sub-\rep s in the tensor product
$\bar R_{\Lb_3}{\otimes}\bar R_{\Lb_4}$, then the matrix $P$ is diagonal.

Decoupling of the system \erf{94} leads to linear \deq s 
of order $\md$ for the components $\calg_{\L_1\L_2\L_3\L_4;i}$.
The solutions to these equations can be given in terms of contour integrals
which generalize the contour integral formula for hypergeometric functions.
This is in principle straightforward, but in practice it is rather tedious,
and accordingly it has been fully worked out only for the case
of $\gB\eq\sltwo$ and for some specific isolated \four s for which the
order $\md$ of the \deq\ is small.

\sect{Coset \CFTS}l\label{s25}%
The well-established \rep\ theory of \aff s makes \wzwts\ amenable to
detailed study. The {\em coset construction\/} exploits the 
\rep\ theory of affine algebras to investigate also more complicated models. 
The basic idea \Cite{goko2} is to consider embeddings 
$\gPP\emb\g$, where both Lie algebras are current \alg s at some level(s) \kv,
and to analyze the difference of the two Virasoro algebras \virg\ and 
\virgP\ (which are obtained via the Sugawara formula \erf{sug} 
for \g\ and \gP, \resp), i.e.\ the \lie\ with generators 
  \be  L_m^{\gM/\hM}:= L_m^\gM - L_m^\hM \,.  \ee

When the embedding of \gP\  into \g\ is induced by an embedding
$\gbP\emb\gb$ of the respective horizontal sub\alg s, then
these combinations commute with every generator $J^a_n$ of $\hM$, 
$[L^{\gM/\hM}_m,\tan]=0$, implying that
  $   [L^{\gM/\hM}_m, L^{\gM/\hM}_n]=  [L^\gM_m, L^\gM_n] - [L^\hM_m, L^\hM_n]$.
As a consequence the generators $L^{\gM/\hM}_n$ span a Virasoro algebra 
with central element $C_{\gM/\hM} = C_\gM -C_\hM$, called the {\em\cvira}.
It is, however, still an open problem whether for {\em any\/} embedding
$\gbP\emb\gb$ this prescription of a \vira\ can be complemented in such a way
that one arrives at a consistent \cft\ -- called the {\em coset theory\/} and
briefly denoted by `\,\ggp\,' -- and if so, whether that theory is unique.
To decide these questions, one must in particular construct
the (maximally extended) chiral algebra \wc\ of \ggp, as well as
the spectrum of the theory, i.e.\ tell which modules -- of
the coset chiral \alg\ \wc, or at least of the \cvira\ \virc\ -- appear.

As it turns out, to obtain the spectrum of the \ct\ is a somewhat delicate 
issue. While by construction the generators of the \cvira\ act on the chiral 
state space of the \wzwt\ based on \g, i.e.\ on the direct sum
${\cal H}_\gM = \bigoplus_\Lambda \hill$ of all inequivalent
irreducible highest weight modules $\hill$ of \g\ at level \kv,
${\cal H}_\gM$ is {\em not\/} the state space \hc\ of the coset theory.
The crucial observation is that due to $[\hM,\virc]\eq0$,
retaining the full state space ${\cal H}_\gM$ would imply that
the coset theory would possess (infinitely many) primary fields of zero \cdim\ 
other than the identity field, namely all the currents $J^a(z)$ of the 
subalgebra $\hM$ and all their \gP-descendants. 
In a field theoretic language this means that --
in the full theory, not just accidentally in some approximation to it --
the vacuum is not unique.\,%
\futnote{Note that at the level of the fusion ring, the vacuum 
gives the unital element, which must be unique.}
This disaster is avoided by requiring that these fields act trivially on \hc,
i.e.\ by imposing the gauge principle that
${\cal H}_\gM$-vectors which differ only by the action of $\U(\hM)$
merely provide different descriptions of the same physical
situation and hence represent one and the same vector in \hc. In short, the 
elements of \hc\ are $\U(\hM)$-{\em orbits\/} of vectors in ${\cal H}_\gM$ 
rather than individual vectors of ${\cal H}_\gM$.

Concretely, the restriction to $\U(\hM)$-orbits is implemented by decomposing
the \g-modules \hl\ into \gP-modules \hLP\ as\,%
\futnote{Here and below I denote quantities referring to the subalgebra \gP\
by the same symbol as the corresponding quantities for \g,
but with a prime added. In particular
$\L$ and $\L'$ stand for integrable highest weights of \g\ and \gP, \resp.}
  \be  \hl=\Bigoplus{\,\LambdaP} \hll\otimes\hLP\,, \ee
and the {\em branching spaces\/} $\hll$ appearing here
are natural candidates for the irreducible modules of the coset
theory. In terms of characters, this corresponds to regarding
the {\em branching functions} $\bll$ which appear in the decomposition
  \be \chii^\gM_\L(\tau) = \Sum{\lP}\bll(\tau)\, \chii^{\hM}_{\lP}(\tau)
  \labl{cc}
of the characters $\chii^\gM_\L$ of \g\ with respect to the characters 
$\chii^\hM_\LambdaP$ of $\hM$ as the characters of the coset theory. 

It follows immediately from \erf{cc} that
branching functions have a definite behavior under modular
transformations, namely the same as the irreducible characters of the tensor
product $\gM\oplus\hM^*$ of the \wzwt\ based on \g\ and a putative \cft\
(called the {\em complement\/} of the $\hM$ theory \Cite{scya6}) which 
behaves just like the \wzwt\ based on the \aff\ \gP\ except that, as indicated 
by the notation `\,*\,', it carries the complex conjugate \rep\ of \slz.
(Note that if $S$ and $T$ generate a unitary \rep\
of \slz, then so do $S^*$ and $T^*$.)

\sect{Field Identification}m\label{s26}%
As it turns out, even after fixing the gauge symmetry \gP\ there still remain
severe problems; they can be attributed to the fact that in general the 
redundancy symmetry of the coset theory is larger than the obvious gauge
symmetry \gP. It is instructive to observe this in
the example of the critical Ising model, which can
be realized with $\hM\eq A_1\Untw$ at level 2 embedded into 
$\gM\eq A_1\Untw{\oplus}A_1\Untw$ at levels 1. 
This model is a rational theory with $c\eq\onehalf$ 
(more generally, by taking levels $m{+}2$, 1 and $m{+}1$, \resp,
one obtains the whole minimal series \erf{mimo} of \vir);
hence its spectrum follows already from the representation theory of the
Virasoro algebra. Namely, there are three \pf s, and comparing their \cdim s 
$\cD\eq\bar\cD$ with those of the WZW primaries one finds that each of 
them can be realized by two distinct pairs $\Ilabel$ of weights:
  \be  \begin{tabular}{lrcc} \hline &&& \\[-.98em] \multicolumn{2}{c}
  {Field}         & $\cD$  & $\Ilabel$ \\&&&\\[-1.12em] \hline &&&\\[-.88em]
  identity field  &~$\!\!${\bf1} &  0     & $(0,0;0)$~,~$(1,1;2)$ \\[.25em]
  order parameter &~$\!\!\sigma$ &$\;\ 1/16\;\ $&$(0,1;1)$~,~$(1,0;1)$ \\[.25em]
  energy operator &~$\!\!\psi$   &  1/2   & $(1,1;0)$~,~$(0,0;2)$ \\[.25em]
  \hline \end{tabular} \Labl IM
Thus each field expected from the \rep\ theory of \vir\ seems to appear 
twice. On the other hand, half of the possible pairs $\Ilabel$, namely those
with $\L_1+\L_2+\L'\in2\zet{+}1$, do not correspond to any field of the
theory at all; inspection shows that these pairs possess vanishing branching 
functions.

This situation generalizes for arbitrary \cts. 
First, typically there are selection rules, i.e.\ the
branching functions for certain pairs $\Ilabel$ vanish. This observation is
in itself not too disturbing, as one just has to make sure to find
all selection rules.\,%
\futnote{But still, while empirically for most coset theories
all selection rules come from conjugacy class selection rules for the 
embedding $\hb\emb\gB$, so far no prescription for
enumerating all selection rules for every coset theory is known. 
Another possible reason for the vanishing of branching functions is
the occurrence of additional null states in the Verma modules. This indeed
happens for conformal embeddings, for which by definition \Cite{babo,scwa} 
the coset central charge vanishes, and presumably also for the closely
related maverick \Cite{dujo2} cosets. For all conformal embeddings,
the level of \g\ is $\kv\eq1$, while $\kv\eq2$ for all known maverick cosets.}
However, along with each selection rule there also comes a redundancy, i.e.\
non-vanishing branching functions for distinct pairs $\Ilabel$ turn 
out to be identical. In particular the putative vacuum sector
seems to occur several times. By the same argument which forced us to divide
out the action of the subalgebra \gP, one learns that this degeneracy
cannot be interpreted as the multiple appearance of a corresponding primary
field in the spectrum (in particular, it would then
not be possible to obtain the required modular transformation properties).
Rather, the correct interpretation is that a primary field of the coset theory
is not associated to an individual pair $\Ilabel$, but rather to an
appropriate equivalence class $\Ilabl$ of such pairs. This prescription has
been termed {\em field identification\/} (which is a bit of a misnomer, since
it is pairs of labels rather than \con fields that get identified).

Both selection rules and field identification can be described conveniently
via the concept of the {\em identification group\/} $\Gid$ \Cite{scya6}.
The elements $\JJ$ of $\Gid$ are simple currents of the tensor product 
theory $\g{\oplus}\hM^*$ of zero \cdim\ $\cd\J\mi\cd{\J'}$; they act on
pairs $\Ilabel$ via the fusion product. The selection rules are 
equivalent to the vanishing of the {\em monodromy charges\/} 
  \be  Q_\jj(\Ilabel)\df Q_\J(\L)-Q_{\J'}(\L') \ee
of any allowed branching function \wrt all $\JJ\iN\Gid$;
here $Q_\J(\L)$ is the combination $Q_\J(\L)=\cd\L+\cd\J-\cd{\J\star\L}$
(see equation \erf{moch}) of conformal weights. Moreover,
the equivalence classes in the field identification are precisely the orbits 
  \be  \ilabl:= \{\, \jlabel \mid \Mu\eq\J\L,\;\Mu'\eq\J'\L'\
  {\rm for\ some}\ \JJ\iN\Gid \,\}  \ee
of $\Gid$. Provided that all $\Gid$-orbits have a common length, taking one 
branching function out of each $\Gid$-orbit and combining them diagonally
yields a modular invariant spectrum. In this case the modular \smat\ is given by
  \be  S^{\gM/\hM}_{\ilabl,\jlabl} =S^{}_{\L,\Mu}\,(S'_{\LambdaP,\MuP})
  ^*_{}  \,,  \ee
where $\Ilabel$ and $\Jlabel$ are arbitrary representatives of the
orbits $\Ilabl$ and $\Jlabl$, \resp, i.e.\ is obtained by restricting the 
\smat\ $S\!\otimes\!(S')^*$ of $\g\!\oplus\!\hM^*$ to orbits.

\sect{Fixed Points}n\label{s27}%
The next degree of generality, and difficulty, is reached when $\Gid$-orbits
with different sizes appear. (Of course, all sizes are divisors of the size
of the orbit through $(0;\!0)$, on which $\Gid$ acts freely,
i.e.\ of the order $N\eq|\Gid|$ of $\Gid$.) Orbits of non-maximal size are 
referred to as {\em fixed points}. As soon as fixed points are present, taking 
precisely one representative out of each orbit $\Ilabl$ does not give a modular
invariant spectrum. On the other hand, the sesquilinear combination
  \be  Z = \sum_{[\Lambda;\lP] \,{:}\, Q=0}
  |\Gstab| \cdot |\!\!\! \sum_{(\J;\J')\in\GId/\Gstab} \!\!\!
  b_{(\J;\J')\star(\Lambda;\lP)}  \,|^2  \labl Z
of (non-zero) branching functions {\em is\/} modular invariant. 
Here $\Gstab$ denotes the {\em stabilizer subgroup\/} $\Gstab\subseteq
\Gid$ which consists of those elements of $\Gid$ that leave
$\Ilabel$ invariant. Now because of $G_{(0;0)}\eq\{(0;\!0)\}$,
$|\Gid|^2$ copies of the identity orbit $[(0;\!0)]$ appear in \erf Z, so that
one would like to divide \erf Z by this factor. 
Regarding the branching functions \bll\ that contribute to \erf Z
as irreducible \ggp-characters would then
lead to the conclusion that fractional multiplicities occur in $Z$, which does
not allow for any sensible interpretation of $Z$ as a partition function.

The difficulty in interpreting $Z$ \erf Z constitutes the {\em fixed point 
problem\/} of coset theories.
This is indeed a severe problem, as it seems to prevent one
from gaining control over the coset theory \ggp\ by completely understanding it
in terms of the underlying \wzwts\ based on \g\ and \gP.
The solution to this problem
(see \Cite{fusS4} or, for a condensed exposition, \Cite{fusS7};
earlier work is summarized in \Cite{scya6})
shows, however, that \ggp\ is in fact
still fully controlled by the affine \alg s \g\ and \gP, provided that one 
implements additional novel structures associated to \kma s,
the so-called {\em twining characters\/} and {\em \olie s\/}.\
The basic observation is that the coset
modules $\hll$ for fixed points are {\em not\/} irreducible and hence
a {\em fixed point resolution\/} into irreducible subspaces must be performed.
(For the full \twodim\ theory obtained by combining its two chiral 
halves, this means that {\em less\/} states are present than one would naively
expect.) These submodules turn out to be the simultaneous eigenspaces of 
certain finite order automorphism of \hll\ that are associated to every
$\JJ\iN\Gstab$ and hence are labelled by the group-characters $\Psi$ of 
$\Gstab$; their characters can be related to the branching 
functions for a coset construction of \olie s via Fourier transformation
\wrtt abelian group of $\Gstab$-characters.
This results in particular in a formula which expresses the \smat\ of \ggp\ as 
  $$   \hsp{-.25} S^{\gM/\hM}_{(\ilabl,\Psi),(\jlabl,\tilde\Psi)} \eq
  \Frac{|\GId|}{|\Gstab|\cdot|\Gstabb|} \hsp{-1.67}\sum_{\scriptstyle\jj\in
  \hsp{1.65} \atop \scriptstyle \hsp{1.4}\Gstab\cap\Gstabb}\hsp{-2.3}
  \overline{\Psi(\JJ)}\, S^{\jjj}_{\ilabel,\jlabel} \tilde\Psi(\JJ)  
  $$  \mbox{}\\[-2.4em] \be {} \ee
in terms of the group characters $\Psi$ of stabilizer subgroups of
\Gid\ and of certain modular transformation
matrices $S^{\jjj}$ which are products of \smats\ of the relevant \olie s.

In closing this last regular \sec, I would like to reiterate the remark made
earlier that while in principle
it is desirable to work with observable quantities only,
the use of non-observable objects often proves to be convenient. The situation 
at hand provides an outstanding example for such an approach, in that extremely
non-trivial effects can be detected without even knowing precisely what
the observables \wc\ (except for the subalgebra \virc) look like.

\lecT{Addenda}{17.6}{12.35}

\sect{Omissions}Y\label{s28}%
Lack of time forced me to omit a number of topics which should
definitely occur in a decent introduction to \cft\ and to \kma s.
To partially compensate this shortcoming, I give a brief list of some
of these issues and point out a few relevant references.
Concerning citations, I decided to sacrifice
historical accuracy for up-to-dateness, so I usually refer either to reviews
or to the most recent publications I am aware of, which can serve as a
guide to further literature. (This remark applies likewise to most of
the references that I already mentioned in the main text.)
\blec1
\nXt more on the \rep\ theory of the \vira\ \Cite{gape3,Scho,Sche2};
\nXt classification and properties of chiral algebras \Cite{Bosc,Watt};
\nXt vertex operator algebras \Cite{geni,Kac};
\nXt more on chiral blocks and chiral vertex operators \Cite{Algs,Mose,crgr};
\nXt \cft\ and higher genus Riemann surfaces \Cite{Mose,bave,fewi3};
\nXt the $C^*$-\alg ic approach to \cft\ \Cite{frrs2,gafr,wass2,bock3};
\nXt `logarithmic' \con theories \Cite{gaKa,floh5}.
\blec2
\nXt more on duality, crossing symmetry, and polynomial equations 
     \Cite{Algs,Mose};
\nXt the \rep\ theory and the classification of fusion rings 
     \Cite{FRke,jf24,ehol3};
\nXt the \role\ of quantum groups, or more general, quantum symmetries
     \Cite{Algs,scho3,fugv3,reck2};
\nXt classification of modular invariants \Cite{krSc,gann10,fusS2,garw};
\nXt free fermion \Cite{Gool,FUch} and orbifold \Cite{bant4,kaTo2,bohs} \cfts;
\nXt Galois symmetry \Cite{bcldb,fusS2,fuSc3};
\nXt more on the classification of $c=1$ and related \cfts\ \Cite{fupt,floh}
     and on\\{}${}$\hsp{.67} the Coulomb gas description of $c\,{<}\,1$
     theories \Cite{Fust,Sche2}.
\blec3
\nXt more on \lie s with triangular decomposition \Cite{KAc3,FUha,MOpi,FUsc};
\nXt realizations of af\-fi\-ne \lie\ \rep s via free fermions or via
     vertex\\{}${}$\hsp{.67} operators \Cite{Gool,KAc3,dogm3,geni3}, and through
     more complicated free field constructions \Cite{defe};
\nXt character formul\ae\ and combinatorial identities \Cite{KAc3,borC4};
\nXt generalized \kma s and their \role\ in string theory \Cite{geni,hamo}.
\blec4
\nXt the affine-Virasoro construction which generalizes the Sugawara 
     formula \Cite{Hkoc,haob3};
\nXt Lagrangian formulations of WZW and \cts\ \Cite{hori,rhed2,schW3};
\nXt Knizhnik\hy Zamo\-lod\-chi\-kov\hy Ber\-nard equations 
     \Cite{fewi3,faga3,ivaN2};
\nXt \wzwts\ which have fractional level \Cite{maSw,fgpp2,pery3} or are
     based on non-reductive\\{}${}$\hsp{.67} \lie s \Cite{fist3,arpa};
\nXt \hamred\ of \wzwts\ \Cite{fortw3,gape3,blls};
\nXt the relevance of \kze s to moduli spaces of flat connections, 
     \\{}${}$\hsp{.67} to quasi-Hopf \alg s and to integrable systems
     \Cite{ivaN2,alrs2}.

\sect{Outlook}Z\label{s29}%
It is often claimed (implicitly or explicitly) that in \cft\ all interesting 
problems are already solved. This statement is \bullshit\ (as is e.g.\ 
indicated by the large number of recent publications quoted above);
some examples of open questions are the following.
\nxt
A complete understanding of the \role\ of modular invariance, in particular in
view of the presence of similar structures in arbitrary rational \twodim\ 
quantum field theories (cf.\ e.g.\ \Cite{rehr4,bave,muge});
\nxt
a mathematically rigorous proof of the \vf\ for general \cfts, not just for
(diagonal) \wzwts;
\nxt
structural information on fusion rules of non-quasirational \twodim\
field theories (e.g.\
existence of vector space bases rather than only topological bases), and
the possible description of such theories as deformations or limits of 
quasirational theories;
\nxt
the quantum field-theoretic interpretation of non-unitary, and in particular
of logarithmic \cfts\ (from a quantum field theoretic perspective it is
puzzling that non-unitary theories can be treated in much the same setting 
as unitary ones; e.g.\ \h\ is often fully reducible, even though 
positivity of an inner product can no longer be invoked);
\nxt
the classification of modular invariants for all \wzwts\ -- as well as deeper
insight into its guiding principles (note that even the $A$-$D$-$E$ pattern that
arises for $\gb\eq\sltwo$ essentially emerges in a technical rather than
a conceptual manner);
\nxt
the \role\ of `\asis' that appear in the differential equations for chiral 
blocks \Cite{jf21} as possible parameters in the classification of \rcfts; 
\nxt
simple current and Galois symmetries of braiding and fusing matrices, and
a complete analysis of the behavior of fusing matrices under
`tetrahedral' transformations in the presence of \frc s larger than one;
\nxt
a more substantial classification of modular fusion rings;
\nxt
an efficient algorithm for solving the polynomial equations for prescribed
fusion rules;
\nxt
a description of \cft\ on surfaces with boundaries \Cite{card,prss3}
at the same level of sophistication as for theories on closed surfaces;
\nxt
a complete description of the \sps\ for arbitrary coset theories, in
particular a characterization and enumeration of all maverick cosets.

\newpage\label{S.X}
\noindent{\large\bf\thetoce~~Glossary}\\[1.5ex] \footnotesize
\noindent
In the following list of keywords the quoted numbers refer to the \sec(s)
in which the term is\\ introduced or plays a \role.

\noindent\begin{tabular}{lr}
\glO16 {\aff}
\glo3  {annihilation operator}
\glo2  {anti-holomorphic chiral \alg}
\glO14 {\asi}
\glO11 {$A$-type modular invariant}
\glO20 {BGG resolution}
\Glo56 {bootstrap}
\glo9  {braiding matrix}
\glo1  {$C^*$-\alg ic approach}
\glO15 {\carm}
\glO15 {\csa}
\glO13 {center of a \cft}
\Glo2{16} {central element}
\glO17 {central extension}
\glo9  {channel}
\GlO1020 {character}
\glo2  {chiral \alg}
\GLo79{14} {chiral block}
\glo2  {chiral half}
\glo7  {\cvo}
\Glo24 {conformal anomaly}
\glo3  {\cdim}
\glO26 {conformal embedding}
\glo5  {conformal field}
\glO11 {conformal spin}
\glo3  {conformal weight}
\glo7  {conjugate field}
\Glo56 {contour integral}
\glo6  {contraction}
\glO15 {coroot}
\glo7  {\corfu}
\glO25 {coset construction}
\glO25 {coset theory}
\glo3  {creation operator}
\glo1  {critical points}
\glo9  {crossing symmetry}
\glo7  {cross-ratio}
\glO22 {current}
\glO21 {current \alg}
\glO20 {denominator identity}
\glO16 {derivation}
\glo5  {descendant}
\glO11 {diagonal modular invariant}
\glO21 {dual Coxeter number}
\glO15 {\dyd}
       \end{tabular}\hsp{3.1}\begin{tabular}{lr}
\glo5  {\emt}
\glo3  {energy operator}
\glo5  {family}
\glo5  {field}
\glO26 {field identification}
\GlO1327 {fixed point resolution}
\glo5  {formal variable}
\glo5  {Fourier\hy Laurent coefficient}
\Glo9{14} {\fpf}
\glO12 {free boson}
\glO15 {fundamental weight}
\glo9  {fusing matrix}
\glo8  {fusion \alg}
\Glo8{13} {fusion ring}
\glo8  {\frc}
\glo8  {\furu s}
\glO15 {generalized \carm}
\glo5  {generating function}
\Glo35 {grade}
\glO12 {Heisenberg \alg}
\glo9  {hexagon equation}
\Glo3{19} {\hwm}
\glo3  {\hwv}
\glo2  {holomorphic chiral \alg}
\glO26 {identification group}
\glo5  {identity field}
\glo7  {insertion point}
\glo3  {\ihwm}
\GlO1516 {\kma}
\glO23 {\kpf}
\glO23 {\kwf}
\glO24 {\kze}
\glo5  {Laurent coefficient}
\glo2  {\lie}
\glo2  {locally nilpotent}
\glO17 {loop \alg}
\glo2  {L\"uscher\hy Mack theorem }
\glO26 {maverick coset}
\glO19 {maximal weight}
\Glo4{12} {minimal model}
\glo4  {M\"obius transformation}
\glo5  {mode}
\glO11 {modular group}
\glO11 {modular \smat}
\glO11 {modular \tmat}
       \end{tabular}\newpage\noindent\begin{tabular}{lr}
\glO11 {modular transformation}
\glo3  {module}
\glO13 {monodromy charge}
\glo4  {naturality}
\Glo56 {normal ordering}
\Glo3{19} {null vector}
\glo2  {observable}
\glo6  {\opa}
\glo6  {\opc}
\glo6  {\ope}
\glO12 {orbifold}
\glO13 {order of a simple current}
\glo1  {particle contents}
\GlO1011 {partition function}
\glo9  {pentagon equation}
\glo1  {perturbation theory}
\glo3  {physical state}
\glo9  {polynomial equations}
\glo5  {\pf}
\glo4  {projective transformation}
\glo7  {projective Ward identity}
\glo1  {quantum field theory}
\glo5  {\qpf}
\glo4  {quasi-primary vector}
\glo8  {\qrcft}
\glo6  {radial ordering}
\Glo48 {\rcft}
\glO14 {Riemann monodromy problem}
\glO14 {Riemann scheme}
\glO15 {root}
\glo5  {secondary field}
\glo4  {sector}
\glO13 {simple current}
\glO13 {simple current extension}
\glO15 {simple root}
       \end{tabular}\hsp{3.1}\begin{tabular}{lr}
\glO11 {\smat}
\glo4  {spectrum}
\glo3  {spectrum condition}
\glO27 {stabilizer subgroup}
\glo5  {state-field correspondence}
\glo3  {state space}
\Glo14 {string theory}
\glO21 {Sugawara form}
\glo4  {superselection sector}
\glO15 {symmetrizable \kma}
\glO11 {T-matrix}
\Glo3{18} {triangular decomposition}
\GlO1617 {twisted \aff}
\glO17 {twisted boundary conditions}
\glo3  {unitary \rep}
\GlO1617 {untwisted \aff}
\Glo56 {vacuum expectation value}
\glo4  {vacuum sector}
\glo4  {vacuum state}
\Glo3{19} {Verma module}
\Glo56 {\voa}
\glo2  {\vira}
\glO10 {Virasoro-specialized character}
\glo7  {Ward identity}
\Glo3{19} {weight}
\glO15 {weight space}
\glO10 {weight system}
\glO15 {Weyl group}
\glO20 {Weyl\hy Kac character formula}
\glO20 {Weyl vector}
\Glo12 {Wightman field theory}
\glo2  {Witt \alg}
\glO21 {\wzwt}
\glo3  {zero mode sub\alg}
\glo{}{}
\end{tabular}

\vfill \noindent
{\small{\bf Acknowledgements.}\\[.3em] I am grateful to K.\ Blaub\"ar, 
B.\ Ha\ss ler, C.\ Schweigert and P.\ Vecserny\'es\\ for suggesting various
improvements to the manuscript.\\[4ex]{}}

 \def\wb{\,\linebreak[0]} \def\wB {$\,$\wb}
 \newcommand\Erra[3]  {\,[{\em ibid.}\ {#1} ({#2}) {#3}, {\em Erratum}]}
 \newcommand\BOOK[4]  {{\em #1\/} ({#2}, {#3} {#4})}
 \newcommand\inBO[7]{{\sl #7}, in:\ {\em #1}, {#2}\ ({#3}, {#4} {#5}), p.\ {#6}}
 \renewcommand\J[5]   {{\sl #5}, {#1} {#2} ({#3}) {#4} }
 \newcommand\Prep[2]  {{\sl #2}, preprint {#1}}
 \def\jf    {J.\ Fuchs}
 \def\jF    {Fuchs}
 \def\adma  {Adv.\wb Math.}
 \def\anip  {Ann.\wb Inst.\wB Poin\-car\'e}
 \def\comp  {Com\-mun.\wb Math.\wb Phys.}
 \def\foph  {Fortschr.\wb Phys.}
 \def\ijmp  {Int.\wb J.\wb Mod.\wb Phys.\ A}
 \def\inma  {Invent.\wb math.}
 \def\joag  {J.\wB Al\-ge\-bra\-ic\wB Geom.}
 \def\jodg  {J.\wb Diff.\wb Geom.}
 \def\jomp  {J.\wb Math.\wb Phys.}
 \def\mpla  {Mod.\wb Phys.\wb Lett.\ A}
 \newcommand\npbF[5] {{\sl #5}, \nupb\ {#1} [FS{#2}] ({#3}) {#4}}
 \def\npbp  {Nucl.\wb Phys.\ B (Proc.\wb Suppl.)}
 \def\nuci  {Nuovo\wB Cim.}
 \def\nupb  {Nucl.\wb Phys.\ B}
 \def\phlb  {Phys.\wb Lett.\ B}
 \def\phrd  {Phys.\wb Rev.\ D}
 \def\prep  {Phys.\wb Rep.}
 \def\rmap  {Rev.\wb Math.\wb Phys.}
 \def\slnm  {Lecture\wB Notes\wB in\wB Mathematics}
 \def\somd  {Sov.\wb Math.\wb Dokl.}
 \newcommand\gxxi[2] {{\sl #2}, {#1}, to appear in the \icgtmp{XXIth}, {H.-D.\
            Doebner and W.\ Scherer, eds.} }
 \renewcommand\gxxi[2] {{#1}, to appear in the \icgtmp{XXIth}, {H.-D.\ Doebner
            and W.\ Scherer, eds.} }
 \newcommand\icgtmp[1] {{\rm Proceedings of the {#1} International
            Colloquium on} {\em Group Theoretical Methods in Physics}}
 \newcommand\kniz[2] {\inBO{The Physics and Mathematics of Strings}
            {L.\ Brink et al., eds.} \WS\Si{1990} {{#1}}{{#2}} }
 \newcommand\ntqf[2] {\inBO{New Trends in Quantum Field Theory}
            {A.Ch.\ Ganchev, R.\ Kerner and I.\ Todorov, eds.}
            {Heron Press}{Sofia}{1996} {{#1}}{{#2}} }
 \newcommand\phgt[2] {\inBO{Physics, Geometry, and Topology}
            {H.C.\ Lee, ed.} \PL\NY{1990} {{#1}}{{#2}} } 
 \newcommand\Suse[2] {\inBO{The Algebraic Theory of Superselection
            Sectors.\ Introduction and Recent Results} {D.\ Kastler,
            ed.} \WS\Si{1990} {{#1}}{{#2}} }
\def\A      {Algebra}
\def\and    {and }
\def\And    {~$\!$and$\!$~}
\def\AnD    {~$\!$and }
\def\AND    { and }
\def\AP     {{Academic Press}}
\def\AW     {{Addi\-son\hy Wes\-ley}}
\def\Be     {{Berlin}}
\def\Ca     {{Cambridge}}
\def\Cfts   {Conformal field theories}
\def\CS     {Chern\hy Si\-mons\ }
\def\cua    {current algebra}
\def\CUP    {{Cambridge University Press}}
\def\Dif    {Differential}
\def\dimn   {dimension}
\def\eq     {equation}
\def\etal   {~$\!$et~$\!$al.$\!$~}
\def\etaL   {~$\!$et~$\!$al. }
\def\etAL   {~$\!$et al.\ }
\def\ETAL   { et al.\ }
\def\fts    {field theories}
\def\hopf   {Hopf algebra}
\def\ide    {identification}
\def\JW     {{John Wiley}}
\def\mimo   {minimal model}
\def\modinv {modular invarian}
\def\nn     {$N\,{=}\,2$\ }
\def\NY     {{New York}}
\def\P      {~$\!$}
\def\parfu  {partition function}
\def\PL     {{Plenum}}
\def\qhe    {quantum Hall effect}
\def\qzn    {quantization}
\def\Si     {{Singapore}}
\def\SV     {{Sprin\-ger Verlag}}
\def\stc    {statistic}
\def\susic  {supersymmetric\ }
\def\sym    {symmetry}
\def\syms   {sym\-me\-tries}
\def\va     {Vira\-soro algebra}
\def\vop    {vertex operator}
\def\WS     {{World Scientific}}
\def\WZ     {Wess\hy Zu\-mino\ }
\def\wzw    {WZW}
\def\WZW    {Wess\hy Zu\-mi\-no\hy Wit\-ten }
\newcommand\Bi[3]    {\bibitem{#2}}
\newcommand\BI[3]    {\bibitem{#2}}
\newcommand\au[2]    {#1.\ #2,}
\newcommand\av[2]    {#1.\ #2}
\newcommand\aU[2]    {#1.\ #2,}
\newcommand\Au[3]    {#1.#2.\ #3,}
\newcommand\Av[3]    {#1.#2.\ #3}
\newcommand\AU[3]    {#1.#2.\ #3,}
\newcommand\yr[1]    {}
\def\Jf    {J.\ Fuchs, }
\def\tai   {to appear in }

\end{document}